\documentclass[11pt]{article}
\usepackage{amssymb,amsmath,pstricks}
\usepackage{graphics,graphicx}
\usepackage{setspace}

\renewcommand{\textit}{}
\makeatletter\@addtoreset {equation}{section}\makeatother

\def\bexa{\begin{example}}\def\eexa{\eex\end{example}}
\def\brem{\begin{remark}}\def\erem{\eex\end{remark}}
\def\bthm{\begin{theorem}}\def\ethm{\end{theorem}}
\def\blem{\begin{lemma}}\def\elem{\end{lemma}}
\def\bcor{\begin{corollary}}\def\ecor{\end{corollary}}
\def\bdefi{\begin{definition}}\def\edefi{\end{definition}}
\def\beq{\begin{equation}}\def\eeq{\end{equation}}

\newcommand{\R}{{\mathbb R}}\newcommand{\C}{{\mathbb C}}
\newcommand{\N}{{\mathbb N}}\renewcommand{\P}{{\mathbb P}}
\newcommand{\T}{{\mathbb T}}
\newcommand{\Z}{{\mathbb Z}}

\def\hA{\hat{\bf A}}\def\bA{{\bf A}}\def\hB{\hat{\bf B}}\def\hb{\hat{\bf b}}
\def\Ah{\hat{A}}
\def\xrev{X_{{\rm rev}}}
\def\hJ{\hat{{\bf J}}}\def\bJ{{\bf J}}
\def\hN{\hat{{\bf N}}}\def\hF{\hat{{\bf F}}}
\def\hR{\hat{{\bf R}}}

\def\CT{{\cal T}}
\def\CA{{\cal A}}\def\CD{{\cal D}}  
\def\CF{{\cal F}}
\def\CO{{\cal O}}
\def\CN{{\cal N}}
\def\CX{{\cal X}}

\def\Bsm{{\scriptscriptstyle B}}

\def\ga{\gamma}\def\om{\omega}
\def\noi{\noindent}\def\ds{\displaystyle}
\def\pa{{\partial}}

\def\Lti{\tilde{L}}
\newcommand{\bi}{\begin{itemize}}\newcommand{\ei}{\end{itemize}}
\newcommand{\ben}{\begin{enumerate}}\newcommand{\een}{\end{enumerate}}
\newcommand{\bce}{\begin{center}}\newcommand{\ece}{\end{center}}
\newcommand{\reff}[1]{(\ref{#1})}
\newcommand{\mbf}[1]{{\mathbf{#1}}}\newcommand{\ov}[1]{{\overline {#1}}}
\newcommand{\nonu}{\nonumber}
\newcommand{\spr}[1]{\left\langle #1 \right\rangle}

\def\eps{\varepsilon}
\def\ra{\rightarrow}\def\Ra{\Rightarrow}

\newcommand{\barr}{\begin{array}}\newcommand{\earr}{\end{array}}
\newcommand{\bpm}{\begin{pmatrix}}\newcommand{\epm}{\end{pmatrix}}
\newcommand{\ba}{\begin{array}}\newcommand{\ea}{\end{array}}
\def\dd{\, {\rm d}}\def\ri{{\rm i}}

\def\er{{\rm e}}

\def\om{\omega}\def\Om{\Omega}

\def\del{\delta}
\def\supp{{\rm supp}}
\def\ba{\begin{array}} \def\ea{\end{array}}
\def\bgat{\begin{gather}} \def\egat{\end{gather}}
\def\eps{\varepsilon} 
\def\phih{\hat{\phi}}\def\psih{\hat{\psi}}

\def\eex{\hfill\mbox{$\rfloor$}}
\def\phiti{\tilde{\phi}}
\def\sig{\sigma}\def\al{\alpha}

\def\wt{\widetilde}
\def\vtp{\vec{\wt{\phi}}}

\newcommand{\bspm}{\left(\begin{smallmatrix}}\newcommand{\espm}{\end{smallmatrix}\right)}

\def\psils{\tilde\psi_{{\rm LS}}^{(0)}}
\def\etals{\tilde\eta_{{\rm LS}}^{(0)}}

\def\etalsv{\vec{\tilde{\eta}}_{{\rm LS}}^{(0)}}
\def\nab{\nabla}\def\eex{\hfill\mbox{$\rfloor$}}
\def\etls{\eta_{{\rm LS}}^{(0)}}

\newtheorem{theorem}{Theorem}[section]
\newtheorem{definition}[theorem]{Definition}

\newtheorem{lemma}[theorem]{Lemma}

\newtheorem{remark}[theorem]{Remark}
\newtheorem{corollary}[theorem]{Corollary}
\newtheorem{example}[theorem]{Example}

\newenvironment{proof}{
    \noindent {\it Proof.}}{\hfill$\Box$
}

\begin{document}

\title{\bf Coupled Mode Equations and Gap Solitons for the 2D Gross-Pitaevskii
  equation with a non-separable periodic potential}

\author{Tom\'{a}\v{s} Dohnal$^1$ and Hannes Uecker$^2$\\
  {\small $^1$ Institut f\"{u}r Angewandte und Numerische Mathematik 2,
    Universit\"{a}t Karlsruhe, Germany}\\ {\small $^2$ Institut f\"{u}r
    Mathematik, Carl von Ossietzky Universit\"{a}t Oldenburg, Germany} }

\date{\today}
\maketitle

\begin{abstract}
  Gap solitons near a band edge of a spatially periodic nonlinear PDE can be
  formally approximated by solutions of Coupled Mode Equations (CMEs). Here we
  study this approximation for the case of the 2D Periodic Nonlinear
  Schr\"{o}dinger / Gross-Pitaevskii Equation with a non-separable 
  potential of
  finite contrast. We show that unlike in the case of separable potentials
  [T. Dohnal, D. Pelinovsky, and G. Schneider, J. Nonlin. Sci. {\bf 19}, 95--131 (2009)]
  the CME derivation has to be carried out in Bloch rather than physical
  coordinates. Using the Lyapunov-Schmidt reduction we then give a rigorous
  justification of the CMEs as an asymptotic model for reversible
  non-degenerate gap solitons and even potentials and provide $H^s$ estimates for this
  approximation. The results are confirmed by numerical examples including
  some new families of CMEs and gap solitons absent for separable potentials.
\end{abstract}

\section{Introduction}

Coherent structures, like gap solitons, in nonlinear periodic wave propagation
problems are important both theoretically and in applications. 
Typical examples include optical waves in photonic
crystals and matter waves in Bose-Einstein condensates loaded onto optical
lattices. A standard model in these contexts is the 
Nonlinear Schr\"{o}dinger/Gross-Pitaevskii equation
with a periodic potential, which applies in Kerr-nonlinear photonic crystals
\cite{SK02,Efre03,KEVT06,MK01,FTNSKDK06} as well as in Bose-Einstein condensates loaded onto
an optical lattice \cite{Gross63,Pitaev61,OK03}. Here we consider the case of two
spatial dimensions and without loss of generality take the potential
$2\pi$--periodic in both directions and, hence, consider
\begin{equation}\label{E:GP}
  \ri E_t=-\Delta E +V(x)E+\sigma|E|^2 E, \quad V(x_1{+}2\pi,x_2)
  =V(x_1,x_2{+}2\pi)=V(x), \ x\in\mathbb{R}^2,\ t\in\R, 
\end{equation}
with $E=E(x,t)\in\C$, $\sig=\pm 1$ and $V\in H^m_{{\rm loc}}(\R^2), m>1, m \in \R$.

We are interested in stationary \textit{gap solitons} (GSs) $E(x,t) =
\phi(x)e^{-\ri  \omega t}$. Thus $\phi$ solves
\begin{equation}\label{E:GP_stat}
  (-\Delta +V(x)-\omega)\phi +\sigma |\phi|^2\phi=0, 
\end{equation} 
where soliton is understood in the sense of a solitary wave, which means that
$|\phi(x)|\ra 0$ exponentially as $|x|\ra 0$. 
Necessarily, then $\omega$
has to lie in a gap of the essential spectrum of the operator $L:=-\Delta
+V(x)$, hence the name ``gap soliton.'' 
From the phenomenological and experimental
point of view multidimensional GSs have 
been widely studied in the context of both photonic crystals 
\cite{AJ98,MK01,Efre03,FSECh03,DA05,FTNSKDK06} and Bose-Einstein condensates 
\cite{AAG07,OK03,BMS03}.

The essential spectrum of $L$ is given by the so called band structure, 
and for our analysis we choose $\omega$ close to a band edge, 
i.e., $\omega
= \omega_* + \eps^2 \Omega, \ 0<\eps \ll 1$, where $\omega_*$ is an edge of a
band gap and $\Omega$ has a sign chosen so that $\omega$ lies inside the gap.
Using a multiple scales expansion one may formally derive coupled mode
equations (CMEs) to approximate envelopes of the gap solitons near gap
edges. CMEs are a constant coefficient problem formulated in slowly varying
variables. They are, therefore, typically more amiable to analysis and 
also cheaper for
numerical approximations compared to the original system \eqref{E:GP_stat}.
The multiple scales approach has been used both for the Gross-Pitaevskii and
Maxwell equations with infinitesimal, i.e. $\CO(\eps)$, contrast in the
periodicity $V(x)$ \cite{ACD95,AJ98,SdSE02,AFI04,AP05,DA05} as well as with
finite contrast \cite{dSSS96,SY07,DPS08}. The main difference in the
asymptotic approximation of the two cases is that for infinitesimal contrasts
the expansion modes are Fourier waves while for finite contrast they are Bloch
waves. However, in dimension two and higher sufficiently
large (finite) contrast is necessary to generate band gaps due to 
overlapping of bands in the corresponding homogeneous medium. 
The only exception is the
semi-infinite gap of the band structure of $L$ of the Gross-Pitaevskii
equation. As a result, gap solitons in finite gaps of the Gross-Pitaevskii
equation and in any gap of Maxwell systems in dimensions two and
higher can only exist for finite contrast structures.

Localized solutions of CMEs \textit{formally} yield gap solitons 
of the original system. 
However, the formal derivation of the CMEs, 
discarding some error at higher order in $\eps$, does
not imply that all localized solutions of the CMEs yield gap solitons.  
For this 
we need to estimate the error in some function space and to show the
persistence of the CME solitons under perturbation of the CMEs. A famous 
result concerning non--persistence is the non-existence of breathers 
in perturbations of the sine--Gordon equation, e.g., \cite{De93}. 
On the other hand, GSs are known to exist in every band gap of 
$L$, see, e.g.~\cite{Stuart95,Pankov05}. The proofs, however, 
are based on variational methods and do not relate GSs to 
solutions of the CMEs. 

\begin{figure}[h!]
  \begin{center}
    \includegraphics[height=5.4cm,width=7.5cm]{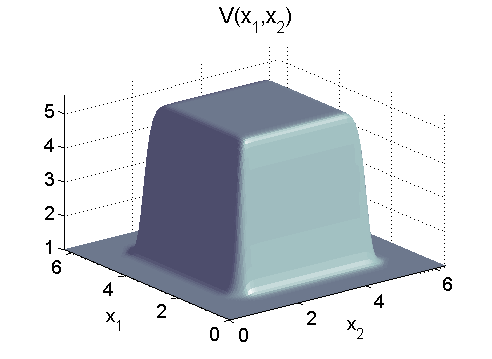}
  \end{center}
  \caption{The periodic potential $V$ in \eqref{E:V} over the Wigner-Seitz
    cell.}
  \label{F:V}
\end{figure}

A rigorous justification of the CMEs has been given for
\eqref{E:GP_stat} in 1D in \cite{PSn07}, and in 2D in \cite{DPS08}, 
but only for the case of a separable potential
$$
V(x_1,x_2)=W_1(x_1)+W_2(x_2).
$$
Here we transfer this result to not necessarily separable potentials, 
where we need some minimal smoothness, namely 
$V\in H^m_{{\rm loc}}(\R^2), m>1$.  As an
example we choose
\begin{equation}\label{E:V}
  V(x_1,x_2)=1+(\eta-1)W(x_1)W(x_2), \quad
  W(s)=\frac{1}{2}\left[\tanh\left(7\left(s-\frac{2\pi}{5}\right)\right)+\tanh\left(7\left(\frac{8\pi}{5}-s\right)\right)\right], 
\end{equation}
which represents a square geometry with smoothed-out edges (Fig. \ref{F:V}). 
We choose the contrast
$\eta$ in $V$ so that two finite gaps appear in the band structure of the
corresponding linear eigenvalue problem. 
One main difference between the separable and non-separable
case lies in the fact that for the non-separable case band edges may be
attained at wavenumbers not within the set of vertices of the first
irreducible Brillouin zone. Then the CME derivation and
justification is \textit{impossible} to carry out \textit{in physical
  variables} and has to be performed in \textit{Bloch variables}. 
This case occurs at least at one band edge of the potential
\eqref{E:V}, and the presented CMEs corresponding to this edge have, to our
knowledge, not been studied before. Similarly, the GSs which we show to
bifurcate from this edge are new.


In \S\ref{S:band_str} we discuss in detail the band structure for \reff{E:V}
and the associated Bloch eigenfunctions, together with their symmetries. Then
in \S\ref{S:derive_CME} we first give the formal derivation of the CME in
physical space, reporting a failure in one case where the band edge is
attained simultaneously at four wave numbers outside the set of vertices of
the first Brillouin zone, and present a general CME derivation in Bloch
variables.  The existence of gap solitons is proved in \S\ref{S:justification}
based on the existence of special (namely reversible and 
non-degenerate, see below) localized solutions of 
the CMEs, in the following sense. 

\noi
{\bf Function spaces and notation.}  
For $m\in\N$, the  Sobolev spaces $H^m(\R^2)$ are classically defined as 
$H^m(\R^2):=\{u\in L^2(\R^d): \pa_x^\al u\in L^2(\R^2)\text{ for } 
|\al|\le m\}$, with norm $\|u\|_{H^s}=\left(\sum_{|\al|\le m}\|\pa_x^\al u\|^2_{L^2}\right)^{1/2}$, 
where $\pa_x^\al u$ denotes the distributional derivative, see, e.g., 
\cite{AdamsFour03}. 
Then, for $s=m\in\N$, Fourier transform 
\begin{gather}\label{E:FT}
\hat{\phi}(k):= (\CF \phi)(k):=\frac{1}{(2\pi)^2}\int_{\R^2}\phi(x)e^{-\ri k\cdot x}\dd x, \quad
\phi(x)=(\CF^{-1}\hat{\phi})(k):=\int_{\R^2}\hat{\phi}(k)e^{\ri k\cdot x}\dd k, 
\end{gather}  
is an isomorphism from $H^s(\R^2)$ to 
\beq\label{figo}
L^2_s(\R^2):=\{\hat\phi\in L^2(\R^2): \|\hat{\phi}\|_{L^2_s}:=\|(1+|k|)^s\hat\phi\|_{L^2}<\infty\}, 
\quad\text{i.e. } C_1\|\hat{\phi}\|_{L^2_s}\le \|\phi\|_{H^s}\le 
C_2\|\hat{\phi}\|_{L^2_s}.
\eeq 
From the applied point of view we could restrict to integer $s$. However, 
since our analysis is strongly based on Fourier transform, it is 
conceptually cleaner to use a definition of Sobolev spaces based on 
$L^2_s$ with arbitrary $s\ge 0$. 
Thus, henceforth we use $H^s(\R^2):=\CF^{-1}L^2_s(\R^2)$ as definition 
for $0\le s\in \R$. This also gives a very simple proof of the 
Sobolev embedding theorem $\|\phi\|_{C^k}\le C\|\phi\|_{H^s}$ for 
$k<s-1$, see Lemma \ref{sob-lem} below. 

\noi{\bf Main result.} 
Let $s\geq 2$ and 
$V\in H^{\lceil s\rceil-1+\delta}_{{\rm loc}}(\R^2), \delta>0$, 
where $\lceil s\rceil$ is the smallest integer larger than or equal to $s$, 
and $V$ even in $x_1,x_2$. Let $\bA=(A_1,\ldots,A_N)$ 
be a  reversible non-degenerate localized 
solution of the CMEs with $\bA\in [H^q(\R^2)]^N$ for all $q\ge 0$. 
Then for $\omega = \omega_*+\eps^2\Omega$ with $\eps^2$ sufficiently small
there exists a GS $\phi_{GS}$ for \reff{E:GP_stat}, such that  
$\phi_{GS}\in H^s(\R^2)$, and $ \phi_{GS}$ 
can be approximated by 
\beq \label{as1} 
\eps\phi^{(0)}=\eps \sum_{j=1}^N A_j(\eps x)u_{n_j}(k^{(j)};x), 
\end{equation}
where $u_{n_j}(k^{(j)};x)\in H^{\lceil s\rceil+1+\delta}_{{\rm loc}}(\R^2)$ are the pertinent 
Bloch waves, $j=1,\ldots,N$. In detail, we prove 
\beq\label{as2}
\|\phi_{GS}-\eps\phi^{(0)}\|_{H^s(\R^2)}\le C\eps^{2/3}, 
\eeq
where the estimate can be improved in special cases, see below. 

Note that $\|\eps\phi^{(0)}\|_{L^\infty(\R^2)}=\CO(\eps)$ 
but $\|\eps\phi^{(0)}\|_{L^2(\R^2)}=\CO(1)$ such that the error 
in \reff{as2} is indeed smaller than the approximation. The
proof is based on a Lyapunov--Schmidt reduction and analysis of suitable
extended CMEs.  In \S\ref{S:num} we give some numerical illustrations and
verify convergence of the asymptotic coupled mode approximation. 

\brem{\rm 
The (apparent) lack of an estimate 
$\|\phi_{GS}-\eps\phi^{(0)}\|_{L^\infty(\R^2)}\le C\eps^{1+\beta}$ 
with $\beta>0$ is 
a disadvantage of our analysis. It is due to the fact that we work 
in $L^2$--spaces in Fourier resp.~Bloch variables, while a 
direct $L^\infty$ estimate in physical variables would require working 
in $L^1$--spaces in Fourier resp.~Bloch variables.  
This is not possible due to a technical obstacle, see 
\cite[\S8]{DPS08}. On the other hand, 
Hilbert spaces $L^2$ are also more natural spaces to work in since 
they allow direct transition from physical to Bloch variables and back. 
Moreover, localization in $x$ in the sense of decay to $0$ for 
$|x|\ra\infty$ follows directly in spaces of integrable functions.
Note also that based on the formal asymptotics, instead of $\eps^{2/3}$ 
one can expect the 
convergence rate $\eps^1$ in $H^s$ in \eqref{as2} which is the approximate rate 
observed in our numerical examples. Finally, in Remark \ref{fr2rem} we 
explain how  the long wave  modulational form of 
the formal asymptotics allows to obtain an $\CO(\eps^{1+\beta})$ 
convergence of the error in $L^\infty$ from the $\CO(\eps^\beta)$ 
convergence in $H^s$. 
However, a completely rigorous calculation is lengthy and therefore 
here we content ourselves with \reff{as2}.  
}\erem
\brem{\rm  Time-dependent CMEs have been justified in 1D
for infinitesimal \cite{GWH01,SU01} and finite \cite{BSTU06} contrast, and in 2D for finite contrast under the condition of a separable potential in 
\cite[\S7]{DPS08}. 
Here justification means that non--stationary solutions of \reff{E:GP} 
can be approximated by CME dynamics over long but finite intervals. 
Given the analysis below, this result of \cite{DPS08} can be immediately transfered 
to our non--separable case.
}\erem

\section{Band structure and Bloch functions}\label{S:band_str}

\begin{figure}[htpb]
  \centering 
\includegraphics[height=3.5cm,width=3.5cm]{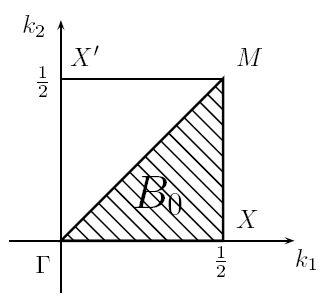}
  \caption{The first irreducible Brillouin zone $B_0$ for the two-dimensional
    potential $V$.} \label{F:B_ir}
\end{figure}

Let $\omega_n(k), n \in \N$, denote the spectral bands and $u_n(k;x)$ the
corresponding Bloch functions of the operator $L:=-\Delta+V(x)$, where $k$
runs through the first Brillouin zone $\T^2=(-1/2,1/2]^2$. This means that
$(\omega_n(k), u_n(k;x))$ is an eigenpair of the quasiperiodic eigenvalue
problem
\begin{equation}\label{E:qp_problem}
  \begin{split}
    &Lu_n(k;x) = \omega_n(k)u_n(k;x), \ x \in \P^2:=[0,2\pi)^2,\\
    &u_n(k;(2\pi,x_2)) = \er^{\ri 2\pi k_1} u_n(k;(0,x_2)), \ u_n(k;(x_1,2\pi)) =
    \er^{\ri 2\pi k_2} u_n(k;(x_1,0)).
  \end{split}
\end{equation}
The Bloch functions $u_n(k;x)$ can be rewritten as
\begin{gather}
  u_n(k;x)=\er^{\ri k\cdot x}p_n(k;x), \quad \text{where $p_n$ is 
$2\pi$-periodic in both $x_1$ and $x_2$, and fulfills}\label{E:Bloch_form}\\
\Lti(k;x)p_n(k;x):=[(\ri\pa_{x_1}{-}k_1)^2+(\ri\pa_{x_2}{-}k_2)^2+V(x)]p_n(k;x)
=\om_n(k)p_n(k;x).  \label{lbloch}
\end{gather}
For each
$k\in \T^2$ the operator $\tilde{L}(k;\cdot)$ is elliptic and self adjoint in
$L^2(\P^2)$, which immediately yields the existence of infinitely many real
eigenvalues $\om_n(k), n\in\N$ with $\om_n(k)\ra\infty$ as $n\ra\infty$. 
The spectrum of $L$ equals $\bigcup_{n\in \mathbb{N}, k\in \T^2}\omega_n(k)$,
see Theorem 6.5.1 in \cite{Eastham}. Moreover, if $V$ satisfies $V(-x_1,x_2) = V(x_1,-x_2)=V(x)$ and $V(x)=V(x_2,x_1) \ \forall x \in \R^2$, the $\omega_n(k)$
can be recovered from their values in the irreducible Brillouin zone $B_0$,
see Fig.~\ref{F:B_ir}. 

From \reff{E:Bloch_form} we also note that for  $x\in\P^2$ 
and $n\in \Z^2$ we have 
\begin{gather}
u_n(k;(x_1+2n_1\pi,x_2+2n_2\pi))=\er^{2\pi\ri n \cdot k}u_n(k;x). 
\label{uqp}
\end{gather}

Gaps in the spectrum of $L$ have to be confined by extrema of bands. Unlike in
the case of the separable potential $V(x_1,x_2)=W_1(x_1)+W_2(x_2)$ the 
extrema of $\omega_n$
within $B_0$ do not have to occur only at $k = \Gamma,X$ and $M$ but may occur
anywhere throughout $B_0$. 
Thus we need to solve \eqref{E:qp_problem} for all
$k \in B_0$.
\begin{figure}[h!]
  \begin{center}
    \includegraphics[scale=0.4]{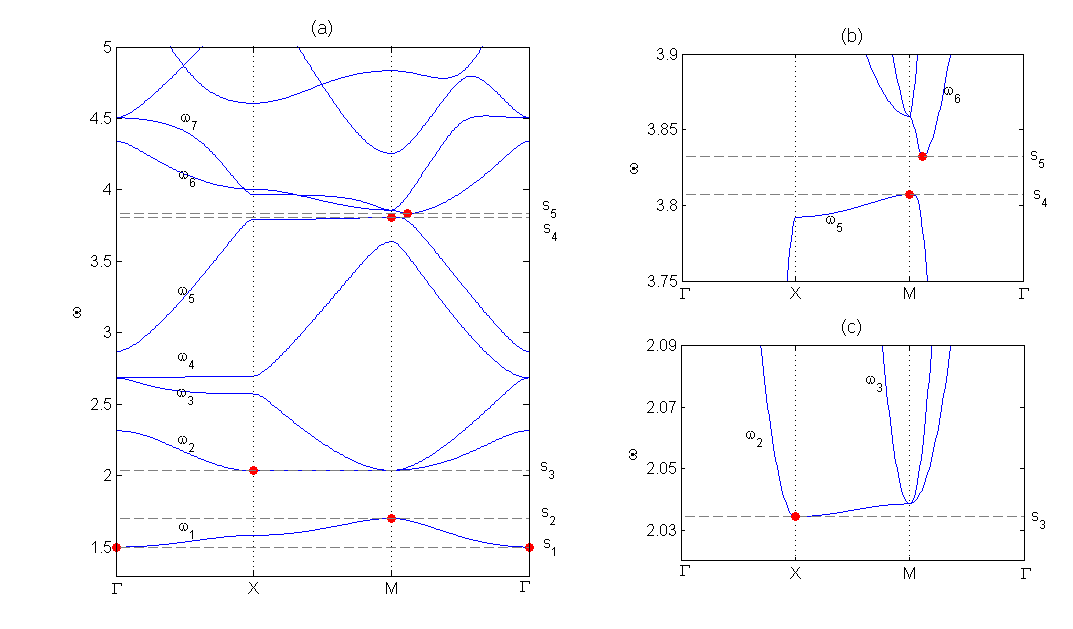} 
  \end{center}
  \caption{(a) Band structure of $L$ with $\eta =5.35$ along $\pa B_0$. Red
    dots label band extrema at gap edges $s_1, \ldots, s_5$. (b) Detail in the
    second finite gap. (c) Detail near the edge $s_3$ showing that $\omega_2$
    is not flat for $k$ between $X$ and $M$.}
  \label{F:band_str_bdry}
\end{figure}

\begin{figure}[h!]
  \begin{center}
    \includegraphics[height=9cm]{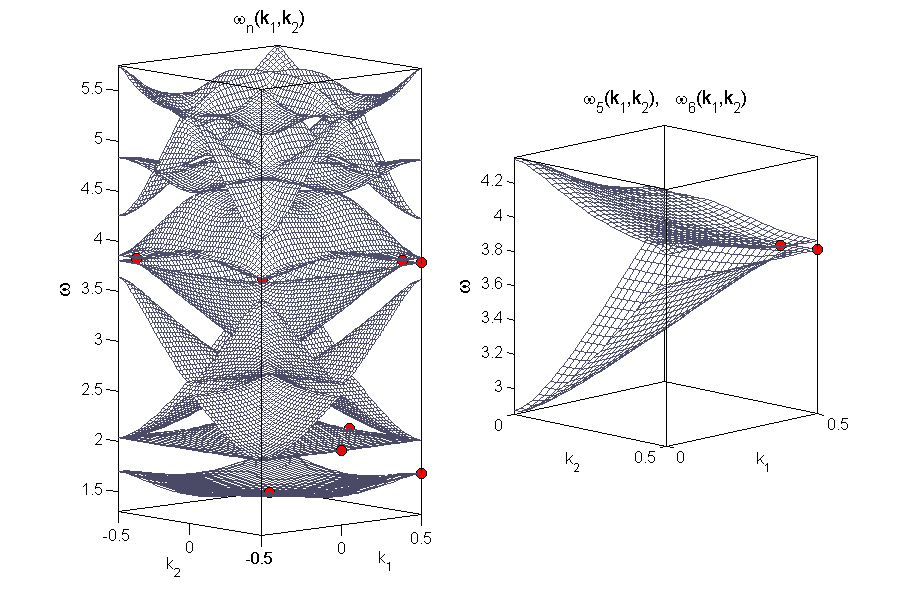}
  \end{center}
  \caption{Band structure of $L$ with $\eta =5.35$. On the right a detail of
    $\omega_5$ and $\omega_6$ near the second finite gap.}
  \label{F:band_str}
\end{figure}
In the example \eqref{E:V} we choose the contrast $\eta$ so that
two finite band gaps are open. Our computations show that this happens, for
instance, at $\eta =5.35$, which we select. The band structure of $L$ is
computed in a 4th order centered finite-difference discretization. For reasons
of tradition we plot in Fig. \ref{F:band_str_bdry} the band structure along
$\pa B_0$. In Fig. \ref{F:band_str} we plot the first few bands over
$B$. Though not true in general \cite{HKSW07}, in our case the extrema of the
first 6 bands fall on $\pa B_0$. The dots in Figs.\,\ref{F:band_str_bdry}
and \ref{F:band_str} label those band edge extrema which also mark gap
edges. One of these extrema in Fig.\,\ref{F:band_str_bdry} (corresponding to 4 extrema in Fig. \ref{F:band_str}) falls out of the vertex set
$\left\{\Gamma, X, M\right\}$. We also label in Fig. \ref{F:band_str_bdry} the
first 7 bands $\omega_1, \ldots, \omega_7$ and the gap edges $s_1,s_2, \ldots,
s_5$. The edge values with six converged decimal places are
\[s_1\approx 1.502064, \ s_2 \approx 1.702299, \ s_3 \approx 2.034433, \ s_4
\approx 3.807113, \ \text{and} \ s_5 \approx 3.832442.\]

For any corresponding value of $k$ each gap edge eigenvalue of
\eqref{E:qp_problem} is simple because none of the edge-defining extrema
belongs to more than one band. We now combine this with symmetries of the
problem to find symmetries of the Bloch functions, which will be needed in the
derivation of the CME. In the rest of this section we assume 
$\|u_n(k;\cdot)\|_{L^2(\P^2)}=1$, where we are, of course, still free to 
multiply any mode $u_n$ by a phase factor $\er^{\ri a}, \ a\in \R$. See also 
Remark \ref{srem}. 

First, due to evenness of $V(x)$ in \eqref{E:V} in both variables we have 
\begin{gather}
\begin{split}
  u_n((-k_1,k_2);(x_1,x_2))&=\er^{\ri a_1}u_n((k_1,k_2);(2\pi-x_1,x_2)),\\
  u_n((k_1,-k_2);(x_1,x_2))&= \er^{\ri a_2} u_n((k_1,k_2);(x_1,2\pi-x_2)),\\ 
  \omega_n(-k_1,k_2)&=\omega_n(k_1,-k_2)=\omega_n(k).
\end{split} \label{E:sym1}
\end{gather}
for some $a_1,a_2 \in \R$.  Note that when $(-k_1,k_2)\doteq (k_1,k_2)$, where
$k\doteq l$ reads ``$k$ congruent to $l$'' and means $k=l+m$ for some $m\in
\Z^2$, a renormalization of the phase 
\textit{cannot} be used in general to obtain $a_1=0$
because $u_n((k_1,k_2);(\pi,x_2))=0 \ \forall x_2\in \P$ is
possible. Similarly, when $(k_1,-k_2)\doteq (k_1,k_2)$, we cannot generally achieve $a_2=0$ because $u_n((k_1,k_2);(x_1,\pi))=0 \ \forall x_1\in \P$ is
possible.

Next, the symmetry $V(x_1,x_2)=V(x_2,x_1)$ implies 
\beq
  u_n((k_1,k_2);(x_1,x_2)) = \er^{\ri a}u_n((k_2,k_1);(x_2,x_1)), \qquad 
  \omega_n(k_1,k_2)=\omega_n(k_2,k_1). \label{E:sym3}
\eeq
for some $a\in\R$. Similarly to the case of symmetry \eqref{E:sym1}, when $k_1 \doteq  k_2$, one \textit{cannot}, in general, apply renormalization to achieve $a=0$ because $u_n((k_1,k_1);(x_1,x_1))=0 \ \forall x_1\in \P$ is possible. 

Finally, since $L$ is real, $\overline{u_n(k;x)}$ satisfies \eqref{E:qp_problem} with
the factors in the boundary conditions replaced by $\er^{-\ri 2\pi k_1}$ and 
$\er^{-\ri 2\pi k_2}$. Thus
\begin{gather}
  u_n(-k;x)=\ov{u_n(k;x)},\qquad 
  \omega_n(-k)=\omega_n(k). \label{E:sym2}
 \end{gather}
 Note that unlike in \eqref{E:sym1} and \eqref{E:sym3} no exponential factor
 appears in \eqref{E:sym2}. This is because for the conjugation symmetry
 \eqref{E:sym2} such a factor $\er^{\ri a}$ can be easily removed via
 multiplication by $\er^{-\ri a/2}$.

 \brem\label{srem}{\rm If, e.g., $(-k_1,k_2)$ is not
   congruent to $(k_1,k_2)$, we can, for instance, multiply
   $u_n((-k_1,k_2);\cdot)$ by $\er^{\ri a_1}$ and obtain
   $u_n((-k_1,k_2);(x_1,x_2))=u_n(k;(2\pi-x_1,x_2))$. However, one will
   generally not be able to simultaneously ensure also
   $u_n((k_1,k_2);(x_1,x_2)) = u_n((k_2,k_1);(x_2,x_1))$ in \reff{E:sym3} and
   therefore we stick to the factors in \reff{E:sym1} and \reff{E:sym3}.
 }\erem

\begin{figure}[h!]
  \begin{center}
    \includegraphics[height=7.5cm,
    width=11cm]{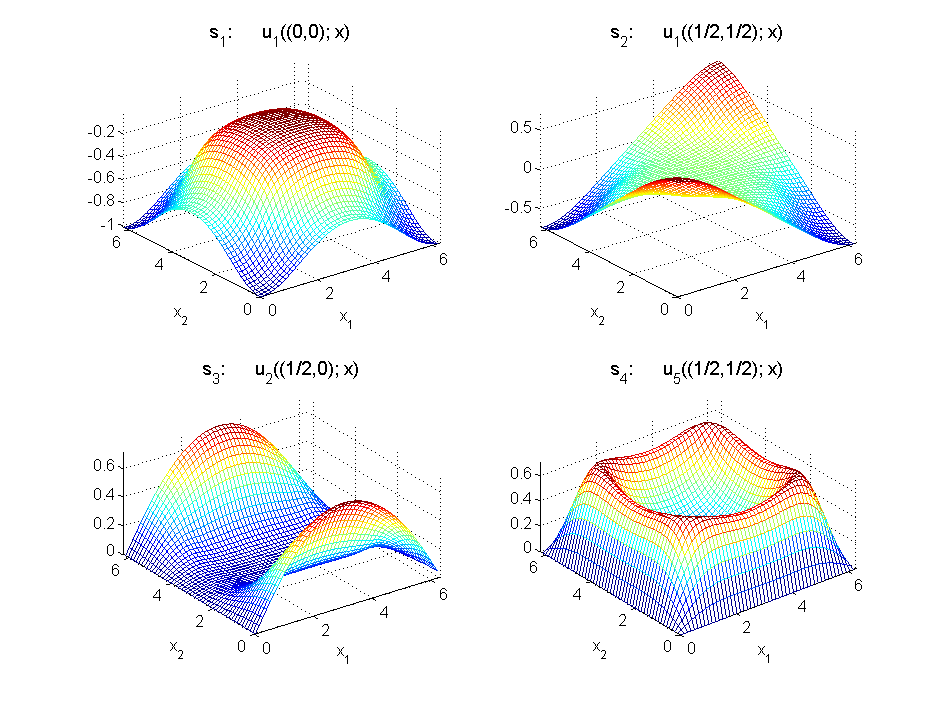}
     \includegraphics[height=3.6cm,
    width=10.5cm]{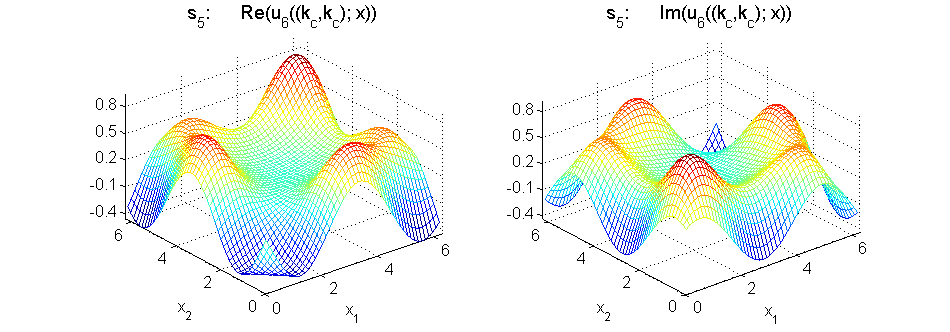}
  \end{center}
  \caption{Bloch functions at gap edges $s_1, s_2 \ldots, s_5$.}
  \label{F:Bloch_waves}
\end{figure}

Let us consider implications of the above three symmetries \eqref{E:sym1}, \eqref{E:sym3} and \eqref{E:sym2} for our example \eqref{E:V}
and plot the gap edge Bloch functions in Fig. \ref{F:Bloch_waves}. Each edge
$s_1,s_2$ and $s_4$ is attained only at a single extremum within $B$, namely
at $k=\Gamma, M$ and $M$ respectively. The corresponding Bloch functions are
$u_1((0,0);x)$, $u_1((1/2,1/2);x)$ and $u_5((1/2,1/2);x)$ respectively, which are 
all real due to \reff{E:sym2}.  The edge
$s_3$ is attained by extrema at $k=X$ and $X'$ with the Bloch functions
$u_2((1/2,0);x)$ and $u_2((0,1/2);x)$.  Referring to \reff{E:sym3} only
$u_2((1/2,0);(x_1,x_2))$ is plotted, which is again real due 
to  \reff{E:sym2}. Finally, the edge $s_5$ is attained
by 4 extrema, namely at $k=(k_c,k_c), (-k_c,k_c), (-k_c,-k_c)$ and
$(k_c,-k_c)$, where the numerically computed value, converged to 6 decimal
places, is $k_c \approx 0.439028$. The corresponding Bloch functions are
$u_6((k_c,k_c);x)$, $u_6((-k_c,k_c);x)$, $u_6((-k_c,-k_c);x)$ and
$u_6((k_c,-k_c);x)$.  Due to \eqref{E:sym1} and \eqref{E:sym2} and because $k_c\notin \{0,1/2\}$, we can normalize the Bloch functions so that 
$u_6((-k_c,k_c);(x_1,x_2))= u_6((k_c,k_c);(2\pi-x_1,x_2))$, 
$u_6((k_c,-k_c);(x_1,x_2)) =u_6((k_c,k_c);(x_1,2\pi-x_2))$,
$u_6((-k_c,-k_c);(x_1,x_2)) = u_6((k_c,k_c);(2\pi-x_1,2\pi-x_2))
 = \ov{u_6((k_c,k_c);(x_1,x_2))}$. 
Thus it suffices to plot only $u_6((k_c,k_c);(x_1,x_2))$. In addition, \eqref{E:sym3} and the fact that $u_6((k_c,k_c);(x_1,x_1))$ is not identically zero imply $u_6((k_c,k_c);(x_1,x_2))=u_6((k_c,k_c);(x_2,x_1))$. The Bloch waves $u_6((k_c,k_c);x)$ and $u_6((-k_c,-k_c);x)$ are, therefore, symmetric about the diagonal $x_1=x_2$.

Fig. \ref{F:Bloch_waves} shows that all the Bloch functions at
$s_1,s_2, \ldots, s_4$ are either even or odd in each variable. This 
actually follows from \eqref{E:sym1} and the fact that these gap edges occur
at $k\in \Sigma=\{\Gamma,X,X',M\}$. As each coordinate of any $k\in \Sigma$ is either $0$
or ${1\over 2}$, the eigenvalue problem \eqref{E:qp_problem} is real and we can choose $a_1,a_2 \in \{0,\pi\}$ in \eqref{E:sym1}. The choice $a_1=a_2=0$ is, however, in general impossible as 
explained after \eqref{E:sym1}. 
Taking, for instance, $k_1={1\over 2}$, we have
\[\begin{split}
  u_n((1/2,k_2);x) &= \pm u_n((-1/2,k_2);(2\pi-x_1,x_2))=\pm u_n((1/2,k_2);(2\pi-x_1,x_2))\\
  &= \pm e^{\ri 2\pi \frac{1}{2}}u_n((1/2,k_2);(-x_1,x_2)) = \mp
  u_n((1/2,k_2);(-x_1,x_2)),
\end{split}
\]
where the second equality follows from $1-$periodicity of $u_n$ in each
$k-$coordinate and the third equality from the quasi-periodic boundary
conditions in \eqref{E:qp_problem}. Similarly, we get
$u_n((0,k_2);(x_1,x_2))=\pm u_n((0,k_2);(-x_1,x_2))$.
Therefore, we have the following 
\blem\label{L:sym_S} Suppose $V(x)$ is even
in the variable $x_j$ for some $j\in \{1,2\}$. If $k_j \in\{0,1/2\}$ and $\omega_n(k)$, as an eigenvalue of \eqref{E:qp_problem} has geometric multiplicity 1,
then $u_n(k;x)$ is either even or odd in $x_j$.\elem

\section{Formal asymptotic derivation of Coupled Mode
  Equations}\label{S:derive_CME}

Gap solitons in the vicinity of a given band edge are expected to be
approximated by the Bloch waves at the band edge modulated by slowly varying
spatially localized envelopes. 
The governing equations for the envelopes, called Coupled
Mode Equations (CMEs), can be derived by a formal asymptotic procedure. 
 Here we are interested in gap solitons
$E(x,t)=\phi(x)e^{-\ri \omega t}$ with $\omega = \omega_* + \eps^2 \Omega, \
0<\eps \ll 1$, where $\omega_*$ is an edge of a given band gap of a fixed
($\mathcal{O}(1)$) width, and $\Omega$ has a sign chosen so that $\omega$ lies
inside the gap. The leading order term in the asymptotic expansion of the
spatial profile $\phi$ is expected to be
\begin{equation}\label{E:general_asympt}
  \phi(x) \sim \eps \sum_{j=1}^N A_j(\eps x)u_{n_j}(k^{(j)};x),
\end{equation}
where $\left\{u_{n_j}(k^{(j)};x)\right\}_{j=1}^N$ are the Bloch waves at
$\omega =\omega_*$ and \reff{uqp} is used for $x\not\in\P^2$. 
We assume: 

\noindent{\bf Assumption A.1} \quad The band structure defined by
\eqref{E:qp_problem} has a gap with an edge (lower/upper) defined by $0<N <
\infty$ extrema (maxima/minima) of the bands $\omega_n(k)$.  
The extrema occur for
bands $\omega_{n_j}(k), j =1, \ldots N$ at the corresponding points 
$k^{(j)} \in B$, where $k\mapsto \omega_{n_j}(k)$ is analytic in 
$k$ locally near $k^{(j)}$. 

\noindent{\bf Assumption A.2}  The quadratic form $\pa_{k_1}^2\om_{n_j}(k^{(j)})x^2 + 2\pa_{k_1}\pa_{k_2}\om_{n_j}(k^{(j)})xy+\pa_{k_2}^2\om_{n_j}(k^{(j)})y^2$ defined by the Hessian of $\omega_{n_j}$ at $k=k^{(j)}$ is (positive or negative) definite.

\brem\label{R:resonance}{\rm a) Analyticity of the $n_j$-th band near $k^{(j)}$ holds if 
$\om_{n_j}(k^{(j)})$ is simple, see \cite{W78}. \\
b) The definiteness in A.2 ensures that the extremum of $\omega_{n_j}$ at $k=k^{(j)}$ is quadratic and that the resulting CMEs are of second
order. Unlike in the separable case \cite{DPS08} it is possible that
$\pa_{k_1}\pa_{k_2}\omega_{n_j}(k^{(j)}){\neq}0$, which then leads to CMEs
with mixed second order derivatives.\\
c) The Bloch waves $u_{n_j}(k^{(j)};\cdot), 
j{=}1, \ldots, N$ defined by the extrema are
called ``resonant''.\\
d)  Assumptions A.1 and A.2 are satisfied by the potential \eqref{E:V} with
$\eta = 5.35$ at all the gap edges $s_1, \ldots, s_5$.}\erem

\brem {\rm The approximation \eqref{E:general_asympt} with the same
  $\eps$-scaling applies also to gap solitons in an $\mathcal{O}(\eps^2)$-wide
  gap which closes at $\omega =\omega_*$ as $\eps \rightarrow 0$ in such a way
  that the plane $\omega=\omega_*$ at $\eps=0$ is not intersected by any band but is
  tangent to bands at $N$ extremal points. $u_{n_1}, \ldots, u_{n_N}$ are then
  the resonant Bloch waves at $\omega=\omega_*$ at $\eps=0$. Such a case was
  studied in \cite{DPS08} for a separable periodic potential.

 
  The above discussion is not limited to the case of the Gross-Pitaevskii
  equation but applies to general differential equations with periodic
  coefficients, as it depends only on the band structure. A typical example is
  Maxwell's equations with spatially periodic coefficients. } \erem

We now give the derivation of CMEs under the assumptions A.1 and
A.2. For the example \eqref{E:V} with $\eta=5.35$ we first review 
the derivation in physical variables $\phi(x)$ near $\om=s_3$, 
then comment on an obstacle for this calculus near $\om=s_5$, 
and therefore present a derivation in the general case in the so
called Bloch variables which avoids this obstacle. Finally, we apply this
general procedure to all the five gap edges of the example \eqref{E:V}.

\subsection{CME derivation in Physical Variables $\phi(x)$}
\label{S:deriv_phys}
The ansatz in physical variables is
\begin{gather}\label{E:ansatz_phys}
\begin{split}
  \phi(x) &= \eps\phi^{(0)}(x)+\eps^2\phi^{(1)}(x)
  +\eps^3\phi^{(2)}(x)+\mathcal{O}(\eps^4),\\
  \eps\phi^{(0)}(x)&= \eps\sum_{j=1}^NA_j(y)u_{n_j}(k^{(j)};x), \qquad \omega = \omega_*+\eps^2\Omega, \qquad y= \eps x, \qquad 0<\eps \ll 1.
\end{split}
\end{gather} 
To review the derivation of the CMEs we choose $\omega_*=s_3$ 
for the example \eqref{E:V} with $\eta=5.35$.

\subsubsection{CMEs near the gap edge $\omega =s_3$}\label{S:e3}

At the edge $s_3$ we have $N=2, n_1=n_2=2, k^{(1)} = X$ and $k^{(2)} = X'$,
i.e.~the two resonant Bloch waves are $v_1(x):=u_2(X;x)$ and
$v_2(x):=u_2(X';x)$. Using \eqref{uqp}, Lemma \ref{L:sym_S} and \reff{E:sym2}, 
we have that $v_1$ is odd and $2\pi-$antiperiodic in $x_1$ and 
even and $2\pi-$periodic in $x_2$. Opposite symmetries hold for $v_2$.  
Moreover, \reff{E:sym2} implies that $v_1$ and $v_2$ are real. 
We normalize the Bloch functions
$v_{1,2}$ over their common period $[-2\pi,2\pi]^2$ so that
$\|v_j\|_{L^2([-2\pi,2\pi]^2)}=1, \ j=1,2$.

Substituting \eqref{E:ansatz_phys} in \eqref{E:GP_stat} leads to a hierarchy
of problems at distinct powers of $\eps$, each of which we try to solve within
the space of functions $4\pi$-periodic in both $x_1$ and $x_2$, invoking the
Fredholm alternative (see e.g.~chapter 3.4 of \cite{Stak79}) where necessary.
At $\CO(\eps)$ we have the linear eigenvalue problem $[L-s_3]v_j(x)=0, \
j=1,2$. At $\CO(\eps^2)$ we have
$$[L-s_3] \phi^{(1)} =
2\left(
  \pa_{y_1}A_1\pa_{x_1}v_1+\pa_{y_1}A_2\pa_{x_1}v_2+\pa_{y_2}A_1\pa_{x_2}v_1+\pa_{y_2}A_2\pa_{x_2}v_2
\right). 
$$
By differentiating the eigenvalue problem \eqref{E:qp_problem} with respect to
$k_j, j\in \{1,2\}$ and evaluating at $n=2,k=X=(1/2,0)$, we find that
\begin{gather}\label{E:gen_Bloch_fn_eq} 
[L-s_3]v_1^{(x_j)}(x) = 2\pa_{x_j}v_1, 
\end{gather} and
similarly $[L-s_3]v_2^{(x_j)}(x) = 2\pa_{x_j}v_2$, where
$$v_1^{(x_j)}(x) = -\ri(\pa_{k_j} p_2(X;x))e^{\ri  X
  \cdot x}\text { and } v_2^{(x_j)}(x) = -\ri(\pa_{k_j} p_2(X';x))e^{\ri  X' \cdot
  x}
$$ 
are called generalized Bloch functions \cite{PSK04}. Thus $ \phi^{(1)}=
\pa_{y_1}A_1v_1^{(x_1)}{+}\pa_{y_1}A_2v_2^{(x_1)}
{+}\pa_{y_2}A_1v_1^{(x_2)}{+}\pa_{y_2}A_2v_2^{(x_2)}$. 
\eqref{E:gen_Bloch_fn_eq} implies that $v_n^{(x_j)}(x)$ is odd/even in $x_j$ 
if $v_n(x)$ is even/odd in $x_j$ respectively.

At $\mathcal{O}(\eps^3)$ we obtain the CMEs. We have
$$\begin{array}{rl}
  [L-s_3] \phi^{(2)} = &\Omega(A_1v_1+A_2v_2) 
  + \Delta_{y_1,y_2}A_1 v_1+\Delta_{y_1,y_2}A_2 v_2\\
  &+2\left[\pa^2_{y_1}A_1\pa_{x_1}v_1^{(x_1)} 
    + \pa^2_{y_1}A_2\pa_{x_1}v_2^{(x_1)}
    +\pa^2_{y_2}A_1\pa_{x_2}v_1^{(x_2)} 
    + \pa^2_{y_2}A_2\pa_{x_2}v_2^{(x_2)}\right.\\
  & \left. \qquad +\pa_{y_1}\pa_{y_2}A_1\pa_{x_1}v_1^{(x_2)}
    +\pa_{y_1}\pa_{y_2}A_2\pa_{x_1}v_2^{(x_2)}
    +\pa_{y_1}\pa_{y_2}A_1\pa_{x_2}v_1^{(x_1)}
    +\pa_{y_1}\pa_{y_2}A_2\pa_{x_2}v_2^{(x_1)} \right]\\
  &-\sigma\left[ \sum_{j=1}^2|A_j|^2A_j v_j^3
    +2|A_1|^2A_2v_1^2v_2+2|A_2|^2A_1v_2^2v_1 + A_1^2\bar{A_2} v_1^2v_2 +
    A_2^2\bar{A_1} v_2^2v_1\right], 
\end{array}
$$ 
and the Fredholm alternative requires the right hand side to be
$L^2(-2\pi,2\pi]^2$-orthogonal to $v_1$ and $v_2$, the two generators of
$\text{Ker}(L^*-s_3)$. Taking the inner product, we see that the terms $\langle v_1,v_2\rangle$ 	and $\langle v_2,v_1\rangle$ in the inner product
vanish due to orthogonality of Bloch waves. Many additional terms vanish due
to odd or $2\pi$-antiperiodic integrands (in at least one variable). Namely,
in the inner product of the right hand side with $v_1$ the integrals $\langle
\pa_{x_1}v_2^{(x_1)},v_1\rangle, \langle \pa_{x_2}v_2^{(x_2)},v_1\rangle,
\langle \pa_{x_1}v_1^{(x_2)},v_1\rangle$ and $\langle
\pa_{x_2}v_1^{(x_1)},v_1\rangle$ vanish due to odd
integrands and the integrals  $\langle v_2^3,v_1\rangle, \langle
v_1^2v_2,v_1\rangle, \langle v_1^2v_2,v_1\rangle$ $\langle \pa_{x_1}v_2^{(x_2)},v_1\rangle$ and
$\langle \pa_{x_2}v_2^{(x_1)},v_1\rangle$ due to $2\pi-$antiperiodic
integrands. An analogous discussion applies for the orthogonality with the
respect to $v_2$. The remaining terms have to be set to zero, which leads to
the CMEs for the envelopes $A_1$ and $A_2$:
\begin{equation}\label{E:CME_s3_phys}
  \begin{split}
    \Omega A_1 + \al_1 \pa_{y_1}^2A_1+\al_2\pa_{y_2}^2A_1-\sigma \left[\ga_1|A_1|^2A_1+\ga_2(2|A_2|^2A_1+A_2^2\bar{A_1})\right]=&0,\\
    \Omega A_2 + \al_2 \pa_{y_1}^2A_2+\al_1\pa_{y_2}^2A_2-\sigma
    \left[\ga_1|A_2|^2A_2+\ga_2(2|A_1|^2A_2+A_1^2\bar{A_2})\right]=&0,
  \end{split}
\end{equation}
$$
\barr{ll}
\al_1=1+2\ds\int_{-2\pi}^{2\pi}\int_{-2\pi}^{2\pi}v_1\pa_{x_1}v_1^{(x_1)}dx,
\quad &\al_2=1+2\ds
\int_{-2\pi}^{2\pi}\int_{-2\pi}^{2\pi}v_1\pa_{x_2}v_1^{(x_2)}dx,\\[2mm]
\ga_1=\ds \int_{-2\pi}^{2\pi}\int_{-2\pi}^{2\pi}v_1^4dx \qquad \text{ and }
&\ga_2=\ds\int_{-2\pi}^{2\pi}\int_{-2\pi}^{2\pi}v_1^2v_2^2dx.  \earr
$$

\subsubsection{CMEs near the gap edge $s_5$}\label{S:e5}

At $\omega_*=s_5$ we have $N=4$. The resonant Bloch waves are
$v_1:=u_6((k_c,k_c);x)$, $v_2:=u_6((-k_c,k_c);x)$, $v_3:=u_6((-k_c,-k_c);x)$
and $v_4:=u_6((k_c,-k_c);x)$.  Analogously to \S\ref{S:e3} the
asymptotic expansion needs to be carried out in the space of functions
periodic over the common period of $v_1, \ldots, v_4$. The Bloch functions are
then pairwise orthogonal over this domain. However, if $k_c$ is not rational 
then the Bloch waves are not periodic but only quasi-periodic. Therefore, unlike
in the case of a separable $V(x)$ \cite{DPS08}, where always
$k_c\in\{0,1/2\}$, in the non-separable case in general the derivation in
physical variables is impossible.

\subsection{CME Derivation in Bloch Variables $\tilde{\phi}(k;x)$}

An alternative to the derivation in \S\ref{S:deriv_phys} is to transform
the problem to Bloch variables. The advantage is that the linear
eigenfunctions are then all $2\pi-$periodic in each $x-$coordinate. The
orthogonalization domain is, therefore, always $\P^2$.

\subsubsection{General Case}
The Bloch transform $\CT$ is formally defined by 
\begin{gather}\label{E:Bloch_tr}
\begin{split}
  \tilde{\phi}(k;x) &= (\CT \phi)(k;x) = \sum_{m\in \Z^2} e^{\ri m\cdot
    x}\hat{\phi}(k+m),\quad \phi(x) = (\CT^{-1} \tilde{\phi})(x) =
  \int_{\T^2}e^{\ri k\cdot x} \tilde{\phi}(k;x)dk,
\end{split}
\end{gather} 
where $\hat{\phi}(k)$ denotes the Fourier transform. 
$\CT$ is an isomorphism from $H^s(\R^2,\C)$ to $L^2(\T^2,H^s(\P^2,\C))$, 
$\|\tilde{\phi}\|^2_{L^2(\T^2,H^s(\P^2,\C))}
=\int_{\T^2}\|\tilde{\phi}(k;\cdot)\|^2_{H^s(\P^2)}dk$, cf., e.g.,\cite{RS}, 
and by construction we have
\begin{align}
  &\tilde{\phi}(k;(x_1+2\pi,x_2)) = \tilde{\phi}(k;(x_1,x_2+2\pi))
=\tilde{\phi}(k;x), \label{E:Bloch_2pi_per}\\
  &\tilde{\phi}((k_1+1,k_2);x) = e^{-\ri x_1}\tilde{\phi}(k;x), \quad
  \tilde{\phi}((k_1,k_2+1);x) =
  e^{-\ri x_2}\tilde{\phi}(k;x) \label{E:Bloch_1_per_in_k}.
\end{align}
Multiplication in physical space corresponds to convolution in Bloch space,
i.e., \begin{gather} (\CT (\phi\psi))(k;x) =
\int_{\T^2}\tilde{\phi}(k-l;x)\tilde{\psi}(l;x)dl =: (\tilde{\phi} *_{\Bsm}
\tilde{\psi})(k;x), \end{gather} where \reff{E:Bloch_1_per_in_k} is used if $k-l
\notin \T^2$.  However, if $g$ is $2\pi-$periodic in both $x_1$ and $x_2$,
then $(\CT (gu))(k;x) =g(x)(\CT u)(k;x).$

In order to choose a suitable asymptotic ansatz for $\tilde{\phi}(k;x)$, note first that the Bloch transform $\CT$ of the ansatz \eqref{E:ansatz_phys}  for $\eps \phi^{(0)}(x)$ is
\beq\label{E:ansatz_transf} \eps \tilde{\phi}^{(0)}(k;x) =
\frac{1}{\eps}\sum_{j=1}^N p_{n_j}(k^{(j)};x) \sum_{m \in
  \Z^2}\hat{A}_j\left(\frac{k-k^{(j)}+m}{\eps}\right)e^{\ri m \cdot x}
\eeq 
with $k\in \T^2, x\in \P^2$. As 
$\hat{A}_j(p)$ is localized near
$p=0$, we approximate $\hat{A}_j\left(\frac{k-k^{(j)}+m}{\eps}\right)$
by $\chi_{D_j}(k+m)\hat{A}_j\left(\frac{k-k^{(j)}+m}{\eps}\right)$,
where $\chi_{D_j}(k)$ is
the characteristic function of the set
\beq\label{E:D_j}
D_j = \{k\in \R^2: |k-k^{(j)}|<\eps^r\}
\eeq
and
\begin{gather}
0<r<\frac{2}{3}.\label{rcon1}
\end{gather} 
The reason for \reff{rcon1} will be explained in \S\ref{S:diff_CME}.

\begin{figure}[htpb]
    \centering
     \includegraphics[height=5cm,width=10.5cm]{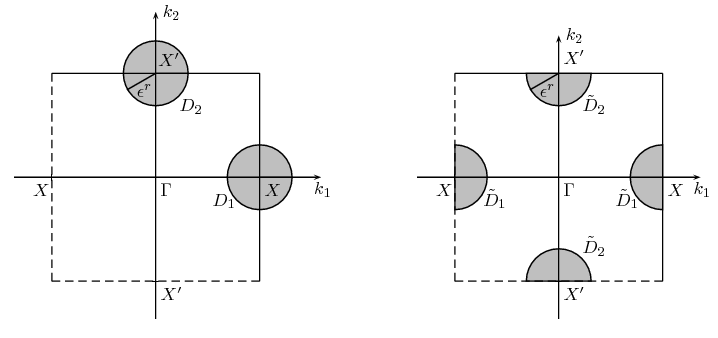}
      \caption{Sets $D_j$ and $\tilde{D}_j$ for $k^{(j)}=X$ and $k^{(j)}=X'$ (as in the example
        \eqref{E:V} with $\eta=5.35$ at $\omega_* = s_3$).}
      \label{F:k_sp_decomp}
\end{figure}

Below we will also use periodically wrapped versions $\tilde{D}_j$ of these neighborhoods, i.e.
\begin{gather}
\tilde{D}_j:=\{k\in\T^2 : |k{-}k^{(j)}|<\eps^r\text{ modulo }\doteq\} \label{E:Dtil_j}
\end{gather} 
where `modulo $\doteq$' means equal modulo $1$ in each component, see 
Fig.~\ref{F:k_sp_decomp} for an example. 


Note that $k+m \in D_j $ with $k\in \T^2$ is possible only for 
$m \in\{m\in \Z^2: 0 \leq m_1,m_2 \leq 1\}$. We define the set of $m-$values for which $k+m\in D_j$ for some $k\in
\T^2$ by $M_j:=\{m\in \Z^2: k+m\in D_j \ \text{for some} \ k\in
\T^2\}$. In fact, for small $\eps$ 
only the following cases occur: $M_j = \left\{
\bspm 0\\0\espm, 
\bspm 1\\0\espm\right\}$ if $k_1^{(j)}=1/2$ and $k^{(j)}\neq (1/2,1/2)$, $M_j =
\left\{\bspm 0\\0\espm, 
\bspm 0\\1\espm\right\}$ if $k_2^{(j)}=1/2$ and $k^{(j)}\neq
(1/2,1/2)$, $M_j = \left\{\bspm 0\\0\espm, \bspm 1\\0\espm,\bspm 0\\1\espm,\bspm
  1\\1\espm\right\}$ if $k^{(j)}=(1/2,1/2)$, and $M_j=\left\{\bspm 0\\0\espm\right\}$ 
if $k^{(j)}\in \text{int}(\T^2)$.

Thus we are lead to the following asymptotic ansatz in Bloch variables
\begin{gather}\label{E:ansatz_Bloch}
\begin{split}
  \tilde{\phi}(k;x) &= \frac{1}{\eps}\tilde{\psi}^{(0)}(k;x)
  +\tilde{\psi}^{(1)}(k;x)+\eps\tilde{\psi}^{(2)}(k;x)+\CO(\eps^2),\\
  \tilde{\psi}^{(0)}(k;x)&=
  \sum_{j=1}^Np_{n_j}(k^{(j)};x)\sum_{m\in M_j}\chi_{D_j}(k+m)\hat{A}_j\left(\frac{k-k^{(j)}+m}{\eps}\right)e^{\ri m \cdot x},\\
  \qquad \omega &= \omega_*+\Omega\eps^2, \qquad 0<\eps \ll 1. 
\end{split}
\end{gather} 
The periodic part $p_{n_j}$ of the Bloch
functions $u_{n_j}$ is normalized so that $\|p_{n_j}(k;\cdot)\|_{L^2(\P^2)}=1$.

The difference between the leading order terms in 
\eqref{E:ansatz_transf} and in \eqref{E:ansatz_Bloch} is 
\beq\label{hdef1}
\frac{1}{\eps}\tilde{\psi}^{(0)}(k;x)-\eps \tilde{\phi}^{(0)}(k;x) 
=: \sum_{j=1}^N \tilde{h}_j(k;x)
\eeq
with
\beq\label{hdef2}
\begin{split}
\tilde{h}_j(k;x) = \frac{1}{\eps}p_{n_j}(k^{(j)};x)&
\left[\sum_{m \in M_j}\left(1-\chi_{D_j}(k+m)\right)
\hat{A}_j\left(\tfrac{k-k^{(j)}+m}{\eps}\right)e^{\ri m \cdot x}  \right. \\
& \left. \quad + \sum_{m \in \Z^2\setminus M_j}\hat{A}_j
\left(\tfrac{k-k^{(j)}+m}{\eps}\right)e^{\ri m \cdot x}\right].
\end{split}
\eeq 

We now estimate $\|\tilde{h}_j\|_{L^2(\T^2, H^s(\P^2))}$. In the first 
sum in \eqref{hdef2} we have $|k+m-k^{(j)}|\geq \eps ^r$ while in the second 
sum $|k+m-k^{(j)}|\geq 1$ because $k+m\notin D_j$ 
for all $k\in \T^2$ if $m \in \Z^2\setminus M_j$. By the triangle inequality and 
the substitution $p = (k-k^{(j)}+m)/\eps$ we obtain
\[
\begin{split}
\|\tilde{h}_j\|_{L^2(\T^2, H^s(\P^2))}^2\leq &\sum_{m\in M_j} 
\|p_{n_j}(k^{(j)};\cdot)e^{\ri m\cdot \cdot}\|_{H^s(\P^2)}^2 
\int_{\substack{|p|>\eps^{r-1}\\p\in (\T^2-k^{(j)}+m)/\eps}}
|\hat{A}_j(p)|^2\dd p \\
& + \sum_{m\in \Z^2\setminus M_j} \|p_{n_j}(k^{(j)};\cdot)
e^{\ri m\cdot \cdot}\|_{H^s(\P^2)}^2 
\int_{\substack{|p|>c\eps^{-1}\\p\in (\T^2-k^{(j)}+m)/\eps}}|\hat{A}_j(p)|^2
\dd p\\
\leq &  C\left[\int_{|p|>\eps^{r-1}}|\hat{A}_j(p)|^2\dd p 
+ \int_{|p|>\eps^{-1}}|\hat{A}_j(p)|^2\dd p\right],
\end{split}
\]
where the $H^s$ regularity of $p_{n_j}(k^{(j)};\cdot)$ 
is guaranteed if $V\in H^{s-2}_\text{loc}(\R^2)$. By rewriting the right hand side as
$
C \left[\int_{|p|>\eps^{r-1}}|\hat{A}_j(p)|^2
\frac{(1+|p|)^{2s}}{(1+|p|)^{2s}}\dd p 
+ \int_{|p|>\eps^{-1}}|\hat{A}_j(p)|^2\frac{(1+|p|)^{2s}}{(1+|p|)^{2s}}
\dd p\right]
$
and taking the supremum of $(1+|p|)^{-2s}$ out of the integrals, we have
\beq\label{E:h_est}
\|\tilde{h}_j\|_{L^2(\T^2, H^s(\P^2))}\leq  C(\eps^{s(1-r)}+\eps^{s})
\|\hat{A}_j\|_{L^2_s(\R^2)} \leq C\eps^{s(1-r)}\|\hat{A}_j\|_{L^2_s(\R^2)}.
\eeq
For $r<1$ we thus have that  $\eps^{-1}\psi^{(0)}(x)$ approximates 
$\eps {\phi}^{(0)}(x)$ up to $\CO(\eps^{s(1-r)})$ in the $H^s(\R^2)$ norm. 
Because $\|\eps {\phi}^{(0)}\|_{H^s(\R^2)}=\CO(1)$, this approximation 
is satisfactory.

Applying next $\CT$ to \eqref{E:GP_stat} yields
\begin{gather}\label{E:GP_stat_Bloch} \left[\tilde{L}-\omega\right]\tilde{\phi} + \sigma
\ \tilde{\phi} *_{\Bsm} \tilde{\bar{\phi}} *_{\Bsm} \tilde{\phi}=0, \end{gather} on
$(k;x)\in\T^2\times \P^2$, where we recall from \reff{lbloch} that 
$\tilde{L}(k;x)=(\ri\pa_{x_1}{-}k_1)^2+(\ri\pa_{x_2}{-}k_2)^2+V(x)$. 

Setting $p^{(j,m)}:=\frac{k+m{-}k^{(j)}}{\eps}$, we have
\begin{gather}\label{E:Ltil_expand}
\begin{split}
\tilde{L}&(k;x)=\tilde{L}(k^{(j)}-m+\eps p^{(j,m)};x)\\
&=\tilde{L}(k^{(j)}{-}m;x)
-2\eps\left[(\ri\pa_{x_1}{-}k_1^{(j)}{+}m_1)p_1^{(j,m)}
+(\ri\pa_{x_2}{-}k_2^{(j)}{+}m_2)
  p_2^{(j,m)}\right]
+\eps^2\left[p_1^{(j,m)^2}{+}p_2^{(j,m)^2}\right].  
\end{split}
\end{gather} 
Substituting \eqref{E:ansatz_Bloch} in \eqref{E:GP_stat_Bloch} and 
using \eqref{E:Ltil_expand}, we obtain a hierarchy of equations on  
$x\in \P^2, k\in \T^2$ such that $k+m\in D_j, j \in \{1, \ldots, N\}$. 
Note that the combination of $k\in \T^2$ and $k+m\in D_j$ implies $m\in M_j$. 
The following hierarchy is thus for each $(j,m)\in \{1, \ldots, N\} 
\times M_j$. \\[2mm] 

$\mbf{\CO(\eps^{-1})}:$ $\quad \hat{A}_j(p^{(j,m)})
\left[\tilde{L}(k^{(j)}-m;x) -\omega_*\right](p_{n_j}(k^{(j)};x)e^{\ri m\cdot x})=0, \ $\\[2mm]
which is equivalent to $ \hat{A}_j(p^{(j,m)})e^{\ri m\cdot x}
\left[\tilde{L}(k^{(j)};x)
    -\omega_*\right]p_{n_j}(k^{(j)};x)=0$  and thus holds by definition of 
$\om_*=\omega_{n_j}(k^{(j)})$. 

$\mbf{\CO(1):} \quad \left[\tilde{L}(k^{(j)}-m;x)-\omega_*\right]
\tilde{\psi}^{(1)}(k;x) $ 
\begin{align*}
\qquad
& =2\hat{A}_j(p^{(j,m)})\left[p_1^{(j,m)}(\ri\pa_{x_1}{-}k_1^{(j)}+m_1)+
    p_2^{(j,m)}(\ri\pa_{x_2}{-}k_2^{(j)}+m_2)\right](p_{n_j}(k^{(j)};x)e^{\ri m\cdot x})\\
    &=2\hat{A}_j(p^{(j,m)})e^{\ri m\cdot x}\left[p_1^{(j,m)}(\ri\pa_{x_1}{-}k_1^{(j)})+
    p_2^{(j,m)}(\ri\pa_{x_2}{-}k_2^{(j)})\right]p_{n_j}(k^{(j)};x)
    \end{align*}
for $k+m\in D_j, j\in\{1,\ldots, N\}$. To solve this, we note
that by differentiating $[\tilde{L}(k;x)-\omega_{n_j}(k)]p_{n_j}(k;x)=0$ with 
respect to $k_l, l\in \{1,2\}$ and evaluating at $k=k^{(j)}-m$, we obtain
\begin{gather}\label{E:gen_Bloch_pn} \left[\tilde{L}(k^{(j)}-m;x)-\omega_*\right]\pa_{k_l}
p_{n_j}(k^{(j)}-m;x)=2(\ri\pa_{x_l}{-}k_l^{(j)}+m_l)p_{n_j}(k^{(j)}-m;x).  \end{gather}
Since $p_n(k-m;x) = e^{\ri m \cdot x}p_n(k;x)$ due to \eqref{E:Bloch_1_per_in_k}, we get for $k+m \in D_j$
\begin{gather}\label{E:psi_til_1} 
\tilde{\psi}^{(1)}(k;x) = \sum_{l=1}^2
p_l^{(j,m)}\hat{A}_j(p^{(j,m)})e^{\ri m\cdot x}\pa_{k_l}p_{n_j}(k^{(j)};x).  
\end{gather}

$\mbf{\CO(\eps):}$ We have 
\begin{gather}\label{E:psi2_eq}
\begin{split}
\bigl[\tilde{L}&(k^{(j)}-m;x){-}\omega_*\bigr]\tilde{\psi}^{(2)}(k;x)\\
  =& \ \Omega
  \hat{A}_j(p^{(j,m)})p_{n_j}(k^{(j)};x)e^{\ri m\cdot x}
  +2\left[p_1^{(j,m)}(\ri\pa_{x_1}{-}k_1^{(j)}{+}m_1){+}
    p_2^{(j,m)}(\ri\pa_{x_2}{-}k_2^{(j)}{+}m_2)\right]\tilde{\psi}^{(1)}(k;x) \\
  & - \left(p_1^{(j,m)^2}{+}p_2^{(j,m)^2}\right)\hat{A}_j(p^{(j,m)})p_{n_j}(k^{(j)};x)e^{\ri m\cdot x}{-}\frac{\sigma}{\eps^4} \chi_{D_j}(k+m)(\tilde{\psi}^{(0)}*_{\Bsm}\tilde{\psi}^{(0)}*_{\Bsm}
  \tilde{\bar{\psi}}^{(0)})(k;x)\\
 =& \ \Omega \hat{A}_j(p^{(j,m)})p_{n_j}(k^{(j)};x)e^{\ri m\cdot x}\\
  &-e^{\ri m\cdot x}\sum_{l=1}^2\left[p_{n_j}(k^{(j)};x)-2(\ri\pa_{x_l}{-}k_l^{(j)})
    \pa_{k_l}p_{n_j}(k^{(j)};x)\right]p_l^{(j,m)^2}\hat{A}_j(p^{(j,m)})\\
  &+2e^{\ri m\cdot x}\left[(\ri\pa_{x_1}{-}k_1^{(j)})\pa_{k_2}p_{n_j}(k^{(j)};x)
    +(\ri\pa_{x_2}{-}k_2^{(j)})\pa_{k_1}p_{n_j}(k^{(j)};x)\right]
  p_1^{(j,m)}p_2^{(j,m)}\hat{A}_j(p^{(j,m)})\\
  &-\frac{\sigma}{\eps^4} \chi_{D_j}(k+m)
  (\tilde{\psi}^{(0)}*_{\Bsm}\tilde{\psi}^{(0)}*_{\Bsm}
  \tilde{\bar{\psi}}^{(0)})(k;x)
\end{split}
\end{gather} 
using $\tilde{\psi}^{(1)}$ from \eqref{E:psi_til_1}. 
   
The nonlinear term has the form
\begin{gather}\label{E:NLrity}
\begin{split}
  G_j(k;x):=&\frac{\sigma}{\eps^4}\chi_{D_j}(k+m)(\tilde{\psi}^{(0)}*_{\Bsm}\tilde{\psi}^{(0)}*_{\Bsm}
  \tilde{\bar{\psi}}^{(0)})(k;x) = \frac{\sigma}{\eps^4}\chi_{D_j}(k+m)\left[\sum_{\alpha=1}^N \xi_\alpha *_{\Bsm} \xi_\alpha *_{\Bsm} \xi_\alpha^c\right.\\
    &\left.+ 2\sum_{\substack{\alpha,\beta=1\\\alpha\neq \beta}}^N \xi_\alpha *_{\Bsm} \xi_\beta *_{\Bsm} \xi_\alpha^c
  + \sum_{\substack{\alpha,\beta=1\\\alpha\neq \beta}}^N \xi_\alpha *_{\Bsm} \xi_\alpha *_{\Bsm} \xi_\beta^c
    + \sum_{\substack{\alpha,\beta,\gamma=1\\\alpha\neq \beta,\alpha\neq \gamma, \beta\neq \gamma}}^N \xi_\alpha *_{\Bsm} \xi_\beta
    *_{\Bsm} \xi_\gamma^c\right],
\end{split}
\end{gather} where $\xi_\alpha=\xi_\alpha(k;x) :=
p_{n_\alpha}(k^{(\alpha)};x)\sum_{m\in M_\alpha}\chi_{D_\alpha}(k+m)\hat{A}_\alpha\left(\frac{k+m{-}k^{(\alpha)}}{\eps}\right)e^{\ri m\cdot x}$
and $\xi_\alpha^c=\xi_\alpha^c(k;x) :=
\ov{p_{n_\alpha}}(k^{(\alpha)};x)\sum_{m\in M_\alpha}\chi_{-D_\alpha}(k-m)\hat{\bar{A}}_\alpha\left(\frac{k-m+k^{(\alpha)}}{\eps}\right)e^{-\ri m\cdot x}
$. The last sum or the three last sums in
\eqref{E:NLrity} are absent if $N=2$ or $N=1$ respectively.
$\xi_\alpha *_{\Bsm} \xi_\beta *_{\Bsm} \xi_\gamma^c$ consists of 
terms of the type
\beq\label{E:NL_term}
\begin{split}
g_{noq}(k;x)=&e^{\ri(n+o-q)\cdot x}p_{n_\alpha}(k^{(\alpha)};x)p_{n_\beta}(k^{(\beta)};x)\ov{p_{n_\gamma}}(k^{(\gamma)};x) \int\limits_{\T^2}\int\limits_{\T^2}
\chi_{D_\alpha}(k{-}r{+}n)\hat{A}_\alpha\left(\tfrac{k{-}r{+}n{-}k^{(\alpha)}}{\eps}\right)\times\\
&\qquad \times \ \chi_{D_\beta}(r{-}s{+}o)\hat{A}_\beta\left(\tfrac{r{-}s{+}o{-}k^{(\beta)}}{\eps}\right)\chi_{-D_\gamma}(s{-}q)\hat{\bar{A}}_\gamma\left(\tfrac{s{-}q{+}k^{(\gamma)}}{\eps}\right)\dd s \dd r
\end{split}
\eeq
with $n\in M_\alpha, o \in M_\beta$ and $q\in M_\gamma$.
Clearly, the integration domains can be reduced to $r\in D_{2\eps^r}(k^{(\beta)}-k^{(\gamma)}-o+q)$ and $s\in D_{\eps^r}(-k^{(\gamma)}+q)$. The changes of variables $\tilde{s}:=(s+k^{(\gamma)}-q)/\eps$, and  $\tilde{r} := (r-k^{(\beta)}+k^{(\gamma)}+o-q)/\eps$ yield
\beq\label{E:NL_term_conv}
\begin{split}
g_{noq}&(k;x)=\eps^4 e^{\ri(n+o-q)\cdot x}p_{n_\alpha}(k^{(\alpha)};x)p_{n_\beta}(k^{(\beta)};x)\ov{p_{n_\gamma}}(k^{(\gamma)};x) \times \\
&\int\limits_{D_{2\eps^{r-1}}\cap\frac{\T^2-k^{(\beta)}+k^{(\gamma)}+o-q}\eps}
\int\limits_{D_{\eps^{r-1}}\cap\frac{\T^2+k^{(\gamma)}-q}\eps}\chi_{D_{\eps^{r-1}}}\left(\tfrac{k-(k^{(\alpha)}+k^{(\beta)}-k^{(\gamma)})+n+o-q}{\eps}-\tilde{r}\right)\times\\
&\hat{A}_\alpha\left(\tfrac{k-(k^{(\alpha)}+k^{(\beta)}-k^{(\gamma)})+n+o-q}{\eps}-\tilde{r}\right) \chi_{D_{\eps^{r-1}}}(\tilde{r}-\tilde{s})\hat{A}_\beta(\tilde{r}-\tilde{s})\chi_{D_{\eps^{r-1}}}(\tilde{s})\hat{\bar{A}}_\gamma(\tilde{s})\dd \tilde{s} \dd \tilde{r},
\end{split}
\eeq
where $D_{\eps^{r-1}} = \{p\in\R^2 : |p|<\eps^{r-1}\}$.

Only those combinations of $(n,o,q)$ which produce nonzero values of all the three characteristic functions in \eqref{E:NL_term} for some $k,r,s\in \T^2$ are of relevance. Due to $\chi_{-D_\gamma}(s{-}q)$ we, therefore, require $q-k^{(\gamma)}\in \overline{\T^2}=[-1/2,1/2]^2$, which ensures that $s-q\in -D_\gamma$ is satisfied by some $s\in\T^2$ for any $\eps>0$. The first condition is, thus, 
\beq\label{E:q_cond}
s_0:=q-k^{(\gamma)}\in \overline{\T^2}.
\eeq
Due to $\chi_{D_\beta}(r{-}s{+}o)$ we get the condition $s_0-o+k^{(\beta)}\in \overline{\T^2}$, i.e., 
\beq\label{E:o_cond}
r_0:=s_0-o+k^{(\beta)}\in \overline{\T^2}.
\eeq
Finally, $\chi_{D_\alpha}(k{-}r{+}n)$ enforces $r_0-n+k^{(\alpha)}\in \overline{\T^2}$, i.e., 
\beq\label{E:n_cond}
k_0:=r_0-n+k^{(\alpha)}\in \overline{\T^2}.
\eeq
Statements \eqref{E:q_cond}, \eqref{E:o_cond}, and \eqref{E:n_cond} form the necessary condition 
\beq\label{E:noq_cond1}
s_0:=q-k^{(\gamma)}\in \overline{\T^2}, \qquad r_0:=s_0-o+k^{(\beta)}\in \overline{\T^2}, \quad \text{and} \quad k_0:=r_0-n+k^{(\alpha)}\in \overline{\T^2}
\eeq
for \eqref{E:NL_term} (and thus \eqref{E:NL_term_conv}) not to vanish.

\medskip
Another condition on $(n,o,q)$ appears due to the factor $\chi_{D_j}(k+m)$ in $G_j$. From \eqref{E:NL_term_conv} it is clear that $g_{noq}$ is supported on $k\in D_{\eps^r}( k^{(\alpha)}+k^{(\beta)}-k^{(\gamma)} -n -o+q)$. The factor $\chi_{D_j}(k+m)$ thus annihilates all terms $g_{noq}$ except those for which 
\beq\label{E:noq_cond2} 
  k^{(\alpha)}+k^{(\beta)}-k^{(\gamma)} -n -o+q = k^{(j)}-m.
\eeq
If \eqref{E:noq_cond2} is satisfied, \eqref{E:NL_term_conv} becomes
\beq\label{E:NL_term_conv2}
\begin{split}
g_{noq}&(k;x)=\eps^4 e^{\ri(n+o-q)\cdot x}p_{n_\alpha}(k^{(\alpha)};x)p_{n_\beta}(k^{(\beta)};x)\ov{p_{n_\gamma}}(k^{(\gamma)};x)  \times \\
 &\int\limits_{D_{2\eps^{r-1}}\cap \frac{\T^2-k^{(\beta)}+k^{(\gamma)}+o-q}\eps}\int\limits_{D_{\eps^{r-1}}\cap\frac{\T^2+k^{(\gamma)}-q}\eps}\chi_{D_{\eps^{r-1}}}\left(\tfrac{k-k^{(j)}+m}{\eps}-\tilde{r}\right)\times\\
&\qquad\qquad \hat{A}_\alpha\left(\tfrac{k-k^{(j)}+m}{\eps}-\tilde{r}\right) \chi_{D_{\eps^{r-1}}}(\tilde{r}-\tilde{s})\hat{A}_\beta(\tilde{r}-\tilde{s})\chi_{D_{\eps^{r-1}}}(\tilde{s})\hat{\bar{A}}_\gamma(\tilde{s})\dd \tilde{s} \dd \tilde{r}.
\end{split}
\eeq

\medskip
As a result, the term $A_\alpha A_\beta \bar{A}_\gamma$ will enter the $j-$th equation of the coupled mode system provided there exist $n\in M_\alpha, o\in M_\beta$ and $q\in M_\gamma$ such that \eqref{E:noq_cond1} holds and such that \eqref{E:noq_cond2} holds for some $m\in M_j$. Let us denote the set of $(n,o,q)$ that satisfy \eqref{E:noq_cond1} and \eqref{E:noq_cond2} by $\CA_{\alpha,\beta,\gamma,j,m}$. 

The sum of the terms \eqref{E:NL_term_conv2} over $(n,o,q)\in \CA_{\alpha,\beta,\gamma,j,m}$ yields a double convolution integral over the full discs $\tilde{r}\in D_{2\eps^{r-1}}$ and $\tilde{s}\in D_{\eps^{r-1}}$, i.e.,
\beq\label{E:NL_term_conv_sum}
\begin{split}
&(\xi_\alpha *_{\Bsm} \xi_\beta *_{\Bsm} \xi_\gamma^c)(k;x)=\eps^4 e^{\ri(k^{(\alpha)}+k^{(\beta)}-k^{(\gamma)}-k^{(j)}+m)\cdot x}p_{n_\alpha}(k^{(\alpha)};x)p_{n_\beta}(k^{(\beta)};x)\ov{p_{n_\gamma}}(k^{(\gamma)};x) \times \\  
&\int_{D_{2\eps^{r-1}}}\int_{D_{\eps^{r-1}}}\hspace{-0.4cm}\chi_{D_{\eps^{r-1}}}\left(\tfrac{k-k^{(j)}+m}{\eps}-\tilde{r}\right)
\hat{A}_\alpha\left(\tfrac{k-k^{(j)}+m}{\eps}-\tilde{r}\right) 
\chi_{D_{\eps^{r-1}}}(\tilde{r}-\tilde{s})\hat{A}_\beta(\tilde{r}-\tilde{s})\chi_{D_{\eps^{r-1}}}(\tilde{s})\hat{\bar{A}}_\gamma(\tilde{s})\dd \tilde{s} \dd \tilde{r},
\end{split}
\eeq
where $e^{\ri(n+o-q)\cdot x}$ was replaced by $e^{\ri(k^{(\alpha)}+k^{(\beta)}-k^{(\gamma)}-k^{(j)}+m)\cdot x}$ due to \eqref{E:noq_cond2}. 


We return now to equation \eqref{E:psi2_eq} for $\tilde{\psi}^{(2)}$ on 
$k\in (D_j-m)\cap \T^2$. Its
solvability condition is $L^2(\P^2)$-orthogonality to
$\text{Ker}(\tilde{L}(k^{(j)}-m;x)-\omega_*)=\text{span}\{\cup_{l} p_{n_l}(k^{(j)};x)e^{\ri m\cdot x}$ s.t. $\omega_{n_l}(k^{(j)})=\omega_*\}$. Clearly, the dimension of the kernel is at most $N$. The value $N$ is attained if $k^{(1)}= \ldots =k^{(N)}$.

In the linear terms in \eqref{E:psi2_eq} the factor $e^{\ri m\cdot x}$ is canceled in the inner product with $p_{n_l}(k^{(j)};x)e^{\ri m\cdot x}$ so that the same solvability condition  holds for all $m$. 
The range of $p^{(j,m)}$ is a different section of the disc $D_{\eps^{r-1}}$ for each $m$. The section is an $(1/|M_j|)$-th of the full disc so that these $|M_j|$ conditions build one equation in $p \in D_{\eps^{r-1}}$.

The resulting $N$ equations are CMEs in Fourier variables $p \in D_{\eps^{r-1}}$:
\begin{gather}\label{E:CME_gen_Fourier}
\Omega\hat{A}_j-\left(\frac{1}{2}\pa_{k_1}^2\omega_{n_j}(k^{(j)})p_1^{2}
  +\frac{1}{2}\pa_{k_2}^2\omega_{n_j}(k^{(j)})p_2^{2}
  +\pa_{k_1}\pa_{k_2}\omega_{n_j}(k^{(j)})p_1p_2\right)\hat{A}_j-\hat{\CN}_j=0,
\end{gather} 
$j\in \{1,\ldots, N\}$, where $\hat{\CN}_j(p^{(j,m)}) = \langle
G_j(\eps p^{(j,m)}+k^{(j)}-m;\cdot),p_{n_j}(k^{(j)};\cdot)e^{\ri m\cdot \cdot}\rangle_{L^2(\P^2)}$. 

For sufficiently smooth $A_j$ we can neglect the contribution to $\hat{A}_j$ from $p\in \R^2\setminus D_{\eps^{r-1}}$ or, for simplicity, assume that the 
$\hat{A}_j$ satisfy \eqref{E:CME_gen_Fourier} also there. Equation \eqref{E:CME_gen_Fourier} is then posed on $p\in \R^2$. Performing the inverse Fourier transform yields the CMEs 
\begin{gather}\label{E:CME_gen_iFourier}
\Omega A_j+\left(\frac{1}{2}\pa_{k_1}^2\omega_{n_j}(k^{(j)})\pa_{y_1}^2
  +\frac{1}{2}\pa_{k_2}^2\omega_{n_j}(k^{(j)})\pa_{y_2}^2
  +\pa_{k_1}\pa_{k_2}\omega_{n_j}(k^{(j)})\pa_{y_1}\pa_{y_2}\right)
A_j-\CN_j=0.
\end{gather}
The structure and coefficients in $\CN_j$ for our
example \eqref{E:V} will be discussed in \S\ref{cmebloch-sec}.

\bigskip

In order to make the discussion of the asymptotic hierarchy complete, we need to mention the part of the $k-$domain outside 
the neighborhoods of $k^{(j)}$. For $k \in \T^2$ such that $k+m \in \T^2\setminus D_j$ for all $m \in M_j$ we have $\bigl[\tilde{L}(k^{(j)}-m;x)-\omega_*\bigr]\tilde{\psi}^{(n)}(k;x)=0$ for $n=1,2$ so that $\tilde{\psi}^{(0)}(k;\cdot)\equiv \tilde{\psi}^{(1)}(k;\cdot)\equiv 0$ for such $k$.

The appearance of
second derivatives of the bands $\omega_{n_j}$ in \reff{E:CME_gen_Fourier} is due to the
following 
\blem\label{L:omega_der_Bloch} For any $l,m\in\{1,2\}$
\[
\pa_{k_l}\pa_{k_m}\omega_{n_j}(k^{(j)})=2\delta_{lm}-2\langle
(\ri\pa_{x_m}{-}k_m^{(j)})\pa_{k_l}p_{n_j}(k^{(j)};\cdot)
+(\ri\pa_{x_l}{-}k_l^{(j)})\pa_{k_m}p_{n_j}(k^{(j)};\cdot),
p_{n_j}(k^{(j)};\cdot)\rangle_{L^2(\P^2)},
\]
where $\delta_{lm}$ is the Kronecker delta.  \elem
\begin{proof} This follows from differentiation of
  $[\tilde{L}(k;x)-\omega_{n_j}(k)]p_{n_j}(k;x)=0$ w.r.t.~$k$.
\end{proof}

As the next lemma shows, for even potentials $V(x)$ the mixed derivatives of $\omega_{n_j}$ are zero
whenever $\omega_{n_j}(k^{(j)})$ has geometric multiplicity one and the extremal point $k^{(j)}$ coincides with one of the vertices of
the first irreducible Brillouin zone or of its reflection.
\blem\label{L:cross_deriv_S} Suppose $V(x)$ is even in $x_1$ as well as in $x_2$. Then $\pa_{k_1}\pa_{k_2}\omega_{n_j}(k^{(j)})=
\pa_{k_2}\pa_{k_1}\omega_{n_j}(k^{(j)})=0 \ $  if $k^{(j)}\in \Sigma =
\{\Gamma,X,X',M\}$ and provided $\omega_{n_j}(k^{(j)})$ has geometric multiplicity 1 as an eigenvalue of \eqref{E:qp_problem}.  \elem
\begin{proof} Take $l,m\in\{1,2\}, l\neq m$. As $\ri\pa_{x_m}{-}k_m^{(j)}$ is
  self-adjoint, we have \begin{gather}\label{E:in_prod_gen_Bloch} \langle
  (\ri\pa_{x_m}{-}k_m^{(j)})\pa_{k_l}p_{n_j}(k^{(j)};\cdot),
  p_{n_j}(k^{(j)};\cdot)\rangle_{L^2(\P^2)}= \langle
  \pa_{k_l}p_{n_j}(k^{(j)};\cdot),
  (\ri\pa_{x_m}{-}k_m^{(j)})p_{n_j}(k^{(j)};\cdot)\rangle_{L^2(\P^2)}.  \end{gather}
  Based on \eqref{E:Bloch_form} we have
  $(\ri\pa_{x_m}{-}k_m^{(j)})p_{n_j}(k^{(j)};x) = -\ri \er^{-\ri k^{(j)}\cdot
    x}\pa_{x_m}u_{n_j}(k^{(j)};x)$.  Next, $\pa_{k_l}p_{n_j}(k^{(j)};x) =
  \ri \er^{-\ri k^{(j)}\cdot x}v_{n_j}^{(x_l)}(k^{(j)};x)$, where $v_{n_j}^{(x_l)}(k^{(j)};x)$ is the
  generalized Bloch function \cite{PSK04} solving \begin{gather}\label{E:gen_Bloch_eq}
  \begin{split}
    &[L-\omega_*]u=2\pa_{x_l}u_{n_j}(k^{(j)};x), \qquad u(2\pi,x_2)=\er^{\ri 2\pi
      k_1^{(j)}}u(0,x_2), \quad u(x_1,2\pi)=\er^{\ri 2\pi k_2^{(j)}}u(x_1,0),
  \end{split}
  \end{gather} analogously to \eqref{E:gen_Bloch_fn_eq}. The inner product in
  \eqref{E:in_prod_gen_Bloch} thus becomes $\langle
  -v_{n_j}^{(x_l)}(k^{(j)};\cdot),\pa_{x_m}u_{n_j}(k^{(j)};\cdot)
  \rangle_{L^2(\P^2)}$. Because $k^{(j)}\in \Sigma$, $u_{n_j}$ is even or odd in
  $x_l$ (Lemma \ref{L:sym_S}). From \eqref{E:gen_Bloch_eq} it is clear that
  $v_{n_j}^{(x_l)}(k^{(j)};x)$ has the opposite symmetry (odd or even
  respectively) in $x_l$. Thus, the integrand is odd in $x_l$ and the integral
  vanishes upon shifting the integration domain to $[-\pi,\pi]^2$.
\end{proof}

\subsubsection{CMEs for the Example \eqref{E:V}}\label{cmebloch-sec}
We now calculate the explicit form of the CMEs \reff{E:CME_gen_iFourier} in
the vicinity of the five gap edges in the example \eqref{E:V} with
$\eta=5.35$. It turns out that only few terms are nonzero in the nonlinearity
$\CN_j$ for this case. Of special importance is the edge $\omega_* = s_5$, where $k^{(j)}\notin \Sigma$ and, indeed,
$\pa_{k_1}\pa_{k_2}\omega_{n_j}(k^{(j)})\neq 0$.

In order to numerically evaluate the coefficients $\pa_{k_l}\pa_{k_m}\omega_{n_j}(k^{(j)})$ given in Lemma \ref{L:omega_der_Bloch}, the functions 
$\pa_{k_l}p_{n_j}(k^{(j)};x)$ have to be computed. They are solutions of the singular system \eqref{E:gen_Bloch_pn}
but as the right-hand side is orthogonal to the kernel of $\tilde{L}(k^{(j)};x)-\omega_*$, the BiCG algorithm
can be used as long as the initial guess is orthogonal to the kernel. We work in a 4th order finite difference 
discretization and use an incomplete LU preconditioning for BiCG.

\paragraph{CMEs near $\omega_*=s_1$:}
Only one extremum defines the edge $\omega_*=s_1$, namely the minimum of the
band $\omega_1$ at $k=\Gamma$. Therefore, $N=1,n_1=1$ and $k^{(1)}=\Gamma$.
Because $k^{(1)}\in \text{int} (\T^2)$, we get $M_1=\{(0,0)^T\}$. Thus \begin{gather}\label{E:CME_s1} \left[\Omega +\al
  (\pa_{y_1}^2+\pa_{y_2}^2)\right]A-\sigma\ga|A|^2A=0, \end{gather} where
$\al=\frac{1}{2}\pa_{k_1}^2\omega_1(\Gamma)
=\frac{1}{2}\pa_{k_2}^2\omega_1(\Gamma)$ and $\ga = \langle
p_1(\Gamma;\cdot)^2,p_1(\Gamma;\cdot)^2\rangle_{L^2(\P^2)} =
\|p_1(\Gamma;\cdot)\|_{L^4(\P^2)}^4$. The identity in $\al$ holds due to
\eqref{E:sym3}. The numerically obtained values are $\al \approx 0.62272$ and
$\ga \approx 0.048029$.

\paragraph{CMEs near $\omega_*=s_2$:}

Here the linear problem is characterized by 
$N=1,n_1=1$ and $k^{(1)}=M=(1/2,1/2)$ and we get $M_1=\{(0,0)^T,(1,0)^T,(0,1)^T,(1,1)^T\}$.
The resulting CMEs have the form \eqref{E:CME_s1}. We determine next the coefficient of the nonlinearity $|A|^2A$.
In \eqref{E:NL_term_conv} we have $\alpha=\beta =\gamma=1$ and $k^{(\alpha)}+k^{(\beta)}-k^{(\gamma)}-k^{(j)}=(0,0)^T$. 
We carry out a straightforward sweep through all the possible combinations $(n,o,q,m)$ (performed using a Matlab script) to determine those that satisfy \eqref{E:noq_cond1} and \eqref{E:noq_cond2}. As a result we have
\bi
\item $m=M_1(:,1)=\bspm 0\\0 \espm$: 
$\ds 
(n,o,q)^T \in \left\{\bspm 0 & 0 \\0 &0\\0&0\espm,\bspm 0 & 0 \\1 &0\\1&0\espm, \bspm 0 & 0 \\0 &1\\0&1\espm,\bspm 0 & 0 \\1 &1\\1&1\espm\right\}$
\item $m=M_1(:,2)=\bspm 1\\0 \espm $: 
$\ds (n,o,q)^T \in \left\{\bspm 1 & 0 \\0 &0\\0&0\espm, \bspm 1& 0 \\1 &0\\1&0\espm,\bspm 1 & 0 \\0 &1\\0&1\espm, \bspm 1& 0 \\1&1\\1&1\espm \right\}$
\item $m=M_1(:,3)=\bspm 0\\1 \espm $: 
$\ds (n,o,q)^T \in \left\{\bspm 0 & 1 \\0 &0\\0&0\espm, \bspm 0 & 1 \\1 &0\\1&0\espm, \bspm 0 & 1 \\0 &1\\0&1\espm, \bspm 0 & 1 \\1 &1\\1&1\espm\right\}$
\item $m=M_1(:,4)=\bspm 1\\1 \espm $: 
$\ds (n,o,q)^T \in \left\{\bspm 1 & 1 \\0 &0\\0&0\espm, \bspm 1 & 1 \\1 &0\\1&0\espm, \bspm 1 & 1 \\0 &1\\0&1\espm, \bspm 1 & 1 \\1 &1\\1&1\espm\right\},$
\ei
where we have used $M_j(:,l)$ to denote the $l-$th vector in $M_j$.

The CME coefficients are thus $\al=\frac{1}{2}\pa_{k_1}^2\omega_1(M)
=\frac{1}{2}\pa_{k_2}^2\omega_1(M)$ and $\ga = \langle
p_1(M;\cdot)^2,p_1(M;\cdot)^2\rangle_{L^2(\P^2)} =
\|p_1(M;\cdot)\|_{L^4(\P^2)}^4$. The identity in $\al$ holds due to \eqref{E:sym3}.
Numerically, $\al\approx -1.971217$ and $\ga \approx 0.076442$.

\paragraph{CMEs near $\omega_*=s_3$:}
Here  $N=2,n_1=n_2=2,k^{(1)}=X$ and
$k^{(2)}=X'$. We have thus $M_1=\{(0,0)^T,(1,0)^T\}$ and $M_2=\{(0,0)^T,(0,1)^T\}$.

For $\CN_j$ we sweep again through all the possible combinations $(n,o,q,m)$ for both $j=1$ and $j=2$. The results are summarized in Table \ref{T:s3_NL_terms}.
\medskip

\begin{table}[h!]
\footnotesize
\begin{center}
\begin{tabular}{|c|c|c|c|c|c|c|}
\hline
term  & $\bspm\alpha \\ \beta\\ \gamma\espm$ &$j$ & $k^{(\alpha)}+k^{(\beta)}$& \multicolumn{2}{|c|}{$(n,o,q)^T$ satisfying \eqref{E:noq_cond1} and \eqref{E:noq_cond2}}& coefficient of \\
in $\CN_j$ & & & $-k^{(\gamma)}-k^{(j)}$ & $m=M_j(:,1)$ & $m=M_j(:,2)$ & the term in $\sigma \CN_j$\\
\hline
$|A_1|^2A_1$ & $\bspm 1\\1\\1\espm$ & 1 & $\bspm 0\\0\espm$ & $\bspm 0 & 0 \\0 &0\\0&0\espm, \bspm 0&0\\1&0\\1&0\espm$ & $\bspm 1&0\\0&0\\0&0\espm, \bspm 1&0\\1&0\\1&0\espm$ & $\langle p_2(X,\cdot)^2, p_2(X,\cdot)^2\rangle$\\
\cline{3-7}
& & 2 & $\bspm 1/2\\-1/2\espm$ & / & / & 0\\
\hline  
$|A_2|^2A_2$ & $\bspm 2\\2\\2\espm$ & 1 & $\bspm -1/2\\1/2\espm$ & / & / & 0 \\
\cline{3-7}
& & 2 & $\bspm 0\\0\espm$ & $\bspm 0 & 0 \\0 &0\\0&0\espm, \bspm 0 & 0 \\0 &1\\0&1\espm$  & $\bspm 0 & 1 \\0 &0\\0&0\espm, \bspm 0 & 1 \\0 &1\\0&1\espm$  & $\langle p_2(X',\cdot)^2, p_2(X',\cdot)^2\rangle$\\
\hline  
$|A_1|^2A_2$ & $\bspm 1\\2\\1\espm,$ & 1 & $\bspm -1/2\\1/2\espm$ & /& / & 0\\
\cline{3-7}
& $\bspm 2\\1\\1\espm$ & 2 & $\bspm 0\\0\espm$ & $\bspm 0 & 0 \\0 &0\\0&0\espm, \bspm 1 & 0 \\0 &0\\1&0\espm$& $\bspm 0 & 0 \\0 &1\\0&0\espm, \bspm 1 & 0 \\0 &1\\1&0\espm$ & $2 \langle |p_2(X,\cdot)|^2, |p_2(X',\cdot)|^2\rangle$\\
\hline  
$|A_2|^2A_1$ & $\bspm 1\\2\\2\espm,$ & 1 & $\bspm 0\\0\espm$ & $\bspm 0 & 0 \\0 &0\\0&0\espm, \bspm 0 & 1 \\0 &0\\0&1\espm$& $\bspm 0 & 0 \\1 &0\\0&0\espm, \bspm 0 & 1 \\1 &0\\0&1\espm $ & $2 \langle |p_2(X,\cdot)|^2, |p_2(X',\cdot)|^2\rangle$\\
\cline{3-7}
& $\bspm 2\\1\\2\espm$ & 2 & $\bspm 1/2\\-1/2\espm$ & / &  / & 0 \\
\hline  
$A_1^2A_2^*$ & $\bspm 1\\1\\2\espm$ & 1 & $\bspm 1/2\\-1/2\espm$ & / & / & 0 \\
\cline{3-7}
& & 2 & $\bspm 1\\-1\espm$ & $\bspm 0 & 0 \\1 &0\\0&1\espm, \bspm 1 & 0 \\0 &0\\0&1\espm$ &  $\bspm 0 & 0 \\1 &0\\0&0\espm, \bspm 1 & 0 \\0 &0\\0&0\espm $ & $\langle e^{\ri(1,-1)^T \cdot \cdot}p_2(X,\cdot)^2, p_2(X',\cdot)^2\rangle$ \\
\hline 
$A_2^2A_1^*$ & $\bspm 2\\2\\1\espm$ & 1 & $\bspm -1\\1\espm$ & $\bspm 0 & 0 \\0 &1\\1&0\espm, \bspm 0 & 1 \\0 &0\\1&0\espm$  & $\bspm 0 & 0 \\0 &1\\0&0\espm, \bspm 0 & 1 \\0 &0\\0&0\espm $ & $\langle e^{\ri(-1,1)^T \cdot \cdot}p_2(X',\cdot)^2, p_2(X,\cdot)^2\rangle$ \\
\cline{3-7}
& & 2 & $\bspm -1/2\\1/2\espm$ & / & /  & 0 \\
\hline 
\end{tabular}\caption{Calculation of the nonlinearity terms for the CME near $\omega_*=s_3$.\label{T:s3_NL_terms}}
\end{center}
\end{table}

\medskip

\normalsize
The resulting CMEs are
\begin{equation}\label{E:CME_s3}
  \begin{split}
    \left[\Omega + \al_1 \pa_{y_1}^2+\al_2\pa_{y_2}^2\right]A_1-\sigma \left[\ga_1|A_1|^2A_1+\ga_2(2|A_2|^2A_1+A_2^2\bar{A_1})\right]=&0,\\
    \left[\Omega + \al_2 \pa_{y_1}^2+\al_1\pa_{y_2}^2\right]A_2-\sigma
    \left[\ga_1|A_2|^2A_2+\ga_2(2|A_1|^2A_2+A_1^2\bar{A_2})\right]=&0,
  \end{split}
\end{equation}
where
\begin{gather*}
\begin{split}
  \al_1&=\frac{1}{2}\pa_{k_1}^2\omega_2(X)=\frac{1}{2}\pa_{k_2}^2\omega_2(X'), 
  \quad \al_2=\frac{1}{2}\pa_{k_2}^2\omega_2(X)=\frac{1}{2}\pa_{k_1}^2\omega_2(X'), \\
  \ga_1&=\langle p_2(X;\cdot)^2,p_2(X;\cdot)^2\rangle_{L^2(\P^2)}=\langle p_2(X';\cdot)^2,p_2(X';\cdot)^2\rangle_{L^2(\P^2)}\\
& = \|p_2(X;\cdot)\|^4_{L^4(\P^2)}=\|p_2(X';\cdot)\|^4_{L^4(\P^2)}, \\ 
  \ga_2&=\langle e^{\ri (1,-1)^T\cdot \cdot} p_2(X;\cdot)^2, p_2(X';\cdot)^2\rangle_{L^2(\P^2)}=\langle e^{\ri (-1,1)^T\cdot \cdot}p_2(X';\cdot)^2,p_2(X;\cdot)^2\rangle_{L^2(\P^2)}\\
  &=\langle |p_2(X;\cdot)|^2,|p_2(X';\cdot)|^2\rangle_{L^2(\P^2)}.
\end{split}
\end{gather*}
The identities in $\al_1, \al_2$ and $\ga_1$ hold due to \eqref{E:sym3}.  The
equalities in $\ga_2$ yield $\ga_2\in\R$ and follow from the fact that $u_2(X,x)=e^{\ri x_1/2}p_2(X;x)$ and $u_2(X',x)=e^{\ri x_2/2}p_2(X';x)$ are real. In detail
\begin{align*}
  \int_{\P^2}&e^{\ri x_1} p_2(X;x)^2 e^{-\ri x_2} \overline{p_2(X';x)}^2 \dd x =  \int_{\P^2} u_2(X;x)^2 \overline{u_2(X';x)}^2 \dd x\\
  =\int_{\P^2}& u_2(X;x)^2 u_2(X';x)^2 \dd x =\int_{\P^2} u_2(X';x)^2 \overline{u_2(X;x)}^2 \dd x = \int_{\P^2}e^{\ri x_2} p_2(X';x)^2 e^{-\ri x_1} \overline{p_2(X;x)}^2 \dd x. 
 \end{align*}

The CMEs \eqref{E:CME_s3} are thus identical to those in Sec. \ref{S:e3} derived in physical variables.
Numerically,  $\al_1 \approx
2.599391, \al_2 \approx 0.040561, \ga_1 \approx 0.090082, \ 
\text{and} \ \ga_2 \approx 0.003032$.

\paragraph{CMEs near $\omega_*=s_4$:}
Here $N=1,n_1=5$ and $k^{(1)}=M$. This case is completely analogous to $\omega_*=s_2$. The CMEs are \eqref{E:CME_s1} with
$\al=\frac{1}{2}\pa_{k_1}^2\omega_5(M)=\frac{1}{2}\pa_{k_2}^2\omega_5(M)
\approx -0.300655$ and $\ga = \langle
p_5(M;\cdot)^2,p_5(M;\cdot)^2\rangle_{L^2(\P^2)} =
\|p_5(M;\cdot)\|_{L^4(\P^2)}^4 \approx 0.039755$.

\paragraph{CMEs near $\omega_*=s_5$:}
Here $N=4,n_1=n_2=n_3=n_4=6, k^{(1)}=(k_c,k_c), k^{(2)}=({-}k_c,k_c),
k^{(3)}=(-k_c,-k_c)$ and $k^{(4)}=(k_c,-k_c)$, where $k_c \approx 0.439028$.
This is an important case in our example because $k^{(j)}\notin \Sigma$ here.
Note that because $k^{(j)}\in \text{int} (\T^2)$ for all $j\in \{1, \ldots, N\}$, we have
$M_1=\ldots=M_4=\{(0,0)^T\}$.

We start with the last two sums of $G$ (see
\eqref{E:NL_term}). Terms of the type $\xi_l *_{\Bsm} \xi_l *_{\Bsm} \xi_m^c$ (the
third sum in $G$) do not contribute to the CMEs because $2k^{(l)}-k^{(m)}$ is
not congruent to any $k^{(j)}, j\in \{1,\ldots, 4\}$ for any choice of
$l,m\in\{1,\ldots,4\}, l \neq m$. For example, $2k^{(1)}-k^{(2)}=(3k_c,k_c)$,
which is not congruent to any $k^{(j)}$ since $k_c\notin\{0, 1/2\}$. Only four
terms of the type $\xi_l *_{\Bsm} \xi_m *_{\Bsm} \xi_n^c$ (the last sum in $G$)
contribute to the CMEs, namely $\xi_2 *_{\Bsm} \xi_4 *_{\Bsm} \xi_3^c$ to the
equation for $k\in D_1$, $\xi_1 *_{\Bsm} \xi_3 *_{\Bsm} \xi_4^c$ to the equation for
$k\in D_2$, $\xi_2 *_{\Bsm} \xi_4 *_{\Bsm} \xi_1^c$ to the equation for $k\in D_3$
and $\xi_1 *_{\Bsm} \xi_3 *_{\Bsm} \xi_2^c$ to the equation for $k\in D_4$. This is
because $k^{(2)}+k^{(4)}-k^{(3)}=k^{(1)}, k^{(1)}+k^{(3)}-k^{(4)}=k^{(2)},
k^{(2)}+k^{(4)}-k^{(1)}=k^{(3)}$ and $k^{(1)}+k^{(3)}-k^{(2)}=k^{(4)}$. The
other terms in the last sum in $G$ do not contribute. As an example,
$k^{(1)}+k^{(2)}-k^{(3)}=(k_c,3k_c)$.

Another consequence of $k^{(j)}\notin \Sigma$ is that Lemma \ref{L:cross_deriv_S}
does not apply and mixed derivatives of $A_j$ may appear.  The system of CMEs

thus becomes
\begin{equation}\label{E:CME_s5_lg}
  \begin{split}
    0=&\left[\Omega + \al_1
      (\pa_{y_1}^2{+}\pa_{y_2}^2)+\al_2\pa_{y_1}\pa_{y_2}\right]A_1\\
&-\sigma \bigg[\ga_1|A_1|^2A_1 + 2\big(\ga_2\left(|A_2|^2+|A_4|^2\right)A_1
+\tilde{\ga}_1|A_3|^2A_1+\tilde{\ga}_2 A_2A_4\bar{A}_3\big)\bigg], \\
0=&\left[\Omega + \al_1 (\pa_{y_1}^2{+}\pa_{y_2}^2)-\al_2\pa_{y_1}\pa_{y_2}
\right]A_2\\
&-\sigma \bigg[\ga_1|A_2|^2A_2 + 2\big(\ga_2\left(|A_1|^2+|A_3|^2\right)A_2
+\tilde{\ga}_1|A_4|^2A_2 +\bar{\tilde{\ga}}_2 A_1A_3\bar{A}_4\big)\bigg],\\
0=&\left[\Omega + \al_1
  (\pa_{y_1}^2{+}\pa_{y_2}^2)+\al_2\pa_{y_1}\pa_{y_2}\right]A_3\\
&-\sigma \bigg[\ga_1|A_3|^2A_3 + 2\big(\ga_2\left(|A_2|^2+|A_4|^2\right)A_3
+\tilde{\ga}_1|A_1|^2A_3 +\tilde{\ga}_2 A_2A_4\bar{A}_1\big)\bigg],\\
0=&\left[\Omega + \al_1
  (\pa_{y_1}^2{+}\pa_{y_2}^2)-\al_2\pa_{y_1}\pa_{y_2}\right]A_4\\
&-\sigma \bigg[\ga_1|A_4|^2A_4 + 2\big(\ga_2\left(|A_1|^2+|A_3|^2\right)A_4
+\tilde{\ga}_1|A_2|^2A_4 +\bar{\tilde{\ga}}_2 A_1A_3\bar{A}_2\big)\bigg], 
  \end{split}
\end{equation}
where
$$
\begin{array}{rl}
  \al_1=&\frac{1}{2}\pa_{k_1}^2\omega_6(k_c((-1)^m,(-1)^n))=\frac{1}{2}\pa_{k_2}^2\omega_6(k_c((-1)^p,(-1)^q))
  \ \text{ for any } \ m,n,p,q \in \{0,1\},\\[2mm]
  \al_2 =&
  \pa_{k_1}\pa_{k_2}\omega_6(k_c,k_c)=\pa_{k_1}\pa_{k_2}\omega_6(-k_c,-k_c)=-\pa_{k_1}\pa_{k_2}\omega_6(-k_c,k_c)=-\pa_{k_1}\pa_{k_2}\omega_6(k_c,-k_c),\\[2mm]
  \ga_1=&\langle |p_6(k_c((-1)^m,(-1)^n);\cdot)|^2,|p_6((k_c((-1)^m,(-1)^n);\cdot)|^2\rangle_{L^2(\P^2)}=\|p_6((k_c,k_c);\cdot)\|^4_{L^4(\P^2)} \\ 
  &\text{for any} \ m,n\in \{0,1\},\\[2mm] 
  \ga_2=&\langle|p_6((-k_c,k_c);\cdot)|^2,|p_6((k_c,k_c);\cdot)|^2\rangle_{L^2(\P^2)}=\langle|p_6((k_c,-k_c);\cdot)|^2,|p_6((k_c,k_c);\cdot)|^2\rangle_{L^2(\P^2)}\\[2mm]
  =&\langle|p_6((-k_c,-k_c);\cdot)|^2,|p_6((-k_c,k_c);\cdot)|^2\rangle_{L^2(\P^2)}=\langle|p_6((k_c,-k_c);\cdot)|^2,|p_6((-k_c,-k_c);\cdot)|^2\rangle_{L^2(\P^2)},\\[2mm]
  \tilde{\ga_1}=&\langle|p_6((-k_c,-k_c);\cdot)|^2,|p_6((k_c,k_c);\cdot)|^2\rangle_{L^2(\P^2)}=\langle|p_6((k_c,-k_c);\cdot)|^2,|p_6((-k_c,k_c);\cdot)|^2\rangle_{L^2(\P^2)},\\[2mm]
  \tilde{\ga_2}=&\langle p_6((-k_c,k_c);\cdot)p_6((k_c,-k_c);\cdot),p_6((-k_c,-k_c);\cdot)p_6((k_c,k_c);\cdot)\rangle_{L^2(\P^2)}.
\end{array}
$$
The identities in $\al_1$, $\al_2$ and 
$\ga_1$ are due to \eqref{E:sym1} and the identities in $\ga_2$ due
to \eqref{E:sym1} and \eqref{E:sym3}.

Moreover, $\ga_1=\tilde{\ga_1}$ and $\ga_2=\tilde{\ga_2}$ due to
\eqref{E:sym2}. This also implies $\tilde{\ga}_2=\bar{\tilde{\ga}}_2$. Using
these identities, we arrive at the system
\begin{equation}\label{E:CME_s5}
  \begin{split}
    \left[\Omega + \al_1 (\pa_{y_1}^2{+}\pa_{y_2}^2)+\al_2\pa_{y_1}\pa_{y_2}\right]A_1-\sigma \left[\ga_1\left(|A_1|^2{+}2|A_3|^2\right)A_1+2\ga_2\left((|A_2|^2{+}|A_4|^2)A_1+A_2A_4\bar{A}_3\right)\right]=&0,\\
    \left[\Omega + \al_1 (\pa_{y_1}^2{+}\pa_{y_2}^2)-\al_2\pa_{y_1}\pa_{y_2}\right]A_2-\sigma \left[\ga_1\left(|A_2|^2{+}2|A_4|^2\right)A_2+2\ga_2\left((|A_1|^2{+}|A_3|^2)A_2+A_1A_3\bar{A}_4\right)\right]=&0,\\
    \left[\Omega + \al_1 (\pa_{y_1}^2{+}\pa_{y_2}^2)+\al_2\pa_{y_1}\pa_{y_2}\right]A_3-\sigma \left[\ga_1\left(|A_3|^2{+}2|A_1|^2\right)A_3+2\ga_2\left((|A_2|^2{+}|A_4|^2)A_3+A_2A_4\bar{A}_1\right)\right]=&0,\\
    \left[\Omega + \al_1 (\pa_{y_1}^2{+}\pa_{y_2}^2)-\al_2\pa_{y_1}\pa_{y_2}\right]A_4-\sigma \left[\ga_1\left(|A_4|^2{+}2|A_2|^2\right)A_4+2\ga_2\left((|A_1|^2{+}|A_3|^2)A_4+A_1A_3\bar{A}_2\right)\right]=&0.\\
  \end{split}
\end{equation}

The numerical values of the coefficients are $\al_1 \approx 6.051248, \al_2
\approx 0.096394, \ga_1 \approx 0.039118$ and $\ga_2 \approx 0.029926$.

\section{Justification of the Coupled Mode Equations}
\label{S:justification}
If $\om=\om_*+\eps^2\Om$ is in the band gap, 
then families of solitons, i.e., of smooth exponentially localized 
solitary wave
solutions, are known for many classes of CMEs \cite{SY07}. 
However, as already noted in the introduction, 
the formal derivation of the CMEs
in \S\ref{S:derive_CME}, discarding some error at higher order in $\eps$, does
not imply that localized solutions of the CMEs yield gap solitons of \eqref{E:GP_stat}.  
For this
we need to estimate the error in some function space and show 
persistence of the CME solitons under perturbation of the CME including the
error.  We proceed similarly to \cite{DPS08}. However, 
as function space we choose $H^s(\R^2)$ with $s> 1$, 
in contrast to $\CF^{-1}L^1_s(\R^2)$ in \cite{DPS08}. 
The latter is possible in the separable case but there is 
the technical obstacle of the extension of \cite[(3.7)]{DPS08} 
to the nonseparable case. 
On one hand, $L^1_s(\R^2)$ in Fourier space gives a direct 
pointwise estimate in 
physical space via $\|\phi\|_{C^{ s}}\le C\|\hat{\phi}\|_{L^1_s}$. 
On the other hand, working in Hilbert spaces $H^s$ is conceptually simpler 
since it allows for going back and forth between physical space (for 
the nonlinearity) and Bloch space (for the linear part). Moreover, localization of the resulting gap solitons follows directly in $H^s$ spaces. By 
the embedding $\|\phi\|_{C^k}\leq C \|\phi\|_{H^s}$ for $k<s-1$ pointwise 
estimates are still obtainable although these are typically far from 
optimal.

\subsection{Preliminaries}
We have the asymptotic distribution
\begin{equation}\label{E:bands_asymp}
  C_1 n \leq \omega_n(k) \leq C_2 n \quad \forall n \in \N, \forall k \in \T^2
\end{equation}
of bands $\omega_n(k)$, with some constants $C_1, C_2$ independent of $k$ and
$n$. This follows from the asymptotic ``density of states''(see p. 55 of
\cite{Hoerm_85})
$N(\lambda;k) = a \lambda+\CO(\lambda^{\frac{1}{2}})$ as 
$\lambda \rightarrow \infty$, 
where $N(\lambda;k)$ is the number of eigenvalues $\omega_n(k)$ smaller
than or equal to $\lambda$.

We introduce the diagonalization operator
$$
\CD(k)_{k\in\T^2}: \wt{\phi}(k;x) {\to} \vec{\wt{\phi}}(k),
\ \wt{\phi}_n(k)=\spr{\wt{\phi}(k;\cdot),p_n(k;\cdot)}_{L^2(\P^2)}.
$$
Based on \reff{E:bands_asymp} we may estimate $\CD$.  Similarly to \cite[Lemma
3.3]{BSTU06} we find that $\CD(k)$ for all $k$ is an isomorphism between
$H^s(\P^2)$ and $\ell^2_{s}$, $s\ge 0$, where
$$
\ell^2_s:=\{\vec{\wt\phi}: \|\vtp\|^2_{\ell^2_s}:= \sum_{j\in\N}
|\wt{\phi}_j|^2(1+j)^s<\infty\}.
$$ 
Moreover, $\|\CD(k)\|,\|\CD(k)^{-1}\|\le C$ independent of $k$.  Thus,
$\phiti\mapsto\vtp$ is an isomorphism between $L^2(\T^2,H^s(\P^2))$ and
$L^2(\T^2,\ell^2_{s})$ and therefore we have 
\blem\label{L:norm_equiv} For $s\ge 0$ the map
$\phi\mapsto\vtp=\CD\CT\phi$ is an isomorphism between $H^s(\R^2)$ and
$\CX^s:=L^2(\T^2,\ell^2_{s})$, 
$\|\vtp\|_{\CX^s}^2 = \int_{\T^2}\sum_{n\in \N}|\tilde{\phi}_n(k)|^2(1+n)^{s}\dd k$, 
i.e., we have the equivalence of norms 
\begin{gather}\label{neq} C_1\|\phi\|_{H^s}\le \|\vtp\|_{\CX^s}\le
C_2\|\phi\|_{H^s}.  
\end{gather} \elem
For the sake of convenience, and since the idea of the proof is 
needed below, we also include a simple version of a Sobolev embedding 
theorem.
\blem\label{sob-lem} 
For $\phi,\psi\in H^s(\R^d)$, $s>d/2$, 
we have $\phi,\psi\in C^0$, $\|\phi\|_{C^0}\le C\|\phi\|_{H^s}$, 
and $\|\phi\psi\|_{H^s}\le C\|\phi\|_{H^s}\|\psi\|_{H^s}$. 
\elem
\begin{proof} 
 Let $w(k)=(1+|k|)^{s}$. Then $w^{-1}\in L^2$ for $s>d/2$ and thus 
\begin{gather}
\|\phi\|_{C^0}\le C 
\|\phih\|_{L^1}=C\|w^{-1}\phih w\|_{L^1}\le C\|w^{-1}\|_{L^2}\|\phih\|_{L^2_s} 
\le C\|\phih\|_{L^2_s}.\label{embed}
\end{gather}
Next $w(k)\le w(k-\ell)+w(\ell)$ and thus 
\begin{align}
|w(k)(\phih*\psih)(k)|&\le \int |w(k-\ell)\phih(k-\ell)\psih(\ell)|\dd\ell
+ \int |w(\ell)\phih(k-\ell)\psih(\ell)|\dd\ell. 
\end{align}
Therefore, using Young's inequality $\|\phih*\psih\|_{L^r}\le 
C\|\phih\|_{L^p}\|\psih\|_{L^q}$, $1/p+1/q=1/r+1$, with $r=2, p=2, q=1$ 
we have 
\begin{align}
\|\phi\psi\|_{H^s}&\le C \|\phih*\psih\|_{L^2_s} 
\le C\left(\| |w\phih|*|\psih|\|_{L^2}+\| |\phih|*|w\psih|\|_{L^2}\right)
\notag\\
&\le C\left(\|w\phih\|_{L^2}\|\psih\|_{L^1}+\|w\psih\|_{L^2}\|\phih
\|_{L^1}\right) \le 2C\|\phih\|_{L^2_s}\|\psih\|_{L^2_s}
\le C\|\phi\|_{H^s}\|\psi\|_{H^s}. 
\label{hse} 
\end{align}
\end{proof}

By Lebesgue dominated convergence, $\phi\in H^s$ with $s>1$ also implies 
$\phi(x) \rightarrow 0$ pointwise 
as $|x|\rightarrow \infty$, which shows that 
solutions in $H^s$ are indeed localized. Moreover, using the product rule,   
Lemma \ref{sob-lem} can be generalized to $\|\phi\|_{C^k}\le C\|\phi\|_{H^s}$, 
$k\in \N$, $k<s-d/2$. 
Finally, by Lemma \ref{L:norm_equiv} these statements directly 
transfer to $\CX^s$. 
\blem\label{L:smooth_closed} Let $s>1$ and
$\vec{\wt\phi},\vec{\wt\psi}\in\CX^s$. Then $\vec{\wt{\phi\psi}}\in \CX^s$,
$\phi \in C_b^k(\R^2)$ for $k<s-1$, and $\phi(x) \rightarrow 0$ as
$|x|\rightarrow \infty$.  \elem

\subsection{Justification Step I: The extended Coupled Mode
  Equations}\label{S:diff_CME} 
To justify the general stationary CMEs
\eqref{E:CME_gen_Fourier} as an asymptotic model for stationary gap solitons
near an edge $\omega =\omega_*$, we again consider \reff{E:GP_stat_Bloch},
i.e. 
\begin{gather}\label{E:GP_stat_Bloch_resc} 
\left[\tilde{L}(k;x)-\omega_* - \eps^2
  \Omega\right]\tilde{\phi}(k;x) =-\sigma \ \left(\tilde{\phi} *_{\Bsm}
  \tilde{\bar{\phi}} *_{\Bsm} \tilde{\phi}\right)(k;x).  
\end{gather} In contrast to
the formal derivation of the CME in \S\ref{S:derive_CME} we now want to keep
track of higher order remainders.  We first apply the diagonalization operator
$\CD: \tilde{\phi}(k;x) \rightarrow \vec{\tilde{\phi}}(k), \
\tilde{\phi}_n(k)=\langle
\tilde{\phi}(k;\cdot),p_n(k,\cdot)\rangle_{L^2(\P^2)}$ to get
\begin{gather}\label{E:GP_stat_Bloch_diag} \left[\omega_n(k)-\omega_* - \eps^2
  \Omega\right]\tilde{\phi}_n(k) =-\sigma \tilde{g}_n(k) 
\end{gather} with
$\tilde{g}_n(k) = \langle (\tilde{\phi} *_{\Bsm} \tilde{\bar{\phi}} *_{\Bsm}
  \tilde{\phi})(k;\cdot),p_n(k;\cdot)\rangle_{L^2(\P^2)}. $

Lemma \ref{L:smooth_closed} allows us to consider \eqref{E:GP_stat_Bloch_diag}
in the space $\CX^s$, $s>1$.  The multiplication operator
$\omega_n(k)-\omega_*$ is not invertible since it vanishes at $N$ points $(n,k)=(n_1,k^{(1)}), \ldots,
(n_N,k^{(N)})$. This dictates our generalized Lyapunov-Schmidt
decomposition of the solution $\vec{\wt{\phi}}$, namely
\begin{gather}\label{E:Lyap_Schm} 
\vec{\wt{\phi}}(k) =\eps^{-1}\etalsv(k)+\vec{\wt{\psi}}(k), 
\text{ where } 
\eps^{-1}\etalsv(k)=\eps^{-1
}\sum_{j=1}^N e_{n_j}\sum_{m\in M_j}\hat{B}_j\left(\frac{k+m-k^{(j)}}{\eps}\right) 
\end{gather}
with $\supp\,\hat{B}_j\subset D_{\eps^{r-1}}$, $\wt\psi_{n_j}(k)=0$ for 
$k\in K_c:=\cup\{\wt{D}_l: l\in \{1,\ldots,N\}, n_l=n_j\}$, 
where $\wt{D}_l$ is defined in \eqref{E:Dtil_j} and where $e_{n_j}$ is the unit vector in the $n_j$ direction in $\R^{\N}$.

\brem\label{R:rhoest}{\rm 
Note that in general $\eps^{-1}\etalsv(k)$  is 
not the diagonalization of the leading order term in an ansatz of the form 
\reff{E:ansatz_Bloch} (with $\chi_{D_j}(k+m)\hat{A}_j\left(\tfrac{k+m-k^{(j)}}{\eps}\right)$ replaced by $\hat{B}_j\left(\tfrac{k+m-k^{(j)}}{\eps}\right)$), except at $k=k^{(j)}$,  
since in $\eps^{-1}\etals(k;x)$ we have $p_{n_j}(k;\cdot)$ instead of $p_{n_j}(k^{(j)},\cdot)$.

However, we have 
\beq\label{rho_def}
\eps^{-1}\etals(k,x)
=\eps^{-1}\psils(k,x)+\tilde{\rho}(k,x)
\eeq 
with 
\beq\label{phi0_LS}
\eps^{-1}\psils(k,x)=\eps^{-1}\sum_{j=1}^Np_{n_j}(k^{(j)},x)\sum_{m\in M_j} 
\hat B_j\left(\frac{k+m-k^{(j)}}\eps\right)\er^{\ri m\cdot x}, 
\eeq
and where for $\hat B_j\in L^2_s$ with $s\ge 1$
\beq\label{rhoest2}
\|\tilde{\rho}\|_{L^2(\T^2, H^s(\P^2))}\le C\eps\sum_{j=1}^N\|\hat{B}_j\|_{L^2_s}. 
\eeq 
This follows from writing $k=k^{(j)}-m+(k+m-k^{(j)})$, expanding 
$$
p_{n_j}(k,x)=p_{n_j}(k^{(j)}-m,x)+
\nab_k p_{n_j}(k^{(j)}_\star,x)\cdot(k+m-k^{(j)})
$$
with $k^{(j)}_{\star,l}\in[\min(k_l^{(j)}-m_l,k_l), \max(k_l^{(j)}-m_l,k_l)], \ l=1,2$, 
and using $p_{n_j}(k^{(j)}-m,x)=p_{n_j}(k^{(j)},x)
\er^{\ri m\cdot x}$,  which yields 
\begin{align*}
\tilde{\rho}(k,x)=\eps^{-1}\sum_{j=1}^N\sum_{m\in M_j} 
\hat B_j\left(\tfrac{k+m-k^{(j)}}\eps\right)(k+m-k^{(j)})\cdot
\nab_k p_{n_j}(k^{(j)}_\star,x).
\end{align*} 
To prove \reff{rhoest2}, we may fix some (of the finitely many) $j,m$. 
Since $\om_{n_j}(k)$ are simple for $k\in \tilde{D}_j$, we have 
$\sup_{k\in \tilde{D_j}}\|\nab_k p_{n_j}(k,\cdot)\|_{H^s(\P^2)}\le C$, and 
it remains to estimate 
\begin{align}
\biggl\|\eps^{-1}&
\hat B_j\left(\tfrac{k+m-k^{(j)}}\eps\right)|k+m-k^{(j)}|\biggr\|_{L^2(\T^2)}^2
=\eps^{-2}\int_{k\in \T^2}  
\left|\hat B_j\left(\tfrac{k+m-k^{(j)}}\eps\right)\right|^2 |k+m-k^{(j)}|^2
\dd k\nonu\\
&\le C\eps^2 \int_{K\in \R^2} |\hat{B}_j(K)|^2|K|^2\dd K\le C\eps^2
\|\hat{B}_j\|_{L^2_s}^2\label{rhoest3}
\end{align}
for $s\ge 1$, and this proves \reff{rhoest2}. 
}\erem

With the decomposition \eqref{E:Lyap_Schm} we obtain the 
Lyapunov--Schmidt equations 
\begin{align}
\frac 1 \eps (\om_{n_j}(k)-\om_*-\eps^2\Omega)
\hat{B}_j\left(\frac{k+m-k^{(j)}}{\eps}\right)&
=-\sigma\chi_{\wt{D}_j}(k)\tilde{g}_{n_j}(k),\quad j=1,\ldots,N, 
\ m\in M_j,\label{cpart}\\
(\om_{n}(k)-\om_*-\eps^2\Omega)\wt{\psi}_n(k)&=-\sigma
\left(1-\sum_{j=1}^N\chi_{\wt{D}_j}(k)\del_{n,n_j}\right)\tilde{g}_{n}(k), 
\quad n\in\N. \label{spart}
\end{align}
The goal is to solve \reff{spart} for the 
correction $\vec{\tilde{\psi}}$ as a 
function of ${\hB}=(\hat{B}_j)_{j=1}^N\in L^2(D_{\eps^{r-1}},\C^N)$
and plug this into \reff{cpart}. It turns out that 
the right norm for $\hat{B}_j$ is $\|\hat{B}_j\|_{L^2_s(D_{\eps^{r-1}})}$, 
where we recall
$$
\|\hat{B}_j\|_{L^2_s(D_{\eps^{r-1}})}=
\|(1+|p|)^s\hat{B}_j\|_{L^2(D_{\eps^{r-1}})}. 
$$
Note that 
$L^2(D_{\eps^{r-1}})=L^2_s(D_{\eps^{r-1}})$ as spaces for any $s\ge 0$, 
but below we need the estimate $\|B_j\|_{H^s}\le 
C\|\hat{B}_j\|_{L^2_s(D_{\eps^{r-1}})}$ 
with $C$ independent of $\eps$, cf.\,\reff{figo}. 
\blem\label{nonlin-lem} 
Let $s>1$ and $V\in H^{\lceil s\rceil-1+\delta}_{{\rm loc}}(\R^2), \delta>0$. For ${\hB}\in L_{s}^2(D_{\eps^{r-1}})$ and $\vec{\tilde{\psi}}\in {\CX^s}$, we have 
\beq\label{sne}
\begin{split}
\|\vec{\tilde{g}}\|_{\CX^s}\le C&\left[
\eps^2\left(\sum_{j=1}^N\|\hat{B}_j\|_{L_s^2(D_{\eps^{r-1}})}\right)^3
+\eps^2\left(\sum_{j=1}^N\|\hat{B}_j\|_{L_s^2(D_{\eps^{r-1}})}\right)^2
\|\vec{\tilde{\psi}}\|_{\CX^s}\right.\\
&\quad \left. +\eps\left(\sum_{j=1}^N\|\hat{B}_{j}\|_{L_s^2(D_{\eps^{r-1}})}\right)
\|\vec{\tilde{\psi}}\|_{\CX^s}^2
+\|\vec{\tilde{\psi}}\|_{\CX^s}^3\right].  
\end{split}
\eeq 
\elem

\begin{proof} 
Relying on the norm equivalence in \eqref{figo}, we derive most of the estimates in physical variables. Due to \eqref{E:Lyap_Schm} and \eqref{rho_def} we have
$g=|\phi|^2\phi=
|\eps^{-1}\psi_\text{LS}^{(0)}+\rho+\psi|^2(\eps^{-1}\psi_\text{LS}^{(0)}+\rho+\psi)$.
We need to estimate norms of terms of the types
\beq\label{term_types}
\left(\eps^{-1}\psi_\text{LS}^{(0)}\right)^3, \ \left(\eps^{-1}\psi_\text{LS}^{(0)}\right)^2 f, \  \eps^{-1}\psi_\text{LS}^{(0)}f^2, \ \rho^2\psi, \ \rho\psi^2, \ \rho^3, \text{ and } \psi^3.
\eeq
First note that $\eps^{-1}\psi_\text{LS}^{(0)}(x) = \eps\sum_{j=1}^NB_j(\eps x)u_{n_j}(k^{(j)};x)$.  Below we implicitly use $\|B_j(\eps \cdot)u_{n_j}(k^{(j)};\cdot)\|_{H^s}\leq \|u_{n_j}(k^{(j)};\cdot)\|_{C^{\lceil s \rceil}}\|B_j(\eps \cdot)\|_{H^s}$, which holds by interpolation, see, e.g., \cite[\S4.2]{taylor1}. Next,  $\|u_{n_j}(k^{(j)};\cdot)\|_{C^{\lceil s \rceil}}\leq C \|u_{n_j}(k^{(j)};\cdot)\|_{H^{\lceil s \rceil+1+\delta}} \leq C \|V\|_{H^{\lceil s \rceil-1+\delta}_\text{loc}}$ for all $\delta >0$, where the first inequality holds by Sobolev embedding and the second one by the differential equation.

In estimating all except the first term in \eqref{term_types} we use the Banach algebra property in Lemma \ref{sob-lem}. For the first two terms we need to estimate $\|(\eps B_j(\eps \cdot))^n\|_{H^s}$ for $n=2,3$. We have
\begin{align*} 
\|(\eps B_j(\eps \cdot))^n\|_{H^s}^2&=\int (1+|k|)^{2s}|\CF((\eps B_j(\eps \cdot))^n)(k)|^2\dd k\\ 
& \le C\left[\int |\CF((\eps B_j(\eps \cdot))^n)(k)|^2\dd k+
\eps^{2n-4}\int |k|^{2s}\left|\widehat{B_j^n}\left(\frac k \eps\right)\right|^2\dd k\right]\\
&\le C\left[\|(\eps B_j(\eps \cdot))^n\|_{L^2}^2+\eps^{2n-2+2s}\int |K|^{2s}
|\widehat{B_j^n}(K)|^2\dd K\right]\\
&\le C\left[\eps^{2(n-1)}\|B_j\|_{L^\infty}^{2(n-1)}\|\eps B_j(\eps \cdot)\|_{L^2}^2+
\eps^{2n-2+2s}\|B_j^n\|_{H^s}^2\right]\\
&\le C\left[\eps^{2(n-1)}\|B_j\|_{H^s}^{2(n-1)}\|B_j\|_{L^2}^2+
\eps^{2(n-1)+2s}\|B_j\|_{H^s}^{2n}\right]
\end{align*}
and hence 
\beq\label{fine1}
\|(\eps B_j(\eps \cdot))^n\|_{H^s}\le C\eps^{n-1}\|B_j\|_{H^s}^n \text{ for } n=1,2,3.
\eeq 
Note that for $n\geq 2$ this is much better than the naive estimate  
$\|(\eps B_j(\eps \cdot))^n\|_{H^s}\le C\|\eps B_j(\eps \cdot)\|_{H^s}^n\le 
C\|B\|_{H^s}^n$ based on \eqref{fine1} with $n=1$. Next, for the third term in \eqref{term_types} we get 
\beq\label{fine}
\|\eps B_j(\eps \cdot)f(\cdot)\|_{H^s}
\le C\eps \|B_j\|_{H^s}\|f\|_{H^s}, 
\eeq 
and this together with \reff{fine1} can be used to prove \reff{sne}. 
To show \reff{fine}, we start with 
\begin{align*} 
\|\eps B_j(\eps \cdot)f(\cdot)\|_{H^s}&\le \|\eps B_j(\eps \cdot)f(\cdot)\|_{L^2}+
C\left\|\,|k|^s\left(\frac 1 \eps \hat{B}_j\left(\frac\cdot \eps\right)
*\hat{f}(\cdot)\right)\right\|_{L^2}.
\end{align*}
The first term is estimated as 
$\|\eps B_j(\eps \cdot)f(\cdot)\|_{L^2}\le \eps\|B_j\|_{\infty}\|f\|_{L^2}
\leq \eps\|B_j\|_{H^s}\|f\|_{L^2}$, and for the second we note 
that $w(k)\le \eps w(\frac{k-l}{\eps})+w(l)$ where $w(k)=|k|^s$. 
Thus, similarly to the proof of Lemma \ref{sob-lem}, we have, using Young's 
inequality, 
\begin{align*} 
\biggl\|w(k)&\left(\frac 1 \eps \hat{B}_j\left(\frac\cdot \eps\right)
*\hat{f}(\cdot)\right)\biggr\|_{L^2} 
\le C\left\|\,\left|w\left(\frac \cdot\eps\right)\hat{B}_j\left(\frac \cdot\eps\right)\right|
*|\hat{f}(\cdot)|+\left|\frac 1 \eps \hat{B}_j\left(\frac \cdot\eps\right)\right|
*|w(\cdot)\hat{f}(\cdot)|\right\|_{L^2}\\
\le&C\left[
\left\|w\left(\frac \cdot\eps\right)\hat{B}_j\left(\frac \cdot\eps\right)\right\|_{L^2}
\|\hat{f}\|_{L^1}+\left\|\frac 1 \eps \hat{B}_j\left(\frac \cdot\eps\right)
\right\|_{L^1}\|w\hat{f}\|_{L^2}\right].
\end{align*}
Now 
$\|w\left(\frac \cdot\eps\right)\hat{B}_j\left(\frac \cdot\eps\right)\|_{L^2}
\le \eps\|\hat{B}_j\|_{L_s^2}$, $\left\|\frac 1 \eps \hat{B}_j\left(\frac \cdot\eps\right)
\right\|_{L^1}\le C\eps\|\hat{B}_j\|_{L^2_s},$ and  $\|\hat{f}\|_{L^1} \leq C\|\hat{f}\|_{L^2_s}$ (see \eqref{sob-lem}) yield \reff{fine}.

The 4th, 5th and 7th term in \eqref{term_types} are estimated simply using Lemma \ref{sob-lem} and \eqref{rhoest2}. The 6th term is treated the same way and is bounded by $\eps^3\big(\sum_{j=1}^N\|\hat{B}_j\|_{L_{s}^2(D_{\eps^{r-1}})}\big)^3$ so that it is of higher order than the first term on the right hand side of \eqref{sne}.
\end{proof}

We seek now a small solution of \reff{spart}. First, for $\eps$ sufficiently small we have 
\begin{gather}\label{E:inv_bound} 
\min_{\substack{k\in
      \supp(\vec{\tilde{\psi}})\\n\in\N}}|\omega_n(k)-\omega_*|\geq
  C\eps^{2r}
\end{gather}
due to $\pa_{k_1}\omega_{n_j}(k^{(j)})=\pa_{k_2}\omega_{n_j}(k^{(j)})=0$ and the definiteness of the Hessian of $\omega_{n_j}$ at $k=k^{(j)}$. We rewrite now \reff{spart} as
\[\tilde{\psi}_n(k) = (\omega_n(k)-\omega_*)^{-1}(-\sigma \tilde{g}_n(k)+\eps^2\Omega \tilde{\psi}_n(k))=: \tilde{F}_n(\vec{\tilde{\psi}})(k)\]
and solve $\vec{\tilde{\psi}}=\vec{\tilde{F}}(\vec{\tilde{\psi}})$ on a neighborhood of $0$, namely on $B_{\eps^\eta}:=\{\vec{\tilde{\psi}} \in \CX^s : \|\vec{\tilde{\psi}}\|_{\CX^s}\leq \eps^\eta\}, \eta >0$, via the Banach fixed point theorem. 

Performing similar estimates as in the proof of Lemma \eqref{nonlin-lem} and using \eqref{E:inv_bound}, we get
\[\|F(\psi)-F(\phi)\|_{H^s}\leq C \eps^{-2r}\left[\eps^2+\eps(\|\psi\|_{H^s}+\|\phi\|_{H^s})+\|\psi\|^2_{H^s}+\|\phi\|^2_{H^s}\right]\|\psi-\phi\|_{H^s},\] 
where $C = C(\sum_{j=1}^N\|B_j\|_{H^s(\R^2)},|\Omega|)$. The contraction property thus holds if 
\beq\label{E:contr1}
\eps^{2-2r}+\eps^{1-2r}(\|\psi\|_{H^s}+\|\phi\|_{H^s})+\eps^{-2r}(\|\psi\|^2_{H^s}+\|\phi\|^2_{H^s})<1
\eeq
for all $\vec{\tilde{\psi}},\vec{\tilde{\phi}}\in B_{\eps^\eta}$ and if $\vec{\tilde{F}}$ maps $B_{\eps^\eta}$ to $B_{\eps^\eta}$, i.e., using \eqref{sne}, if 
\beq\label{E:contr2}
\eps^{2-2r}+\eps^{2-2r}\|\vec{\tilde{\psi}}\|_{\CX^s}+\eps^{1-2r}\|\vec{\tilde{\psi}}\|^2_{\CX^s}+\eps^{-2r}\|\vec{\tilde{\psi}}\|^3_{\CX^s}< C \eps^\eta
\eeq
for all $\vec{\tilde{\psi}} \in B_{\eps^\eta}$.

Condition \eqref{E:contr1} is satisfied if $r<1$ and $\eta> \max(2r-1,r)$ and \eqref{E:contr2} holds if $\max(2r-1,r)<\eta<2-2r$ and $r<1$. In combination these yield
\beq\label{E:contr_fin}
\eta\in (r,2-2r), \quad 0<r<\tfrac{2}{3}.
\eeq
Here is the reason for the upper bound in \eqref{rcon1}.

In conclusion we have the existence of $\vec{\tilde{\psi}}$ satisfying
\begin{gather} \label{E:psi_til_est_U}
  \|\vec{\tilde{\psi}}\|_{\CX^s}\leq C
  \eps^{2-2r}\sum_{j=1}^N\|\hat{B}_j\|_{L^2_s(D_{\eps^{r-1}})}  \quad \text{with} \ \ 0<r<2/3.
\end{gather}

We now turn to \reff{cpart}. Plugging \reff{E:Lyap_Schm} into 
\reff{E:GP_stat_Bloch_diag}, truncating over $k\in D_j$ 
and mapping $k\in D_j-m$ to $p\in D_{\eps^{r-1}}$ via $p=\eps^{-1}(k+m-k^{(j)})$ 
yields the so called extended CMEs (eCMEs) in the form
 \begin{gather}\label{E:CME_integr_form}
  \Omega\hat{B}_j-\left(\frac{1}{2}\pa_{k_1}^2\omega_{n_j}(k^{(j)})p_1^2
    +\frac{1}{2}\pa_{k_2}^2\omega_{n_j}(k^{(j)})p_2^2
    +\pa_{k_1}\pa_{k_2}\omega_{n_j}(k^{(j)})p_1p_2\right)\hat{B}_j
  -\hat{Q}_j=\eps^{\tilde{r}}\hat{R}_j(p), 
\end{gather} 
$j=1,\ldots,N$. Here  $\hat{Q}_{j}$ denotes the nonlinear term $\hat{\CN}_j$ in
  \eqref{E:CME_gen_Fourier} with $\hat{A}_i$ replaced by $\hat{B}_i$ and truncated on $p\in D_{\eps^{r-1}}$, i.e.,
  $$\hat{Q}_{j}(p^{(j,m)}) = \frac{\sigma}{\eps^4}\chi_{D_{\eps^{r-1}}}(p^{(j,m)})\langle (\psils *_{\Bsm} \tilde{\bar{\psi}}_{{\rm LS}}^{(0)} *_{\Bsm} \psils)(\eps p^{(j,m)}+k^{(j)}-m;\cdot),p_{n_j}(k^{(j)};\cdot)e^{\ri m \cdot \cdot} \rangle_{L^2(\P^2)}$$
with $\psils$ given in \eqref{phi0_LS}. 
In the persistence step below we need to control the remainder 
$\eps^{\tilde{r}}\hat{R}_j(p)$, where 
we set $\tilde{r}=\min\{r,2-2r,1\}$, and which has the form 
\beq\label{E:remainder}
\eps^{\tilde{r}}\hat{R}_j(p^{(j,m)}) = \eps^{-2} q_j(\eps p^{(j,m)})\hat{B}_j(p^{(j,m)})
+ \eps^{-1}\sigma \chi_{D_{\eps^{r-1}}}(p^{(j,m)})\tilde{g}_{n_j}(\eps p^{(j,m)}+k^{(j)}-m)-\hat{Q}_j(p^{(j,m)}). 
\eeq
The first term in \eqref{E:remainder} comes from the Taylor expansion
 of $\omega_{n_j}$, i.e., 
\beq 
q_j(\eps p):=\om_{n_j}(k^{(j)}+\eps p)-T_2(\om_{n_j}(k^{(j)}))(\eps p)
\eeq 
with $T_2(\om_{n_j}(k^{(j)})$ the second order Taylor polynomial. 
Since $\om_{n_j}$ is analytic near $k^{(j)}$, we have that $q_j$ is 
analytic as well and starts with a cubic polynomial. 
Given $\hat{B}_j\in L^2_s(D_{\eps^{r-1}})$, in the persistence 
step we need to bound the remainder in $L^2_{s-2}(D_{\eps^{r-1}})$. For the cubic terms 
$q_j^{(3)}(\eps p)$ in $q_j(\eps p)$ we get 
\begin{align}\label{rem1e}
\|\eps^{-2} q_j^{(3)}(\eps \cdot)\hat{B}_j(\cdot)\|_{L^2_{s-2}(D_{\eps^{r-1}})}^2&\leq
C\int_{D_{\eps^{r-1}}} \eps^2 |p|^6 |\hat{B}_j(p)|^2(1+|p|^2)^{s-2}\dd p\notag\\
&\le C\int_{D_{\eps^{r-1}}} \eps^2 |p|^2 |\hat{B}_j(p)|^2(1+|p|^2)^{s}\dd p
\le C\eps^{2r}\|\hat{B}_j\|_{L^2_s(D_{\eps^{r-1}})}^2, 
\end{align}
and clearly higher order contributions from $q_j(\eps p)$ gain additional 
powers in $\eps$. 

 The leading order contribution to the remaining two terms in $\eps^{\tilde{r}}\hat{R}_j(p)$ comes from $\vec{\tilde{\psi}}$, which is estimated via \eqref{E:psi_til_est_U} to be $\CO(\eps^{2-2r})$. This is guaranteed to be small since $r<2/3$. The last contribution to $\eps^{\tilde{r}}\hat{R}_j(p)$ comes from the $\tilde{\psi}$-independent terms in $\chi_{D_{\eps^{r-1}}}(p)\tilde{g}_{n_j}(\eps p+k^{(j)}-m)$, i.e., from
\[
\begin{split}
\frac{\sigma}{\eps^4}\chi_{D_{\eps^{r-1}}}(p)&\left[  \langle (\etals *_{\Bsm} \tilde{\bar{\eta}}_{{\rm LS}}^{(0)} *_{\Bsm} \etals)(\eps p+k^{(j)}-m;\cdot),p_{n_j}(\eps p+k^{(j)};\cdot)e^{\ri m \cdot \cdot} \rangle_{L^2(\P^2)} \right. \\
& \left. - \langle (\psils *_{\Bsm} \tilde{\bar{\psi}}_{{\rm LS}}^{(0)} *_{\Bsm} \psils)(\eps p+k^{(j)}-m;\cdot),p_{n_j}(k^{(j)};\cdot)e^{\ri m \cdot \cdot} \rangle_{L^2(\P^2)}\right]. 
\end{split}
\]
This difference is estimated in $L_s^2$ to leading order via $\|\tilde{\rho}\|_{L^2_s}$, which is $\CO(\eps)$ according to \eqref{rhoest2}.

These estimates are not strictly needed for our first main result, 
which follows directly from the above Lyapunov-Schmidt reduction, 
and which {\em assumes} the existence of solutions of the extended CMEs 
\reff{E:CME_integr_form}, 
but the estimates show that we may expect solutions of the extended CMEs 
\reff{E:CME_integr_form} near solutions of the CME, as will be worked out 
in \S\ref{S:persist}. 
\bthm \label{T:diff_CME_ext}
 Let $s{>}1, V\in H_{{\rm loc}}^{\lceil s\rceil-1+\delta}(\R^2)$ for 
some $\delta{>}0$ and let $0<r<{2 \over {3}}$. There
  exist $\eps_0,C_1,C_2>0$ such that for all $0<\eps<\eps_0$ the 
following holds.
  Assume that there exists a solution 
$(\hat{B}_j)_{j=1}^N\in L^2(D_{\eps^{r-1}})$
  of the extended CMEs \reff{E:CME_integr_form} with 
$\|\hat{B}_j\|_{L^2_s(D_{\eps^{r-1}})}\le C_1$. Then \eqref{E:GP_stat}
  has a solution $\phi\in H^s(\R^2)$ with 
  \begin{equation}\label{E:bound_CME_ext}
    \| \phi(\cdot) - \eps\sum_{j=1}^N B_j(\eps \cdot) 
u_{n_j}(k^{(j)};\cdot)\|_{H^s(\R^2)} \leq C_2 (\eps^{2-2r} +\eps), 
  \end{equation}
where $u_{n_j}(k^{(j)};x)$ are the resonant Bloch waves
  and $B_j=\CF^{-1}\hat{B}_j$.
\ethm
\begin{proof} By construction, a solution $(\hat{B}_j)_{j=1}^N$ 
of \reff{E:CME_integr_form} yields via \reff{E:Lyap_Schm} 
a solution $\vec{\wt{\phi}}(k)$ 
of \reff{E:GP_stat_Bloch_resc}. The estimate \reff{E:psi_til_est_U} and the norm equivalence in Lemma
  \ref{L:norm_equiv} yield $\|\phi - \eps^{-1}\eta_\text{LS}^{(0)}\|_{H^s(\R^2)}\leq C \eps^{2-2r}$. And because 
  $\eps\sum_{j=1}^N B_j(\eps x) 
u_{n_j}(k^{(j)};x) = \eps^{-1} \psi_\text{LS}^{(0)}(x)$, we have $\eps^{-1}\eta_\text{LS}^{(0)}(x) - \eps\sum_{j=1}^N B_j(\eps x) 
u_{n_j}(k^{(j)};x)=\rho(x)$, which is bounded in \eqref{rhoest2} and the approximation result \eqref{E:bound_CME_ext} follows.
\end{proof} 

\subsection{Justification Step II: Persistence}\label{S:persist}
The remaining step is to make a connection between solitary waves of the 
CMEs \eqref{E:CME_gen_Fourier} and the eCMEs \eqref{E:CME_integr_form}.
To obtain existence of solutions of the eCMEs \reff{E:CME_integr_form}, 
we show a persistence result of special so called reversible non-degenerate
CME solitons to the eCME. 
This is quite similar to \cite[\S5]{DPS08} but in order to deal with an arbitrary set of extrema $k^{(j)}, j\in \{1,...,N\}$, the definition of 
reversibility needs to be generalized.

The eCMEs \reff{E:CME_integr_form} differ from the CMEs \reff{E:CME_gen_Fourier} 
in three ways: the $\hat{B}_j(p)$ are supported on $D_{\eps^{r-1}}$,  
the convolutions are truncated on $D_{\eps^{r-1}}$, and the remainder 
$\eps^{\tilde{r}}\hat{R}_j(p)$ is included. The idea is to handle 
these differences as perturbations and thus seek a solution 
${\hB}=(\hat{B}_j)_{j=1,\ldots,N}$ of the 
eCMEs near a given solution ${\hA}=(\hat{A}_j)_{j=1,\ldots,N}$ of the CMEs. 
Note that the $\hat{A}_j$ are not supported on $D_{\eps^{r-1}}$ and thus first must 
be truncated. 

We start with a formal 
discussion. We write the CME in abstract form as ${\hF}({\hA})=0$ 
and the eCME as 
\begin{gather}\label{cmea}
\chi_{D_{\eps^{r-1}}}{\hF}(\hB)=\eps^{\tilde{r}}\hR(\hB), 
\end{gather}
assume a solution ${\hA}\in [L^2_q(\R^2)]^N$ for any $q\ge 0$ of the CME, 
and look for solutions 
$\hB\in [L^2_s(D_{\eps^{r-1}})]^N$ of the eCME in the form 
$ \hB=\hA_\eps+\hb$ with $\hA_\eps=\chi_{D_{\eps^{r-1}}}\hA$ and 
$\supp\,\hb\subset D_{\eps^{r-1}}$. This yields 
\beq\label{beq}
\begin{split}
&\hJ_\eps\hb=\hN(\hb),\ p \in D_{\eps^{r-1}} \quad \text{ with }\\
&\hJ_\eps=\chi_{D_{\eps^{r-1}}}D_{\hA}\hF(\hA)\quad \text{and}\quad
\hN(\hb)=\eps^{\tilde{r}}\hR(\hA_\eps+\hb)-(\chi_{D_{\eps^{r-1}}}
\hF(\hA_\eps+\hb)-\hJ_\eps\hb). 
\end{split}
\eeq
Note that in \reff{beq} we may replace $\hN(\hb)$ by 
$\chi_{D_{\eps^{r-1}}}\hN(\hb)$ to display explicitly that 
$p\in D_{\eps^{r-1}}$. 

We have $\hF(\hA_\eps+\hb)-\hJ_\eps\hb=\hF(\hA_\eps)
+(D_{\hA}\hF(\hA_\eps)-\hJ_\eps)\hb+\hat{\bf G}(\hb)$ with $\hat{\bf G}(\hb)$ 
quadratic in $\hb$. Moreover, 
$$
\hF(\hA_\eps)=L^{\text{CME}} \ \chi_{D_{\eps^{r-1}}}\hA-\chi_{D_{\eps^{r-1}}} \hat{{\bf \CN}}(\hA)-(\hat{{\bf \CN}}(\hA_\eps)-\chi_{D_{\eps^{r-1}}}\hat{{\bf \CN}}(\hA))
=-(\hat{{\bf \CN}}(\hA_\eps)-\chi_{D_{\eps^{r-1}}}\hat{{\bf \CN}}(\hA)),
$$
where $L^{\text{CME}}$ denotes the linear part of the operator in \reff{E:CME_gen_Fourier} and $\hat{{\bf \CN}}$ denotes the $N-$dimensional vector of the nonlinear terms in \reff{E:CME_gen_Fourier}. 
$\chi_{D_{\eps^{r-1}}} \hat{{\bf \CN}}(\hA)(p)$ is a sum of convolutions $\hat{A}_{j_1}*\hat{A}_{j_2}*\hat{\bar{A}}_{j_3}$, 
and hence in $\hat{{\bf \CN}}(\hA_\eps)-\chi_{D_{\eps^{r-1}}}\hat{{\bf \CN}}(\hA)$ this yields 
terms of the form  
$$
\int_{p_1}\int_{|p_2|\ge \eps^{r-1}} 
\chi_{D_{\eps^{r-1}}}(p-p_1)\chi_{D_{\eps^{r-1}}}(p_1-p_2)
\Ah_{j_1}(p-p_1)\Ah_{j_2}(p_1-p_2)\hat{\bar{A}}_{j_3}(p_2) \dd p_1\dd p_2,
$$ 
which can be bounded by $C\eps^{q}$ for any $q>0$ in $L^2_s(D_{\eps^{r-1}})$ 
due to the fast decay of $\hA$. Therefore, 
\begin{gather}\label{conve1}
\|\hF(\hA_\eps)\|_{L^2_s}\le
C\eps^{q},\quad\text{and}\quad 
\|(D_{\hA}\hF(\hA_\eps)-\hJ_\eps)\hb\|_{L^2_s}\le C\eps^{q}\|\hb\|_{L^2_s}
\end{gather}
by a similar estimate. 

Thus $\|\hN(\hb)\|_{L^2_{s-2}}\le C(\eps^{\tilde{r}}+\eps^{q}+\eps^{\tilde{r}}\|\hb\|_{L^2_s}+\eps^{q}\|\hb\|_{L^2_s}
+\|\hb\|_{L^2_s}^2)$, 
and this suggests applying the contraction mapping theorem 
to \reff{beq} in the form 
\begin{gather} \label{beqc}
\hb=\hJ_\eps^{-1}\hN(\hb).
\end{gather} 
To discuss $\hJ_\eps^{-1}$, we start with $\bJ:H^{s}(\R^2)\ra H^{s-2}(\R^2)$. 
The continuous spectrum $\sig_c(\bJ)$ of $\bJ$ equals that of 
$L^{\text{CME}}$. Thus, if $\om=\om_*+\eps^2\Om$ 
is in a gap, then $\Om$ and the quadratic forms defined by 
${1 \over 2}\pa_{k_1}^2\om_{n_j}(k^{(j)})p_1^2+{1 \over 2}\pa_{k_2}^2\om_{n_j}(k^{(j)}p_2^2
+\pa_{k_1}\pa_{k_2}\om_{n_j}(k^{(j)})p_1p_2$, $j{=}1,{\ldots},N$ have opposite 
signs such that $\sig_c(\bJ)$ is bounded away from zero. 
 However, the problem is that $\bJ$ has a nontrivial kernel since 
$\text{Ker}~{\bf J}$ contains at least $\pa_{y_1}{\bf A}, \pa_{y_2}{\bf A}$ 
and $\ri {\bf A}$ 
which follows from the translational and phase invariances of 
the CME. For $\hJ_\eps^{-1}:L^{2}_s(D_{\eps^{r-1}})\ra L^2_{s+2}(D_{\eps^{r-1}})$ (if it exists) 
this implies that it cannot be bounded independently of $\eps$. 

The solution is to consider \reff{beqc} in a subspace 
$\xrev\subset L_s^2(D_{\eps^{r-1}})$ 
where $\hJ_\eps^{-1}$ is bounded, and where $\hb\in\xrev$ implies $\hN(\hb)\in\xrev$. 
This can be achieved by symmetries of the problem \reff{E:GP_stat} 
if we assume that $\bJ$ on $H^s(\R^2)$ has only 
$\pa_{y_1} \bA, \pa_{y_2} \bA$ and $\ri\bA$ in its kernel. 

The original problem \reff{E:GP_stat} is equivariant under the symmetries 
\begin{equation}\label{E:rev1}
  \phi(x_1,x_2) = \xi_1 \bar{\phi}(-x_1,x_2) = \xi_2 \bar{\phi}(x_1,-x_2)
\end{equation}
where $\xi_1, \xi_2 = \pm 1$, i.e., $(\xi_1,\xi_2)=(1,1)$ or $(\xi_1,\xi_2)=(1,-1)$ 
or $(\xi_1,\xi_2)=(-1,1)$ or  $(\xi_1,\xi_2)=(-1,-1)$, 
and similarly 
\begin{equation}
  \label{E:rev2} \phi(x_1,x_2) = \xi_1 \bar{\phi}(x_2,x_1) =
  \xi_2 \bar{\phi}(-x_2,-x_1),
\end{equation}
where again $\xi_1, \xi_2 = \pm 1$. 

These symmetries have their counterparts in symmetries of the CMEs. As we show, the reversibility in the following definition firstly guarantees that 
$\hb\in\xrev$ implies $\hN(\hb)\in\xrev$, secondly provides a leading order approximation $\eps \eta^{(0)}$ satisfying \eqref{E:rev1} or \eqref{E:rev2}, and 
lastly (under a non-degeneracy condition) guarantees the existence of a Gross-Pitaevskii solution $\phi$ satisfying \eqref{E:rev1} or \eqref{E:rev2}.

\bdefi\label{recdef}
A solution ${\bf A}$ of \reff{E:CME_gen_iFourier} 
is called reversible if it fulfills 
one of the symmetries in \eqref{E:rev1A} or \eqref{E:rev2A}.
\beq\label{E:rev1A}
A_i(y) = s_1 \bar{A}_{i''}(-y_1,y_2)=s_2 \bar{A}_{i'}(y_1,-y_2) \text{ for all } i \in \{1,\ldots, N\} \text{ with } s_{1,2}=\pm 1 \text{ independent of } i,
\eeq
and where the indices $i'$ and $i''$ are defined by\\[2mm]
\begin{tabular}{llllllll}
$\bullet$ & $k^{(i')} = (-k^{(i)}_1,k^{(i)}_2)$ & if & $k^{(i)}_1<\tfrac{1}{2}$& and & $i'=i$ & if & $k^{(i)}_1=\tfrac{1}{2}$,\\
$\bullet$ & $k^{(i'')} = (k^{(i)}_1,-k^{(i)}_2)$ & if & $k^{(i)}_2<\tfrac{1}{2}$& and & $i''=i$ & if & $k^{(i)}_2=\tfrac{1}{2}$.
\end{tabular}
\beq\label{E:rev2A}
A_i(y) = s_1 \bar{A}_{i'}(y_2,y_1)=s_2 \bar{A}_{i''}(-y_2,-y_1) \text{ for all } i \in \{1,\ldots, N\} \text{ with } s_{1,2}=\pm 1 \text{ independent of } i,
\eeq
and where the indices $i'$ and $i''$ are defined by\\[2mm]
\begin{tabular}{lllllll}
$\bullet$ & $k^{(i')} = (k^{(i)}_2,k^{(i)}_1)$, & & & & & \\
$\bullet$ & $k^{(i'')} = (-k^{(i)}_2,-k^{(i)}_1)$ & if & $k^{(i)}_{1,2}<\tfrac{1}{2}$, & $i''=i$ & if & $k^{(i)}=M$,\\
 & $k^{(i'')} = (-k^{(i)}_2,k^{(i)}_1)$ & if & $k^{(i)}_1=\tfrac{1}{2}, k^{(i)}\neq M$, & $k^{(i'')} = (k^{(i)}_2,-k^{(i)}_1)$ & if & $k^{(i)}_2=\tfrac{1}{2}, k^{(i)}\neq M$.\\
\end{tabular}\\[4mm]
We define the space
\[\xrev =\{\hat{\bf f}\in L^2_s(D_{\eps^{r-1}}): {\bf f} \ satisfies \ \eqref{E:rev1A} \ or \ \eqref{E:rev2A}\}.\]
\edefi

\bdefi
A solution ${\bf A}$ of \eqref{E:CME_gen_iFourier}
is called non-degenerate if 
$\text{Ker}~\bJ=\{\pa_{y_1}{\bf A}, \pa_{y_2}{\bf A},\ri {\bf A}\}.$
\edefi

We explain next that under the assumption $\hA \in \xrev$ we, indeed, have the property
\beq\label{E:inherit} 
\hb \in \xrev \Rightarrow \hN(\hb) \in \xrev.
\eeq
Note that this resembles the question of inheritance of symmetries of the full problem for $\phi$ by the Lyapunov-Schmidt reduction, see
e.g. Prop.\,3.3 in \cite{GS85}. We are, however, using a generalized Lyapunov-Schmidt reduction in which a whole neighborhood of the kernel is isolated. Secondly, we wish to carry symmetries of the scalar problem for $\phi$ over to symmetries of the vector problem for the envelopes $\hB$. We inspect, therefore, problem  \eqref{E:inherit} by hand. For the sake of brevity we present the analysis only for the first symmetry in \eqref{E:rev1A}, i.e.,
\beq\label{E:S1}
A_i(y) = s_1 \bar{A}_{i''}(-y_1,y_2) \quad \text{for all } i\in \{1,\ldots, N\}.
\eeq
This is equivalent to $\hat{A}_i(p) = s_1 \hat{\bar{A}}_{i''}(-p_1,p_2)  \quad \text{for all } i\in \{1,\ldots, N\}$ in Fourier variables. If we denote $S{\bf A}(y) = \pm(\bar{A}_{1''}, \bar{A}_{2''}, \ldots, \bar{A}_{N''})(-y_1,y_2)$, condition \eqref{E:S1} reads ${\bf A}=S {\bf A}$.

In the term $\chi_{D_{\eps^{r-1}}} \hF(\hA_\eps+\hb)$ in $\hN(\hb)$ the operator ${\bf F}$ is the CME operator, which commutes with $S$. In the argument of $\hF$ the function ${\bf A}_\eps$ satisfies \eqref{E:S1} because ${\bf A}$ does and the symmetric cut-off $\hA_\eps = \chi_{D_{\eps^{r-1}}}(p) \hA$ preserves the symmetry. The linear operator $\bJ_\eps$ commutes with $S$ because it is a symmetric cut-off of the linear operator corresponding to CMEs. It remains to discuss the term $\eps^{\tilde{r}}\hR(\hA_\eps+\hb)$ in $\hN(\hb)$. The first term on the right hand side of \eqref{E:remainder} is the band structure with its quadratic Taylor expansion subtracted away, i.e.
\beq\label{E:q_term}
\begin{split}
&\eps^{-2} q_j(\eps p)\hat{B}_j(p) = \\
&\left(\eps^{-2}\omega_{n_j}(k^{(j)}+\eps p-m)-\tfrac{1}{2}\pa_{k_1}^2\omega_{n_j}(k^{(j)})p_1^2-\tfrac{1}{2}\pa_{k_2}^2\omega_{n_j}(k^{(j)})p_2^2-\pa_{k_1}\pa_{k_2}\omega_{n_j}(k^{(j)})p_1p_2\right)\hat{B}_j(p),
\end{split}
\eeq
where we write $p$ instead of $p^{(j,m)}$ and keep in mind that $\text{Range}(p)$ depends on $j$ and $m$. To ensure that the right hand side of \eqref{E:q_term} lies in $X_{\text{rev}}$, we need to assume evenness of $V$ in both $x_1$ and $x_2$. In that case $\omega_n(k) = \omega_n(-k_1,k_2)=\omega_n(k_1,-k_2)$, cf. \eqref{E:sym1}. This implies, first of all, 
\[
\begin{split}
\pa_{k_1}^2\omega_{n_j}(k^{(j)}) &= \pa_{k_1}^2\omega_{n_j}(k^{(j')})=\pa_{k_1}^2\omega_{n_j}(k^{(j'')}),\quad \pa_{k_2}^2\omega_{n_j}(k^{(j)}) = \pa_{k_2}^2\omega_{n_j}(k^{(j')})=\pa_{k_2}^2\omega_{n_j}(k^{(j'')}),\\
\pa_{k_1}\pa_{k_2}\omega_{n_j}(k^{(j)}) &= -\pa_{k_1}\pa_{k_2}\omega_{n_j}(k^{(j')})=-\pa_{k_1}\pa_{k_2}\omega_{n_j}(k^{(j'')}).
\end{split}
\]
Hence, the quadratic part in \eqref{E:q_term} satisfies \eqref{E:S1}, i.e.
\[
\begin{split}
&\left(-\tfrac{1}{2}\pa_{k_1}^2\omega_{n_j}(k^{(j)})p_1^2-\tfrac{1}{2}\pa_{k_2}^2\omega_{n_j}(k^{(j)})p_2^2-\pa_{k_1}\pa_{k_2}\omega_{n_j}(k^{(j)})p_1p_2\right)\hat{b}_j(p) = \\
&\qquad \quad s_1 \left(-\tfrac{1}{2}\pa_{k_1}^2\omega_{n_j}(k^{(j'')})p_1^2-\tfrac{1}{2}\pa_{k_2}^2\omega_{n_j}(k^{(j'')})p_2^2-\pa_{k_1}\pa_{k_2}\omega_{n_j}(k^{(j'')})p_1p_2\right)\bar{\hat{b}}_{j''}(p_1,-p_2)
\end{split}
\]
since $\bar{\hat{b}}_{j''}(p_1,-p_2) = \hat{\bar{b}}_{j''}(-p_1,p_2)$

For the first term in \eqref{E:q_term} note first that 
\beq\label{E:om_rev}
\omega_{n_j}(k^{(j)}+\eps p-m)\hat{b}_j(p) = \omega_{n_j}\left(\bspm k_1^{(j)}\\ -k_2^{(j)}\espm +\eps \bspm p_1\\ -p_2\espm -\bspm m_1\\ -m_2\espm\right)\bar{\hat{b}}_{j''}(p_1,-p_2).
\eeq
If $k_2^{(j)}<1/2$, then $(k_1^{(j)}, -k_2^{(j)})^T = k^{(j'')}$ and $m_2=0$ so that $\omega_{n_j}(k^{(j)}+\eps p-m)\hat{b}_j(p) = \omega_{n_j}(k^{(j'')}+\eps \bspm p_1\\-p_2 \espm- m)\bar{\hat{b}}_{j''}(p_1,-p_2)$,
which is \eqref{E:S1}.

If $k_2^{(j)}=1/2$, then $k^{(j)} = k^{(j'')}$ and we have two cases, either $m_2=0$ or $m_2=1$. For $m_2=0$
we get $-k_2^{(j)}+m_2=-1/2=k_2^{(j)}-1$ so that 
\beq\label{E:rev_bdry}
\omega_{n_j}(k^{(j)}+\eps p-m)\hat{b}_j(p) = \omega_{n_j}\left(k^{(j'')}+\eps \bspm p_1\\-p_2\espm -m''\right)\bar{\hat{b}}_{j''}(p_1,-p_2)
\eeq
 with $m''=(m_1,1)\in M_{j''}=M_j$. Finally, for $m_2=1$ one has $-k_2^{(j)}+m_2=1/2=k_2^{(j)}-0$ so that \eqref{E:rev_bdry} holds with $m'' = (m_1,0)\in M_{j''}=M_j$.
  
The last term on the right hand side of \eqref{E:remainder}  commutes with $S$ because it is a symmetric cut-off of the nonlinear part of the CME operator. 

The term $\tilde{g}_{n_j}(k)$ is more complicated because it involves $p_{n_j}(k;x)$ in the convolutions instead of the $k-$independent $p_{n_j}(k^{(j)};x)$. After solving \eqref{spart} for the regular part $\tilde{\psi}$ in terms of $\hB$, 
we obtain that $\tilde{g}_{n_j}(k)$ consists of the terms
\beq\label{E:I_term}
\hat{I}^{\alpha \beta \gamma j} = \sum_{m\in M_j}\chi_{D_j}(k+m)\langle (\xi_\alpha *_{\Bsm}\xi_\beta *_{\Bsm}\xi_\gamma^c)(k;\cdot), p_{n_j}(k;\cdot)\rangle_{L^2(\P^2)}
\eeq
and of higher order convolutions of the $\xi_i$'s coming from $\tilde{\psi}$, 
which starts with cubic convolutions.  
Here $\xi_\alpha = p_{n_\alpha}(k;x) \sum_{m\in M_\alpha} \chi_{D_\alpha}(k+m)\hat{B}_\alpha\left(\tfrac{k+m-k^{(\alpha)}}{\eps}\right)$ and $\xi_\alpha^c = \overline{p_{n_\alpha}}(k;x)\sum_{m\in M_\alpha} \chi_{-D_\alpha}(k-m)\hat{\bar{B}}_\alpha\left(\tfrac{k-m+k^{(\alpha)}}{\eps}\right)$. It suffices to show reversibility for the simplest convolution type \eqref{E:I_term}.
We have
\[\hat{I}^{\alpha \beta \gamma j} =\sum_{m\in M_j} \sum_{n,o,q\in \CA_{\alpha,\beta,\gamma,j,m}}\tilde{w}_{noqm}^{\alpha \beta \gamma j}(k)\]
with 
\[
\begin{split}
\tilde{w}_{noqm}^{\alpha \beta \gamma j}(k) = &\chi_{D_j}(k+m)\int\limits_{\T^2}\int\limits_{\T^2}
\chi_{D_\alpha}(k{-}r{+}n)\hat{B}_\alpha\left(\tfrac{k{-}r{+}n{-}k^{(\alpha)}}{\eps}\right)
\chi_{D_\beta}(r{-}s{+}o)\hat{B}_\beta\left(\tfrac{r{-}s{+}o{-}k^{(\beta)}}{\eps}\right)\times \\
&\times \ \chi_{-D_\gamma}(s{-}q)\hat{\bar{B}}_\gamma\left(\tfrac{s{-}q{+}k^{(\gamma)}}{\eps}\right) \langle p_{n_\alpha}(k-r;\cdot)p_{n_\beta}(r-s;\cdot)\overline{p_{n_\gamma}}(s;\cdot), p_{n_j}(k;\cdot)\rangle_{L^2(\P^2)}\dd s \dd r.
\end{split}
\]
Employing the same change of variables as in \reff{E:NL_term_conv} and using \reff{E:noq_cond2}, we get
\beq\label{E:mu_term}
\begin{split}
\hat{\mu}^{\alpha \beta \gamma j}_{noqm}(p) & := \tilde{w}_{noqm}^{\alpha \beta \gamma j}(\eps p +k^{(j)}-m) = \chi_{D_{\eps^{r-1}}}(p)\int\limits_{\Omega_1}\int\limits_{\Omega_2}
\chi_{D_{\eps^{r-1}}}(p-\tilde{r})\hat{B}_\alpha(p-\tilde{r})
\chi_{D_{\eps^{r-1}}}(\tilde{r}-\tilde{s})\hat{B}_\beta(\tilde{r}-\tilde{s})\times \\
& \times \ \chi_{D_{\eps^{r-1}}}(\tilde{s})\hat{\bar{B}}_\gamma(\tilde{s}) \langle p_{n_\alpha}(k^{(\alpha)}-n+\eps(p-\tilde{r});\cdot)p_{n_\beta}(k^{(\beta)}-o+\eps(\tilde{r}-\tilde{s});\cdot) \times \\ 
& \times \overline{p_{n_\gamma}}(-k^{(\gamma)}+q+\eps \tilde{s};\cdot), p_{n_j}(\eps p +k^{(j)}-m;\cdot)\rangle_{L^2(\P^2)}\dd \tilde{s} \dd \tilde{r},
\end{split}
\eeq
where $\Omega_1=D_{2\eps^{r-1}}\cap \tfrac{\T^2-k^{(\beta)}+k^{(\gamma)}+o-q}{\eps}$ and $\Omega_2=D_{\eps^{r-1}}\cap\tfrac{\T^2+k^{(\gamma)}-q}{\eps}$.

For the symmetry $S$ we need to consider $\hat{\bar{\mu}}^{\alpha \beta \gamma j}_{noqm}(-p_1,p_2)=\bar{\hat{\mu}}^{\alpha \beta \gamma j}_{noqm}(p_1,-p_2)$. Assuming that ${\bf b}$ (and ${\bf A}$) satisfies \eqref{E:S1}, we get that also ${\bf B}$ does. Thus, using $\bar{\hat{B}}_i\left(\bspm p_1\\-p_2\espm-\tilde{r}\right)=\hat{\bar{B}}_i\left(\bspm -p_1\\p_2\espm+\tilde{r}\right)=s_1\hat{B}_{i''}\left(p-\bspm \tilde{r}_1\\-\tilde{r}_2\espm\right)$ and the change of variables $\tilde{r} \rightarrow (\tilde{r}_1,-\tilde{r}_2)$, $\tilde{s} \rightarrow (\tilde{s}_1,-\tilde{s}_2)$, we get
\beq\label{E:mu_refl}
\begin{split}
\bar{\hat{\mu}}^{\alpha \beta \gamma j}_{noqm}(p_1,-p_2) &= \chi_{D_{\eps^{r-1}}}(p)s_1^3\int\limits_{\Omega_1'}\int\limits_{\Omega_2'}
\chi_{D_{\eps^{r-1}}}(p-\tilde{r})\hat{B}_{\alpha''}(p-\tilde{r})
\chi_{D_{\eps^{r-1}}}(\tilde{r}-\tilde{s})\hat{B}_{\beta''}(\tilde{r}-\tilde{s})\chi_{D_{\eps^{r-1}}}(\tilde{s})\hat{\bar{B}}_{\gamma''}(\tilde{s}) \times \\
& \times \left\langle \overline{p_{n_\alpha}}\left(k^{(\alpha)}-n+\eps\left(\bspm p_1\\-p_2\espm - \bspm \tilde{r}_1 \\ -\tilde{r}_2\espm\right);\cdot\right)\overline{p_{n_\beta}}\left(k^{(\beta)}-o+\eps\left(\bspm \tilde{r}_1 \\ -\tilde{r}_2\espm - \bspm \tilde{s}_1 \\ -\tilde{s}_2\espm \right);\cdot\right) \times \right.\\ 
& \left. \times p_{n_\gamma}\left(-k^{(\gamma)}+q+\eps \bspm \tilde{s}_1 \\  -\tilde{s}_2\espm;\cdot\right), \overline{p_{n_j}}\left(k^{(j)}-m+\eps \bspm p_1 \\ -p_2\espm;\cdot\right)\right \rangle_{L^2(\P^2)}\dd \tilde{s} \dd \tilde{r},
\end{split}
\eeq
where $\Omega_1'=D_{2\eps^{r-1}}\cap \eps^{-1}\left(\T^2-\bspm k_1^{(\beta)}\\ -k_2^{(\beta)}\espm+\bspm k_1^{(\gamma)}\\ -k_2^{(\gamma)}\espm+\bspm o_1\\-o_2\espm-\bspm q_1\\-q_2\espm\right)$, and $\Omega_2'=D_{\eps^{r-1}}\cap\eps^{-1}\left(\T^2+\right.$ $\left. + \bspm k_1^{(\gamma)}\\-k_2^{(\gamma)}\espm-\bspm q_1\\-q_2 \espm\right)$.
Finally, we use the Bloch function symmetry $\overline{p_n}((k_1,-k_2);x)=p_n(k;(2\pi-x_1,x_2))$. In our case this means
\[\overline{p_{n_\alpha}}\left(k^{(\alpha)}-n+\eps\left(\bspm p_1\\-p_2\espm - \bspm \tilde{r}_1 \\-\tilde{r}_2\espm\right);x\right)= p_{n_\alpha}\left(\bspm k^{(\alpha)}_1 \\ -k^{(\alpha)}_2\espm - \bspm n_1\\-n_2\espm +\eps (p-\tilde{r});\bspm 2\pi-x_1\\x_2\espm \right).\]
If $k_2^{(\alpha)}\in (-1/2,1/2)$, then $n_2=0$ and we have for $k^{(\alpha'')}=\bspm k^{(\alpha)}_1 \\ -k^{(\alpha)}_2\espm$ and $n_{\alpha''}=n_{\alpha}$
\[p_{n_\alpha}\left(\bspm k^{(\alpha)}_1 \\ -k^{(\alpha)}_2\espm - \bspm n_1\\-n_2\espm +\eps (p-\tilde{r});\bspm 2\pi-x_1\\x_2\espm \right) = p_{n_{\alpha''}}\left(k^{(\alpha'')}- n +\eps (p-\tilde{r});\bspm 2\pi-x_1\\x_2\espm \right).\]
If $k_2^{(\alpha)}=1/2$, then $n_2\in\{0,1\}$ and $\bspm k^{(\alpha)}_1 \\ -k^{(\alpha)}_2\espm - \bspm n_1\\-n_2\espm  = k^{(\alpha)}-\tilde{n}$ for some $\tilde{n}\in M_\alpha$ with $\tilde{n}\neq n$. Thus 
\[p_{n_\alpha}\left(\bspm k^{(\alpha)}_1 \\ -k^{(\alpha)}_2\espm - \bspm n_1\\-n_2\espm +\eps (p-\tilde{r});\bspm 2\pi-x_1\\ x_2\espm \right) = p_{n_\alpha}\left(k^{(\alpha)}- \tilde{n} +\eps (p-\tilde{r});\bspm 2\pi-x_1\\ x_2\espm \right)\]
so that $\alpha''=\alpha$.

Performing the same analysis for the Bloch functions $p_{n_\beta}, p_{n_\gamma}$ and $p_{n_j}$ in \eqref{E:mu_refl} and using $s_1^3=s_1$, we get
\[\hat{\bar{\mu}}^{\alpha \beta \gamma j}_{noqm}(-p_1,p_2)=s_1\hat{\mu}^{\alpha'' \beta'' \gamma'' j''}_{\tilde{n}\tilde{o}\tilde{q}\tilde{m}}(p),\]
where some of the doubly primed and `tilded' indices may equal the bare ones and where $\tilde{n}\in M_\alpha, \tilde{o}\in M_\beta, \tilde{q}\in M_\gamma,$ and $\tilde{m}\in M_j$. After the sum in $n,o,q$ and $m$, we get
\[\hat{I}^{\alpha \beta \gamma j}(-p_1,p_2)=s_1\hat{\bar{I}}^{\alpha'' \beta'' \gamma'' j''}(p).\]
In conclusion ${\bf N} = S{\bf N}$ and \reff{E:inherit} holds.

With $\text{Ker}~\bJ=\{\pa_{y_1}{\bf A}, \pa_{y_2}{\bf A},\ri 
{\bf A}\}$ the operator $\hJ_\eps$ has an $\CO(1)$ bounded inverse
on $\xrev$, i.e., with a bound independent of $\eps$. 
Moreover,  $\|\hJ_\eps^{-1} \hR\|_{L^2_{s}(D_{\eps^{r-1}})}
\le C\|\hR\|_{L^2_{s-2}(D_{\eps^{r-1}})}$, which is why, 
e.g., estimating $\eps^{-2} q_j^{(3)}(\eps p)\hat{B}_j(p)$ in $L^2_{s-2}$ is 
sufficient, cf.~\reff{rem1e}. Thus, \reff{beqc} can now be solved producing
\bthm \label{p-thm}
Let $s>1$ and let $V$ be even in $x_1$ as well as $x_2$. There exists an $\eps_0>0$ such that for all $0<\eps<\eps_0$ 
the following holds. Let $\om=\om_*+\eps^2\Om$ be in a band gap, 
let $\bA$ be a reversible non-degenerate solution of the CME 
\reff{E:CME_gen_iFourier} 
with $\hA\in L^2_q(\R^2)$ for all $q\ge 0$, 
and let $0<r<2/3$. 
Then there exists a $C>0$ and a solution $\hB$ of the eCME such 
that 
\beq \label{E:BA_diff}
\|\hB-\hA\|_{L^2_s(D_{\eps^{r-1}})}\le C\eps^{\tilde{r}}\qquad 
\tilde{r}=\min\{r,2-2r\}. 
\eeq
\ethm  

\bcor \label{C:est_fin} The solution $\phi$ constructed in Theorems \ref{T:diff_CME_ext} 
and \ref{p-thm} is a localized solution of \reff{E:GP_stat}, it is symmetric according to \reff{E:rev1} or \reff{E:rev2}, and 
\beq\label{E:est_fin}
\|\phi(\cdot)- \eps\sum_{j=1}^N A_j(\eps \cdot) u_{n_j}(k^{(j)};\cdot)\|_{H^s(\R^2)} 
\leq C \eps^{\tilde{r}}.  
\eeq
\ecor
\begin{proof} We first show that the reversibility of $\hB$ and the symmetry of Bloch functions provides a $\eps^{-1}\etls$ that satisfies \reff{E:rev1} or \reff{E:rev2}.
Let us work out explicitly only the first symmetry in  \reff{E:rev1}. Recall that
\[\etls(x) = \int_{\T^2}e^{\ri k\cdot x} \sum_{j=1}^N p_{n_j}(k;x)\sum_{m\in M_j}\chi_{D_j}(k+m)\hat{B}_j\left(\tfrac{k+m-k^{(j)}}{\eps}\right)\dd k.\]
Using $p_{n_j}(k;(-x_1,x_2))=\overline{p_{n_{j''}}}((k_1,-k_2);x)$ with $n_{j''}=n_j$, $\hat{B}_j(p) = s_1\hat{\bar{B}}_{j''}(-p_1,p_2)=s_1\bar{\hat{B}}_{j''}(p_1,-p_2)$ together with the definition
$k^{(j'')}=(k^{(j)}_1,-k^{(j)}_2)$ if $k_2^{(j)}<\tfrac{1}{2}$ and $k^{(j'')}=k^{(j)}$ if $k_2^{(j)}=\tfrac{1}{2}$, we get for $k_2^{(j)}<\tfrac{1}{2}$
\[
\etls(-x_1,x_2) = s_1 \int_{\T^2}\overline{e^{\ri (k_1,-k_2)\cdot x}} \sum_{j=1}^N\overline{p_{n_{j''}}}((k_1,-k_2);x)\sum_{m\in M_j}\chi_{D_j}(k+m)\bar{\hat{B}}_{j''}\left(\tfrac{\bspm k_1 \\-k_2\espm +\bspm m_1\\-m_2 \espm -k^{(j'')}}{\eps}\right)\dd k.
\]
Because $m_2=0 \ \forall m\in M_j$ when $k_2^{(j)}<\tfrac{1}{2}$ and since $M_j=M_{j''}$, and $\chi_{D_j}(k+m)=\chi_{D_{j''}}\left(\bspm k_1\\-k_2 \espm +m\right)$, we have after the change of variables $k \rightarrow (k_1,-k_2)$
\[\etls(-x_1,x_2) = s_1 \int_{\T^2}\overline{e^{\ri k\cdot x}} \sum_{j=1}^N \overline{p_{n_{j''}}}(k;x)\sum_{m\in M_{j''}}\chi_{D_{j''}}(k+m)\bar{\hat{B}}_{j''}\left(\tfrac{k +m-k^{(j'')}}{\eps}\right)\dd k = s_1\overline{\etls}(x).
\]

For $k_2^{(j)}=\tfrac{1}{2}$ we have
\[
\etls(-x_1,x_2) = s_1 \int_{\T^2}\overline{e^{\ri (k_1,-k_2)\cdot x}} \sum_{j=1}^N\overline{p_{n_{j''}}}((k_1,-k_2);x)\sum_{m\in M_j}\chi_{D_j}(k+m)\bar{\hat{B}}_{j''}\left(\tfrac{\bspm k_1 \\-k_2\espm +\bspm m_1\\-m_2 \espm -k^{(j'')} + \bspm 0\\1\espm}{\eps}\right)\dd k
\]
because $\bspm k^{(j)}_1 \\ -k^{(j)}_2\espm = k^{(j'')} - \bspm 0\\1\espm$. Next, using firstly $\bspm m_1\\-m_2 \espm + \bspm 0\\1\espm = \tilde{m}$ with some $\tilde{m}=M_j, \tilde{m}\neq m$, and secondly $\chi_{D_j}(k+m)=\chi_{D_{j''}}\left(\bspm k_1\\-k_2 \espm +\tilde{m}\right)$, we arrive (after the change of variables $k \rightarrow (k_1,-k_2)$) at
\[
\etls(-x_1,x_2) = s_1 \int_{\T^2}\overline{e^{\ri k\cdot x}} \sum_{j=1}^N \overline{p_{n_{j''}}}(k;x)\sum_{\tilde{m}\in M_{j''}}\chi_{D_{j''}}(k+\tilde{m})\bar{\hat{B}}_{j''}\left(\tfrac{k +\tilde{m}-k^{(j'')}}{\eps}\right)\dd k = s_1\overline{\etls}(x).
\]

Next, $\psi$ in 
\eqref{spart} inherits the symmetry, so that $\phi$ is symmetric.
The estimate \eqref{E:est_fin} follows from the triangle inequality with \eqref{E:bound_CME_ext} and \eqref{E:BA_diff}.
\end{proof}

\brem\label{R:formal_rate}
{\rm In \eqref{E:BA_diff} we use $\tilde{r}= \min\{r,2-2r\}$ although below \eqref{E:CME_integr_form} we defined $\tilde{r}:= \min\{r,2-2r,1\}$. This is because 
for $0<r<2/3$ we have $\min\{r,2-2r,1\}=\min\{r,2-2r\}$.

The optimal value of $\tilde{r}$ is $\tilde{r}=2/3$ attained at
  $r=2/3$. Based on
  the formal asymptotic expansion in \eqref{E:ansatz_Bloch} and \eqref{E:psi_til_1}, we see that
  the next order term 
$\tilde{\psi}^{(1)}$ (just like $\tilde{\psi}^{(0)}$) consists of terms of
  the type $\hat{F}\left(\frac{k-k^{(j)}}{\eps}\right)q(k^{(j)};x)$, where $F$
  is an envelope and $q$ a periodic carrier wave. $\psi^{(1)}$, therefore,
  consists of terms $\eps^2F(\eps x) q(k^{(j)};x)e^{\ri 2\pi k^{(j)}\cdot x}$
  and $\|\psi^{(1)}\|_{H^s(\R^2)} = \CO(\eps)$. As a result, the formal
  asymptotics predict $\eps^1$ convergence. 
 Thus, while the estimate \eqref{E:est_fin} 
  guarantees convergence of the CME approximation, it does not appear 
to be sharp. But if  all third derivatives of 
$\omega_{n_j}$ vanish at $k^{(j)}$,
  like for separable potentials \cite{DPS08}, we have $\tilde{r} = \min\{2r,2-2r,1\}$ with $0<r<2/3$ and the optimal value is
  $\tilde{r}=1$ attained at $r=1/2$. It is, however, unclear which non-separable potentials result in vanishing third derivatives of the bands at gap edge extrema.   
  }\erem

\begin{figure}[h!]
  \begin{center}
    \includegraphics[width=8.5cm, height=31mm]{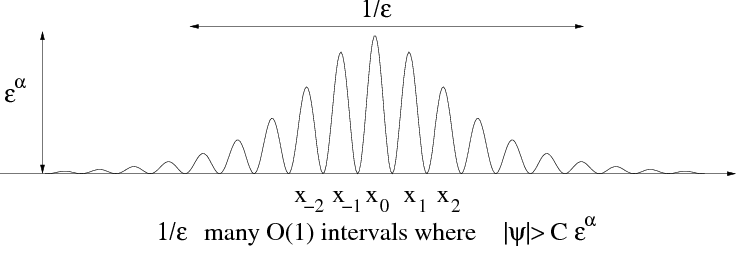}
  \end{center}
  \caption{Sketch of the argument for $\|\psi\|_{H^s}=\CO(\eps^\beta)
\Ra \|\psi\|_{L^\infty}
=\CO(\eps^{\beta+1})$.}
  \label{ffig}
\end{figure}

\brem\label{fr2rem}
{\rm As said in the previous remark, the formal asymptotics predict that 
the error $\psi$ in \eqref{E:Lyap_Schm} has 
the form 
$$
\psi(x)=\eps^2F(\eps x)w(x)
$$ 
with $F\in H^q(\R^2)$ for all $q\ge 0$ 
and $w(x)\in C^{\lceil s\rceil}_b(\R^2)$ with 
$w(2\pi,x_2) = \er^{2\pi\ri k_1} w(0,x_2), \ w(x_1,2\pi) =
    \er^{2\pi\ri k_2} w(x_1,0)$. In this case 
$$
\|\psi\|_{H^s}=\CO(\eps^\beta)\text{ implies }
\|\psi\|_{L^\infty}
=\CO(\eps^{\beta+1}).
$$ 
To see this, assume that $\|\psi\|_{L^\infty}
\ge C_1\eps^\al$ with $\al<1+\beta$, i.e., $|\psi(x_0)|\ge C_1\eps^\al$ 
for some $x_0\in\R^2$. Since $\psi$ is continuous, this implies 
$|\psi(x)|\ge \frac{C_1}{2}\eps^\al$ in some neigborhood of 
$x_0$ of diameter $\delta_1>0$, say. Moreover, since $F$ is continuous, 
$|\eps^2 F(\eps x)|\ge \frac{C_1}{2}\eps^\al$ for $|x|\le \del_2/\eps$. 
Therefore, for $\eps$ sufficiently small,  $|\psi(x)|\ge \frac{C_1}{4}
\eps^\al$ for $x$ in $\CO(\del_1)$ wide neighborhoods of 
$x_j=x_0+(2j_1 \pi, 2j_2\pi)$, where $j_1,j_2\in\Z$ and $|j_1|,|j_2|\le C_2/\eps$, 
see Fig.\ref{ffig} for a 1D sketch. 
Since in 2D there are at least $C_3\eps^{-2}$ such neighborhoods, we obtain 
\begin{align}
\|\psi\|_{H^s}&\ge \frac {C_3 C_1 \eps^{\al} \del}{4\eps}>C\eps^{\al-1}
\end{align}
which contradicts $\|\psi\|_{H^s}\le C\eps^\beta$ as $\eps\ra 0$ 
due to $\al-1<\beta$. 

To make this argument rigorous for the error $\psi=
\phi- \eps\sum_{j=1}^N A_j(\eps \cdot) u_{n_j}(k^{(j)};\cdot)$,
we could split off the next term in the formal asymptotic 
expansion and show that the remainder is of higher order, i.e., 
$\psi=\eps^2\psi^{(1)}+\eps^3\psi_*$. Then we can estimate 
$\|\eps^3\psi_*\|_{H^s}=\CO(\eps^{3-2r})$ using the analysis 
from \S\ref{S:diff_CME}, but here we refrain from these tedious 
calculations. 
}\erem


\section{Numerical Results on Reversible Gap Solitons}
\label{S:num} 
We numerically 
compute some representative cases of gap solitons and their asymptotic
envelope approximations $\eps \phi^{(0)}(x) = \eps \sum_{j=1}^N A_j(\eps
x)u_{n_j}(k^{(j)};x)$. We do not attempt to provide an exhaustive study of possible GS solutions but rather select only 
several cases to corroborate our analysis.
Namely, we select GSs bifurcating from the
edges $s_2$ and $s_5$. The latter case is of particular interest as it
features a situation whose occurrence is impossible for separable potentials $V(x)$.  To
our knowledge this case has not been studied before and the presented GSs are
novel. We also check the reversibility and non-degeneracy conditions 
which are sufficient for persistence, see \S\ref{S:persist}. 
In addition, we compute the
convergence rate in $\eps$, i.e., in the square root of the distance to the gap edge, of the error $\|\phi_{\text{GS}}^{\text{num}}-\eps 
\phi^{(0)}\|_{H^2}$.  

A 4th
order centered finite difference discretization is used for 
\eqref{E:GP_stat}. The computational domain is a square $x\in [-D_{\text{GS}}/2,D_{\text{GS}}/2]^2$
selected large enough so that the asymptotic approximation $\eps
\phi^{(0)}(x)$ of the GS is well-decayed at the boundary and zero Dirichlet
boundary conditions are then used. Equation \eqref{E:GP_stat}
is then solved via Newton's iteration using $\eps \phi^{(0)}$ as the initial
guess. The computational domain is in practice reduced to its quarter using
the corresponding reversibility symmetry.

\subsection{Gap Solitons near $\omega=s_2$}\label{S:numerics_s2}

Near the edge $\omega{=}s_2$ we limit our attention to real, even GSs and to symmetric vortices of charge 1.
As the coupled mode system near $\omega=s_2$ is a scalar nonlinear
Schr\"{o}dinger equation, see \S\ref{cmebloch-sec}, one can search for
solutions of the form $A(y)=R(r)e^{\ri m\theta}$, where
$r=\frac{1}{\sqrt{\alpha}}\sqrt{y_1^2+y_2^2}, \theta = \text{arg}(y_1+\ri y_2),$
and $m\in \N$. We choose $R>0$ and $m=0$ corresponding to the so called 
Townes soliton, and $m=1$ corresponding to a vortex of charge 1.  The function
$R(r)$ satisfies the ODE
\begin{equation}
  R''+\frac{1}{r}R'+\Omega R-\frac{m^2}{r^2}R-\sigma
  \gamma R^3=0, \label{ODE-radial}
\end{equation}
where $R(0) > 0$, $R'(0)=0$ for $m=0$ and $R(0) = 0$, $R'(0) > 0$ for
$m=1$. For $m\neq 0$ the initial-value problem for the ODE (\ref{ODE-radial})
is ill-posed but can be turned into a well-posed one via the transformation
$Q=r^{-m} R(r)$ leading to
\begin{equation}\label{E:Q_equ}
  Q''+\frac{2m+1}{r}Q'+\Omega Q-\sigma \gamma r^{2m}
  Q^3=0
\end{equation}
with $Q(0) > 0$, such that $R(r) \sim r^{|m|}$ as $r\rightarrow 0$. We solve
equation \eqref{E:Q_equ} numerically via a shooting method searching for
$Q(r)$ vanishing as $r\rightarrow \infty$.

For $m=0$ we have the reversibility $A(-y_1,y_2)=A(y_1,-y_2)=A(y)$,
which is the same as \eqref{E:rev1A} with $s_1=s_2=1$ since $A$ is
real. The non-degeneracy condition on ${\bf J}$ in Theorem \ref{p-thm} is known to be satisfied by the positive ground state $A$ \cite{Kwong89,CGNT07} and conditions of this theorem are, therefore, satisfied.

\begin{figure}[!ht]
\begin{center}
    \includegraphics[scale=0.49]{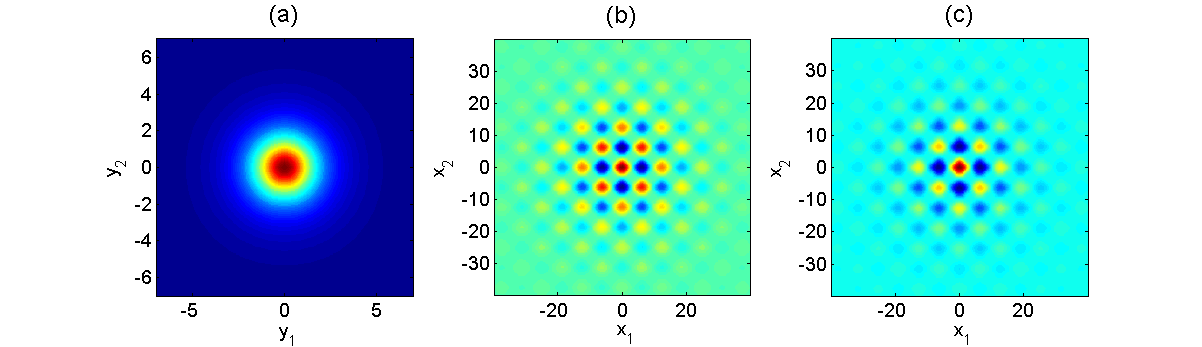}
  \end{center}
  \caption{Profiles of the even real GS at $\omega=s_2+\eps^2 \Omega, \eps =
    0.1, \Omega =1$ . (a) $A(y)$; (b) the corresponding leading-order GS
    approximation $\eps A(y)u_1(M;x)$; (c) the numerically computed GS at
    $\omega=s_2+\eps^2 \Omega$.}
  \label{F:profiles_s2_real}
\end{figure}

\begin{figure}[h!]
  \begin{center}
    \includegraphics[scale=0.44]{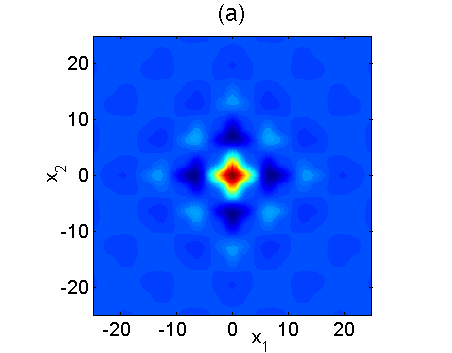}
\quad  \includegraphics[scale=0.44]{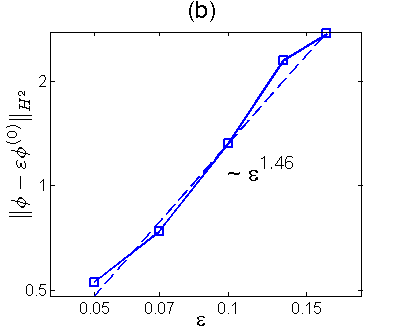}
  \end{center}
  \caption{(a) Profile of a GS corresponding to the even real family that
    bifurcates from $\omega=s_2$ in Fig. \ref{F:profiles_s2_real}. The plotted
    GS is deep inside the gap $(s_2,s_3)$ at $\omega\approx 1.78$
    (corresponding to $\eps \approx 0.28$). (b) $\eps$-convergence of the error 
$\|\phi-\eps\phi^{(0)}\|_{H^2(\R^2)}$.}
  \label{F:profile_s2_real_deep}
\end{figure}

Figure \ref{F:profiles_s2_real} shows the profiles of the envelope $A$, of
the asymptotic approximation $\eps \phi^{(0)}(x)$ and of the GS $\phi(x)$
computed via Newton's iteration on \eqref{E:GP_stat}. A GS deep inside the gap $(s_2,s_3)$ obtained via a homotopy continuation in $\omega$ from 
the $\phi(x)$ in Fig.~\ref{F:profiles_s2_real} is plotted in 
Fig.~\ref{F:profile_s2_real_deep}(a), while (b) shows 
the $\eps$-convergence of the approximation error. 
Here the $\eps^{1.46}$ convergence
 rate is better than the estimate proved in Corollary \ref{C:est_fin} and even 
better than the 
rate $\eps^1$ predicted by formal asymptotics in Rem. \ref{R:formal_rate}.


\begin{figure}[h!]
  \begin{center}
    \includegraphics[scale=0.48]{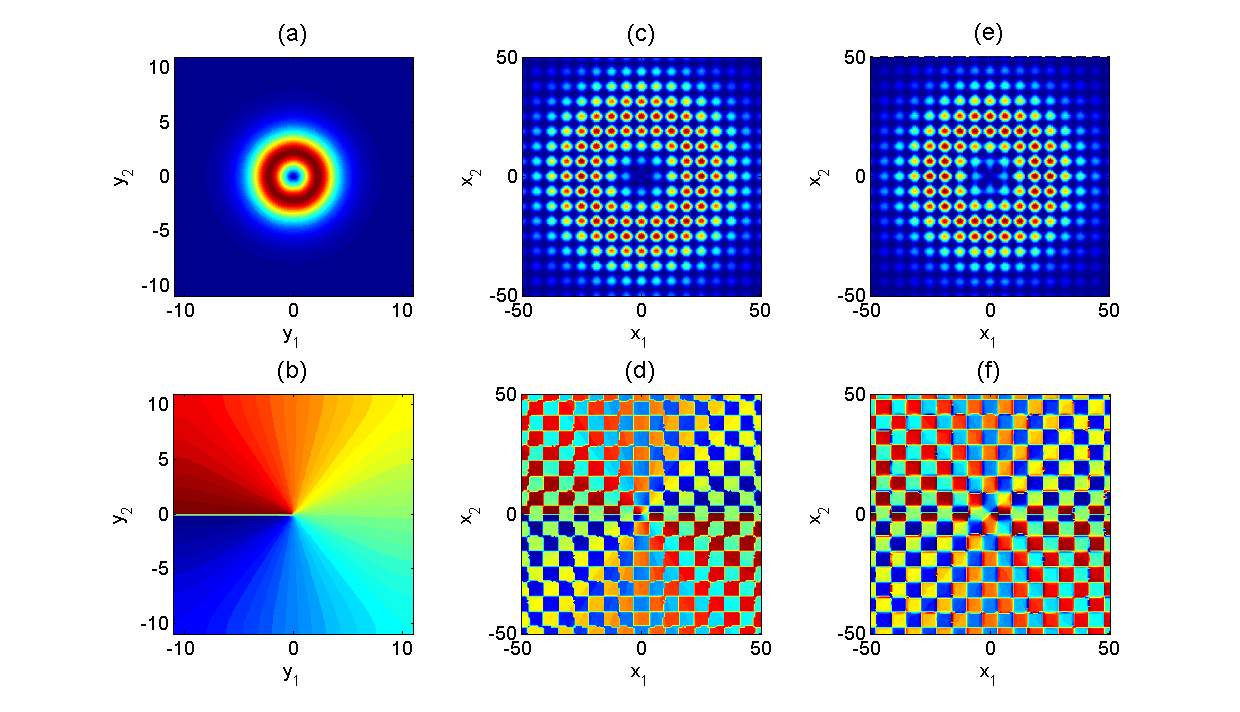}
  \end{center}
  \caption{Profiles of the vortex GS at $\omega=s_2+\eps^2 \Omega, \eps =
    0.09, \Omega =1$ . (a) and (b) modulus and phase of $A(y)$ resp.; (c) and
    (d) modulus and phase of the corresponding leading-order GS approximation
    $\eps A(y)u_1(M;x)$ resp.; (e) and (f) modulus and phase of the numerically
    computed GS at $\omega=s_2+\eps^2 \Omega$ resp.}
  \label{F:profiles_s2_vort}
\end{figure}
\begin{figure}[h!]
  \begin{center}
    \includegraphics[scale=0.46]{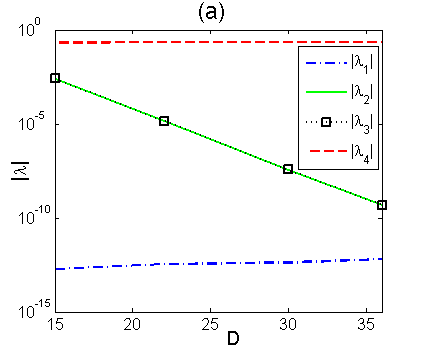}
\quad \includegraphics[scale=0.48]{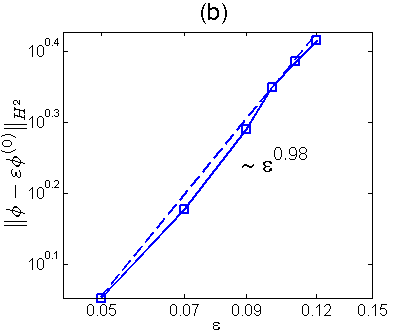}
  \end{center}
  \caption{(a) 
The four smallest eigenvalues of the Jacobian ${\mathbf J}$ in Theorem
\ref{p-thm} at the solution ${\bf A}$ in Fig.~\ref{F:profiles_s2_vort} 
(a-d) for a range of sizes of the computational domain. (b) $\eps$-convergence of the error 
$\|\phi-\eps\phi^{(0)}\|_{H^2(\R^2)}$.}
  \label{F:jac_ker_s2_vort}
\end{figure}

For $m=1$ the solution is complex and we have the reversibility
$A(-y_1,y_2) = -A(y_1,-y_2) = -\bar{A}(y)$, which is \eqref{E:rev1A}
with $s_1=-s_2=-1$. Figure \ref{F:profiles_s2_vort} 
shows the modulus and phase of the envelope $A$, of
the asymptotic approximation $\eps \phi^{(0)}(x) = \eps A(\eps x)u_1(M;x)$ and
of the computed GS. The non-degeneracy of the envelope is illustrated 
in Fig.~\ref{F:jac_ker_s2_vort}(a), which plots the 4 smallest 
eigenvalues (in modulus) of the
Jacobian operator ${\bf J}$ of the CMEs evaluated at the vortex $A$: 
3 eigenvalues converge to zero as the computational domain size grows 
while the fourth one stays bounded away from zero. 
Figure \ref{F:jac_ker_s2_vort}(b) presents 
the $\eps$-convergence of the approximation error
$\|\phi-\eps \phi^{(0)}\|_{H^2(\R^2)}$. The resulting convergence is very
close to $\eps^1$, which is the prediction based on formal asymptotics. 


\subsection{Gap Solitons near $\omega=s_5$}\label{S:numerics_s5}

We limit our attention here to gap solitons with real positive envelopes
satisfying the symmetries $A_1=A_3, A_2=A_4$ and
$A_1(-y_1,y_2)=A_1(y_1,-y_2)=A_1(-y_2,y_1)=A_2(y_1,y_2)$, which 
is \eqref{E:rev2A} with $s_1=s_2=1$ for each $A_j$. 
Such solutions of the CME system
\eqref{E:CME_s5} can be found by first setting $\alpha_2=0$ and computing
radially symmetric positive solutions $A_1=A_2=A_3=A_4=R(r)$, where
$r=\frac{1}{\sqrt{\alpha_1}}\sqrt{y_1^2+y_2^2}$, via a shooting method and
then performing a homotopy continuation in $\alpha_2$ on the system of the
first two equations in \eqref{E:CME_s5} employing the symmetry $A_1=A_3,
A_2=A_4$ up to the original value $\alpha_2= 0.096394$.

We normalize the Bloch functions 
$v_1(x):= u_6((k_c,k_c);x), v_2(x):= u_6((-k_c,k_c);x)$, 
$v_3(x):= u_6((-k_c,-k_c);x)$ and
$v_4(x):= u_6((k_c,-k_c);x)$ so that
\begin{gather}\label{E:s5_Bloch_normaliz} 
v_{2}(-x_1,x_2) = v_1(x_1,x_2), \ v_{3}(x_1,-x_2) = v_2(x_1,x_2) \ \text{and} \ v_{4}(-x_1,x_2) = v_3(x_1,x_2)
\quad (\text{see \S\ref{S:band_str}}).
\end{gather}  
This normalization implies that
$\eps\phi^{(0)}(x)$ is real and even in both variables.
These symmetries are used to reduce the computational domain to one
quadrant and restrict to the real arithmetic. 

\begin{figure}[h!]
  \begin{center}
    \includegraphics[scale=0.49]{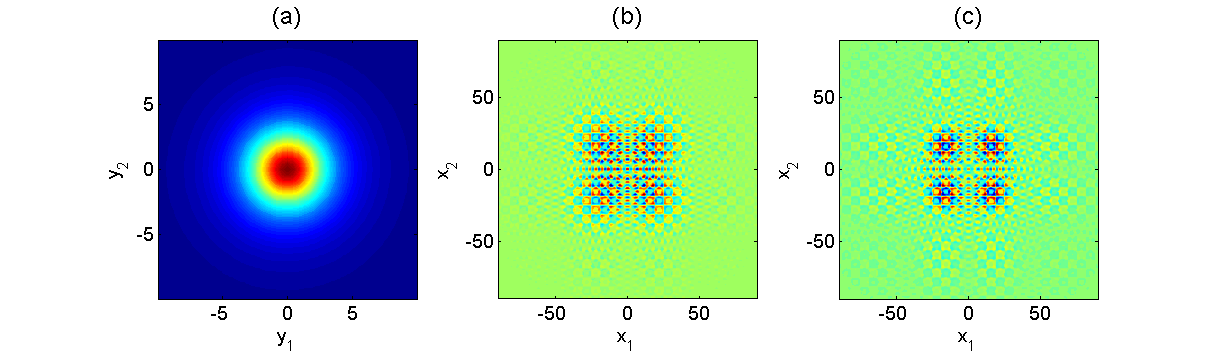}
  \end{center}
  \caption{Profiles of the even, real GS at $\omega=s_5+\eps^2 \Omega, \eps =
    0.1, \Omega =-1$ . (a) $A_1(y)$; (b) the corresponding leading-order GS
    approximation $\eps \sum_{j=1}^4 A_j(y)v_j(x)$; (c) the numerically
    computed GS at $\omega=s_5+\eps^2 \Omega$.}
  \label{F:profiles_s5_even}
\end{figure}

Figure \ref{F:profiles_s5_even} shows the envelope $A_1(y)$, the GS
approximation $\eps \phi^{(0)}$ and the computed GS $\phi$. 
The envelope $A_1(y)$ in Fig.~\ref{F:profiles_s5_even} 
is not radially symmetric due to the mixed derivative
$\pa_{y_1}\pa_{y_2}$ in \eqref{E:CME_s5}, but looks radially 
symmetric because the
coefficient $\alpha_2$ is relatively small ($\alpha_2\approx 0.0964$). 
Profiles of $A_2,\ldots, A_4$ are not plotted as they can be obtained 
from $A_1$ via the above mentioned symmetries. 

\begin{figure}[h!]
  \begin{center}
    \includegraphics[scale=0.48]{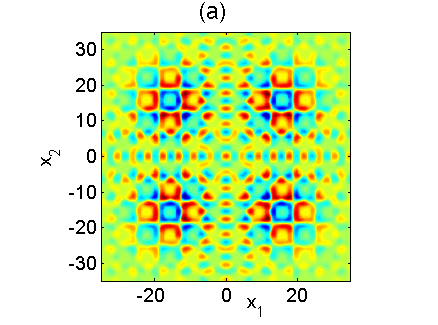}
\quad \includegraphics[scale=0.46]{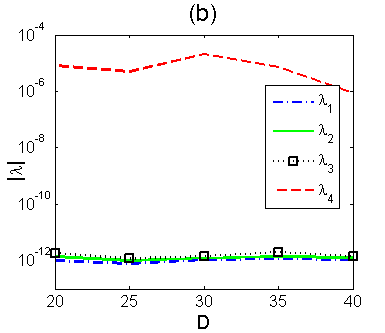}
\quad   \includegraphics[scale=0.46]{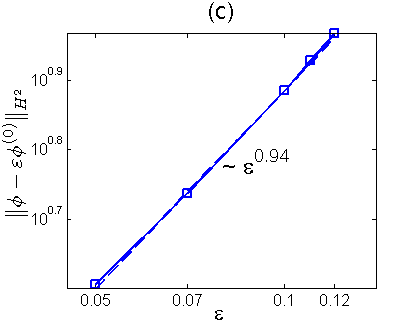}
  \end{center}
  \caption{(a) Detail of the profile in Fig.~\ref{F:profiles_s5_even} (c). 
(b) the non-degeneracy of ${\bf A}$. (c) $\eps$-convergence of the error 
$\|\phi-\eps\phi^{(0)}\|_{H^2(\R^2)}$.}
  \label{F:profile_s5_even_detail}
\end{figure}

A closer look at the structure of $\phi$ near the origin, an illustration 
of  the non-degeneracy of ${\bf A}$, and the $\eps$-convergence of the 
approximation error are 
provided in Fig.~\ref{F:profile_s5_even_detail}. 
The obtained rate is about $\eps^{0.94}$, which is
 once again close to the rate $\eps^1$ predicted by the formal 
asymptotics. 

\section{Conclusions}\label{S:conc}
We have derived systems of Coupled Mode Equations (CME) 
which approximate stationary gap solitons (GSs) of the 2D periodic Nonlinear 
Schr\"{o}dinger Equation/Gross Pitaevskii equation near a 
band edge. In contrast to \cite{DPS08} we do not assume separability 
of the periodic potential $V$. While in the case of a separable 
$V(x)$ \cite{DPS08} the derivation is possible in physical 
variables, here in general it has to be performed in Bloch variables. 
We have rigorously proved via the Lyapunov-Schmidt
reduction that reversible non-degenerate  solitons of the CME yield 
GSs of the Gross-Pitaevskii equation. We have also provided an $H^s(\R^2), \ s>1$ 
estimate on the approximation error showing that it is 
$\CO(\eps^{2/3})$ for GSs with the spectral parameter $\CO(\eps^2)$ 
close to the band edge. 
Our analysis requires some smoothness of $V$,
 namely $V\in H^{\lceil s\rceil-1+\delta}_{{\rm loc}}(\R^2), \delta>0$, 
and, in the persistence step, evenness of $V$.
The analysis has been corroborated
by numerical examples including one which features novel GSs bifurcating 
from a band edge Bloch wave located outside the set of vertices of the first 
Brillouin zone, which is impossible in the case of separable potentials.


\noi
{\bf Acknowledgement.}  The work of T. Dohnal is supported by the Humboldt
Research Fellowship. The authors wish to thank  Dmitry Pelinovsky and 
Guido Schneider for stimulating discussions.

\end{document}